\documentclass[notitlepage,12pt]{article}
\usepackage{fancyvrb, verbatim, listings}                                  
\usepackage[nodisplayskipstretch]{setspace} 
\usepackage[margin=2.54cm]{geometry}                                       
\usepackage[UKenglish,cleanlook]{isodate}                                  
\usepackage{fancyhdr} 
\usepackage[activate={true,nocompatibility},final,tracking=true,kerning=true,spacing=true,factor=1100,stretch=10,shrink=10]{microtype}
\microtypecontext{spacing=nonfrench}                                       
\usepackage{graphicx}                                                      
\usepackage[table,dvipsnames]{xcolor}                                      
\usepackage{float}                                                         
\usepackage[justify]{ragged2e}                                             
\usepackage{booktabs}                                                      
\usepackage[labelfont=bf]{caption} 
\usepackage{subcaption}
\usepackage[shortlabels]{enumitem} \setlist[enumerate]{leftmargin=0pt}     
\usepackage{tikz}
\usetikzlibrary{shapes,decorations,decorations.pathreplacing,arrows,calc,arrows.meta,fit,positioning}
\tikzset{
    auto,node distance =1 cm and 1 cm,semithick,
    state/.style ={ellipse, draw, minimum width = 0.7 cm},
    point/.style = {circle, draw, inner sep=0.04cm,fill,node contents={}},
    bidirected/.style={Latex-Latex,dashed},
    el/.style = {inner sep=2pt, align=left, sloped}
}
\usepackage{mathtools}                                                     
\usepackage{amssymb}                                                       
\usepackage{amsmath}                                                       
\usepackage{dsfont}                                                        
\usepackage{centernot}                                                     
\usepackage{siunitx} 
\usepackage{amsthm}                                                        
\newtheorem{theorem}{Theorem}                                              
\newtheorem{definition}{Definition}                                        
\renewcommand{\vec}[1]{\boldsymbol{\mathit{#1}}}                           
\newcommand{\mat}[1]{\boldsymbol{\mathit{#1}}}                             
\newcommand{\Prob}[1]{\Pr\left( #1 \right)}                         
\newcommand{\Probgiven}[2]{\Pr\left( #1 \, \middle\vert \, #2 \right)} 
\newcommand{\E}[2][]{\mathbb{E}_{#1} \left[ #2 \right]}                    
\newcommand{\Egiven}[3][]{\mathbb{E}_{#1} \left[ #2 \, \middle\vert \, #3 \right]} 
\newcommand{\Var}[2][]{\text{Var}_{#1} \left( #2 \right)}                  
\newcommand{\Cov}[1]{\text{Cov} \left( #1 \right)}                         
\newcommand{\indicator}[1]{\mathds{1}\left\{ #1 \right\}}                  
\newcommand{\partialdiff}[2][]{\frac{\partial#1}{\partial#2}}              
\renewcommand{\hat}[1]{\widehat{#1}}                                       
\renewcommand{\bar}[1]{\overline{#1}}                                      
\renewcommand{\tilde}[1]{\widetilde{#1}}                                   
\newcommand{\indep}{\, \raisebox{0.05em}{\rotatebox[origin=c]{90}{$\models$}} \,}
\definecolor{ao(english)}{rgb}{0.0, 0.5, 0.0}                              
\newcommand{\appendixref}[1]{\hyperref[#1]{Appendix~\ref*{#1}}}

\usepackage[longnamesfirst]{natbib}                                        
\usepackage[backref=page]{hyperref}                                        
\hypersetup{colorlinks=true, linkcolor=blue, citecolor=blue, filecolor=magenta, urlcolor=blue}
\def\sectionautorefname~#1\null{Section~#1\null}                           
\def\subsectionautorefname~#1\null{Subsection~#1\null}                     
\def\subsubsectionautorefname~#1\null{Subsubsection~#1\null}               
\def\equationautorefname~#1\null{Equation~(#1)\null}                       

\author{Senan Hogan-Hennessy\thanks{
    For helpful comments I thank
    Neil Cholli,
    Hyewon Kim,
    Jiwoo Kim,
    Luk\'a\u{s} Laff\'ers,
    Jiwon Lee,
    Douglas Miller,
    Lauri Kytömaa,
    Zhuan Pei,
    Brenda Prallon,
    and
    Evan Riehl.
    I thank seminar participants at
    Cornell University (2025),
    the European Economic Association Congress at Bordeaux Sciences \'Economiques (2025),
    the Econometrics World Congress at Seoul National University (2025),
    and
    the online Causal Data Science Meeting (2025)
    for helpful discussion.
    Some preliminary results previously circulated in an earlier version of the working paper ``The Direct and Indirect Effects of Genetics and Education.''
    Any comments or suggestions may be sent to me at \href{mailto:seh325@cornell.edu}{\nolinkurl{seh325@cornell.edu}}, or raised as an issue on the Github project,
    \url{https://github.com/shoganhennessy/mediation-natural-experiment}.
    } \\
    \vspace{0.1cm}
    Economics Department, Cornell University\footnote{
        Address: Uris Hall \#447, Economics Department, Cornell University NY 14853 USA.
    }
}
\title{Causal Mediation in Natural Experiments}
\date{
    First draft: 12 February 2025 \\
    This version:
    \today
}
\begin{document}
\clearpage
\maketitle
\thispagestyle{empty}
\begin{abstract}
    \noindent
Natural experiments are a cornerstone of applied economics, providing settings for estimating causal effects with a compelling argument for treatment randomisation, but give little indication of the mechanisms behind causal effects.
Causal Mediation (CM) is a framework for sufficiently identifying a mechanism behind the treatment effect, decomposing it into an indirect effect channel through a mediator mechanism and a remaining direct effect.
By contrast, a suggestive analysis of mechanisms gives necessary but not sufficient evidence.
Conventional CM methods require that the relevant mediator mechanism is as-good-as-randomly assigned; when people choose the mediator based on costs and benefits (whether to visit a doctor, to attend university, etc.), this assumption fails and conventional CM analyses are at risk of bias.
I propose an alternative strategy that delivers unbiased estimates of CM effects despite unobserved selection, using instrumental variation in mediator take-up costs.
The method identifies CM effects via the marginal effect of the mediator, with parametric or semi-parametric estimation that is simple to implement in two stages.
Applying these methods to the Oregon Health Insurance Experiment reveals a substantial portion of the Medicaid lottery's effect on subjective health and well-being flows through increased healthcare usage --- an effect that a conventional CM analysis would mistake.
This approach gives applied researchers an alternative method to estimate CM effects when an initial treatment is quasi-randomly assigned, but a mediator mechanism is not, as is common in natural experiments.

\vspace{0.5cm}

\vspace{0.1cm}

\end{abstract}

\newpage
\setcounter{page}{1}
\onehalfspacing
\noindent
Economists use natural experiments to credibly answer social questions, when an experiment was infeasible.
For example, does winning access to health insurance causally improve health and well-being \citep{finkelstein2008oregon}?
Natural experiments provide settings which can answer these questions, but give little indication of how these effects came about.
Causal Mediation (CM) aims to estimate the mechanisms behind causal effects, by estimating how much of the treatment effect operates through a proposed mediator mechanism.
For example, do causal gains from winning access to health insurance come mostly from physical use of healthcare, or plausible psychological gains from no longer having to worry about being uninsured?
This study of mechanisms behind causal effects broadens the economic understanding of social settings studied with natural experiments.
This paper shows that the conventional approach to estimating CM effects is inappropriate in a natural experiment setting, provides a theoretical framework for how bias operates, and develops an approach to correctly estimate CM effects under alternative assumptions.
These methods contrast the current practice in applied economics of providing suggestive evidence of mechanisms, which gives necessary but not sufficient conditions for the mechanisms behind a treatment effect.

In other disciplines --- particularly epidemiology, psychology, and medicine --- CM has become a standard empirical framework for decomposing causal effects into direct and indirect components \citep{imai2010identification}.
I refer to this established approach as ``conventional CM,'' which assumes both the treatment and the mediator mechanism are (quasi-)randomly assigned.
Applied economics has not typically adopted conventional CM, because mediators in economic applications --- such as schooling, labour supply, or healthcare use --- are choice variables decided by individuals' costs and benefits, likely violating this assumption.
Nevertheless, economists often pursue the same goal informally: 
they present descriptive and suggestive evidence on plausible mediating mechanisms, or occasionally test whether a treatment effect remains after controlling for a plausible mediator mechanism.
The first approach gives necessary but not sufficient evidence on the mediating mechanisms, and the second is a conventional CM analysis --- despite rarely being named as such by economists.

This paper starts by considering how conventional CM methods (whether applied explicitly or implicitly) perform in a natural experiment setting.
Conventional CM methods rely on assuming the initial treatment, and the subsequent mediator mechanism, are both quasi-randomly assigned \citep{imai2010identification}.
Assuming the mediator is as-good-as-randomly assigned requires either (1) selection is fully captured by observed control variables, or (2) decisions are effectively random.
While these assumptions can be plausible when a mediator is directly manipulated, or the data include extremely rich control variables, they are not credible in most economic applications; when take-up decisions reflect unobserved costs and benefits, mediator assignment is likely not random.
For example, in the Oregon Health Insurance Experiment, those who got off the wait-list were randomly chosen by a lottery, but made the choice to visit healthcare in the following year of their own free will; this choice considered their own individual costs and benefits, and thus was not a random choice.
I formally derive the selection bias that arises from this non-random assignment, and show through simulations that these biases can be large in settings consistent with standard economic models of selection.
This result explains why conventional CM methods can yield misleading conclusions in applied economics, and motivates the need for a framework consistent with quasi-experimental causal reasoning.

Conventional CM's identifying assumptions are at odds with selection based on costs and benefits, so I import methods grounded in labour economic theory to solve the identification problem.
This approach identifies CM effects via the Marginal Treatment Effect (MTE) of the mediator, and requires three main assumptions.
(1) Mediator take-up must respond only positively to the initial treatment (monotonicity), which implies mediator selection follows a selection model.
(2) Mediator take-up is motivated by mediator benefits.
(3) A valid instrument for mediator take-up must exist, to avoid relying on parametric assumptions on unobserved selection.
While these assumptions are strong, they are plausible in many applied settings.
Mediator monotonicity aligns with conventional theories for selection-into-treatment, and is accepted widely in many applications using an instrumental variables research design.
Selection based on costs and benefits is central to economic theory, and is the dominant concern for judging observational designs that identify causal effects.
Access to valid instrumental variation is a strong condition, though is important to avoid further modelling assumptions; the most compelling example is using variation in mediator take-up costs as an instrument.

Applying the new methods to the Oregon Health Insurance Experiment shows that unobserved selection matters in an analysis of a real-world natural experiment.
A substantial portion of the wait-list lottery's impact on subjective health and well-being is mediated indirectly through extra healthcare usage, after instrumenting for healthcare usage with respondents' usual provider.
A conventional CM analysis would put this indirect mediated share at practically zero, so that my methods expose that negative selection into healthcare usage would be hiding evidence for this mechanism.
These estimates give sufficient evidence that extra healthcare use mediates a sizeable share of the Medicaid-lottery's benefits, though wider confidence intervals underscore the inherent uncertainty on the proportion of the effect operating through the single mediator mechanism of healthcare usage.

The methods I propose for CM analyses are not perfect for every setting: the structural assumptions are strong, and are tailored to selection-into-mediator based on the economic principle of selection based on costs and benefits.
Indeed, this approach provides no safe harbour for estimating CM effects if these structural assumptions do not hold true.
This approach imports insights from the instrumental variables literature, connecting the conventional \cite{imai2010identification} approach to CM with the economics literature on selection-into-treatment and MTEs \citep{vytlacil2002independence,heckman2004using,heckman2005structural,florens2008identification,brinch2017beyond,kline2019heckits}.

Applied economists mostly investigate mediating mechanisms for causal effects with suggestive analyses of mechanisms.
These descriptive analyses are informative, but do not generally identify a mediating pathway without additional assumptions --- see also \citet{blackwell2024assumption,green2010enough}.
A new strand of the econometric literature has arisen in implicit acknowledgement that suggestive evidence of mechanisms, or a conventional approach to CM, can lead to biased inference and needs alternative methods for credible inference.
These include identifying CM effects with overlapping quasi-experimental research designs \citep{deuchert2019direct,frolich2017direct}, functional form restrictions \citep{heckman2015econometric,heckman2013understanding}, partial identification \citep{flores2009identification}, or an hypothesis test of full mediation through observed channels \citep{kwon2024testing} --- see \cite{huber2019review} for an overview.\footnote{
    An alternative method to estimate CM effects is ensuring treatment and mediator quasi-random assignment holds by a running two randomised controlled trials for both treatment and mediator, at the same time.
    This set-up has been considered in the literature previously, in theory \citep{imai2013experimental} and in practice \citep{ludwig2011mechanism}.
}

I develop a framework showing exactly how selection bias contaminates conventional CM estimates when mediator choices are driven by unobserved gains --- settings where none of the existing econometric approaches to CM directly apply.
The recent econometric literature has made important progress in adapting CM to observational settings;
my work builds on this literature by addressing the common case in which the mediator is not quasi-randomly assigned, when selection into a mediator mechanism aligns with economic theory for selection.
\cite{frolich2017direct} is the most similar paper to mine, though I extend the framework for CM from an economic perspective in multiple ways.
First, by linking CM to the MTE literature for a binary mediator, I identify average CM effects rather than complier-specific effects (and provide a different estimation approach).
Second, I show formally how unobserved selection biases conventional CM estimates and propose an alternative identification strategy that retains the natural experiment structure while relaxing mediator ignorability.
Last, this paper provides a rigorous warning to applied economists against uncritically importing conventional CM methods to avoid the selection bias terms derived, and clarifies the conditions and practices to avoid them.

This paper proceeds as follows.
\autoref{sec:lottery} describes the dominant approach in economics for studying mechanisms behind treatment effects, illustrating with data from the Oregon Health Insurance Experiment.
\autoref{sec:mediation} introduces the formal framework for CM, and develops expressions for bias in conventional CM estimates in natural experiment and observational settings.
\autoref{sec:applied} describes this bias in applied settings with (1) a regression framework, (2) a setting with selection based on costs and benefits.
\autoref{sec:selectionmodel} purges bias from CM estimates by identifying CM effects via an MTE approach.
\autoref{sec:controlfun} demonstrates how to estimate CM effects with this approach, with either parametric or semi-parametric methods, and gives simulation evidence.
\autoref{sec:oregon} returns to the Oregon Health Insurance Experiment, providing credible estimates of effects on subjective health and well-being mediated through increased healthcare usage.
\autoref{sec:conclusion} concludes.

\section{Mechanisms in the Oregon Health Insurance Experiment}
\label{sec:lottery}
In the United States, healthcare is generally not provided directly by the government.
Instead, consumers purchase health insurance to fund healthcare expenses, with the government providing insurance only for elderly individuals (Medicare) and for those with low-incomes (Medicaid).
In 2004, the state of Oregon ceased accepting new applications for Medicaid due to budgetary constraints, and did not reopen applications until 2008.
When the state resumed enrolment, 90,000 individuals applied, vastly exceeding the programme's capacity.
Oregon therefore allocated the opportunity to apply for Medicaid via a lottery system among those on the wait-list.
Winning this wait-list lottery significantly increased healthcare usage, plus subjective health and well-being.

\begin{figure}[!htbp]
    \caption{Effects of the Oregon Health Insurance Experiment Wait-list Lottery.}
    \centering
    \includegraphics[width=0.85\textwidth]{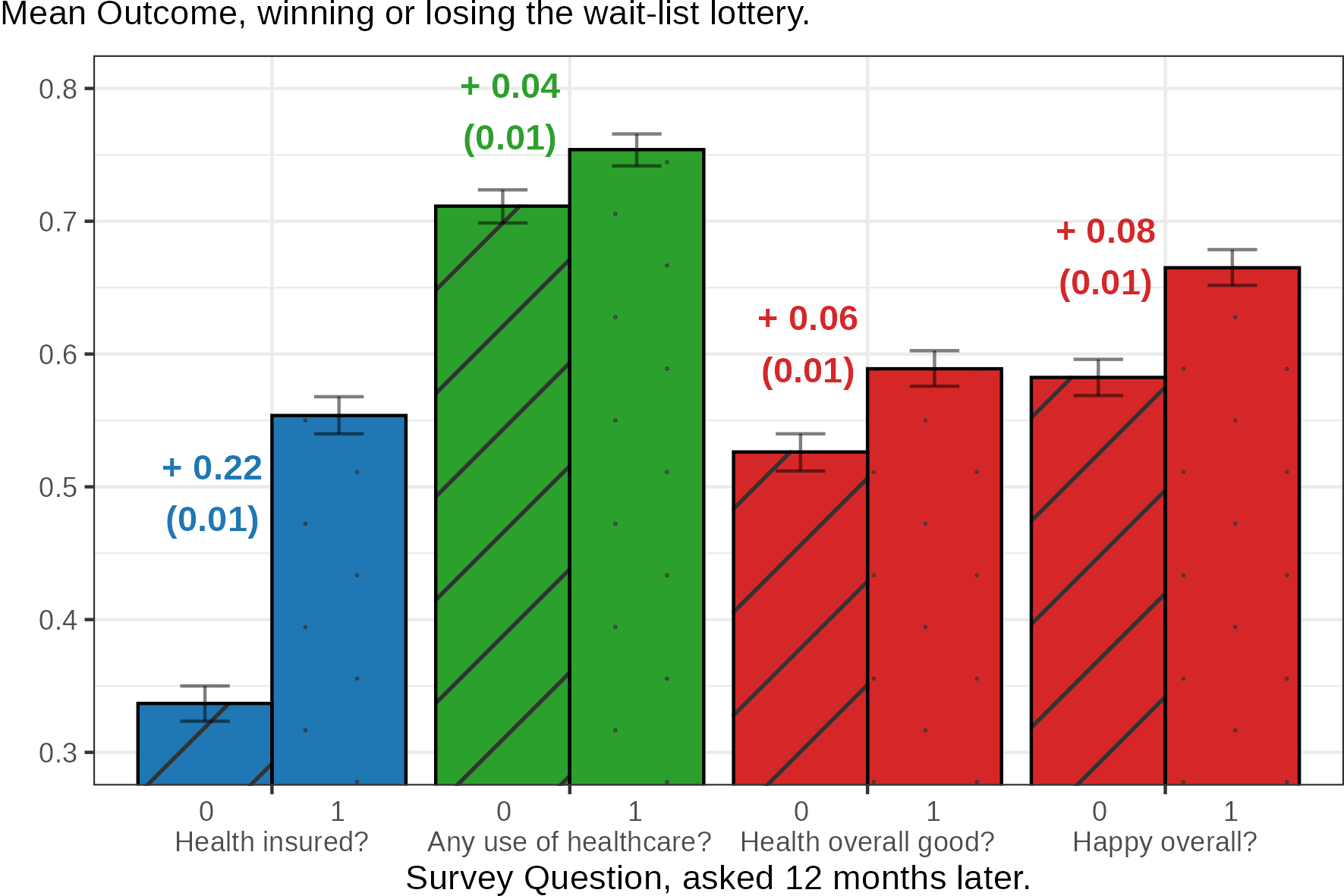}
    \label{fig:healthinsurance-effects}
    \justify
    \footnotesize    
    \textbf{Note:}
    This figure summarises the relevant results of the Oregon Health Insurance Experiment \citep{finkelstein2008oregon}.
    $\Egiven{Y_i}{Z_i = z}$ is the mean outcome, where $z = 0$ refers to the case of losing the wait-list lottery (not given access to Medicaid) and $z = 1$ winning.
    The numbers beside the bars are estimates of the mean difference; winning the Medicaid wait-list lottery increased average health insurance rate by 22 percentage points (pp), with standard errors of the difference reported in brackets.
\end{figure}

Winning the wait-list lottery increased the average health insurance coverage rate by 22 percentage points (pp), and subjective visitation of any healthcare provider in the following 12 months by 4 pp.
In addition, wait-list lottery winners agreed 6 pp more with the question ``In general, would you say your health is excellent, very good, or good'' (hereafter, subjective health),
and 8 pp for ``How would you say things are these days-would you say that you are very  or pretty happy'' (hereafter, subjective well-being).
These numbers are calculated among the
9,957

people eligible for the Oregon wait-list lottery who responded to a survey sent by \cite{finkelstein2008oregon}.\footnote{
    This number restricts to those who gave non-missing answers to all relevant questions, including questions on pre-lottery location of usual healthcare,
    using anonymised data from the Oregon Health Insurance Experiment replication package \citep{icspr2014oregon}.
}
\autoref{fig:healthinsurance-effects} summarises these results.

These results show that winning the wait-list lottery led to large gains in subjective health and well-being.
The economics, medicine, and health policy literatures have primarily focused on the health benefits --- often interpreted as healthcare benefits from new access to government provided health insurance.\footnote{
    \cite{finkelstein2008oregon} use the wait-list lottery as an IV because health insurance is not randomly assigned; this paper focuses on the average effects of winning the wait-list lottery (which is randomly assigned).
}
However, the original authors also noted other benefits, including complete elimination of catastrophic out-of-pocket medical debt among those with new access to Medicaid.
These are plausibly income effects that benefit recipients directly, not only through increased use of healthcare, but also by reducing stress and improving financial security.
These plausible direct effects have not been explored in the applied literature.

Accepted practice in applied economics is to investigate mechanisms behind causal effects with suggestive evidence.
This involves estimating the average causal effect of the wait-lottery on a proposed mediator (healthcare usage) and separately estimating its effect on the final outcomes (subjective health and well-being).
When both estimates are positive, and the mediator precedes the outcome, it is taken as de facto evidence that the mediator transmits the treatment effect.
In the case of the Oregon Health Insurance Experiment, this amounts to concluding that increased healthcare usage mediates the positive effects of winning the lottery on health and well-being.
\autoref{fig:scm-health} illustrates this approach, which is also prevalent in other social science fields --- see \cite{blackwell2024assumption,green2010enough}.

\begin{figure}[!htbp]
    \centering
    \singlespacing
    \caption{Structural Causal Model for Suggestive Evidence of a Mechanism.}
    \label{fig:scm-health}
    \begin{tikzpicture}
        \node[state,thick,ForestGreen] (mediator) at (0,0) {$D_i$};
        \node[state,thick,blue] (treatment) [left=2.5cm of mediator] {$Z_i$};
        \node[state,thick,red] (outcome)   [right=2.5cm of mediator] {$Y_i$};
        \node[color=ForestGreen] [above=0.1cm of mediator] {Healthcare};
        \node[color=blue] [left=0.1cm of treatment] {Wait-list lottery};
        \node[color=red] [right=0.1cm of outcome] {Health \& well-being};
        \path[->, thick] (treatment) edge (mediator);
        \path[->, thick] (treatment) edge[bend right=30] (outcome);
    \end{tikzpicture}
    \justify
    \footnotesize
    \textbf{Note}:
    This figure shows the structural causal model behind a suggestive analysis for effects of the Oregon Health Insurance Experiment, where arrows represent causal effects --- e.g., $Z_i \to D_i$ means $Z_i$ affects $D_i$ with no reverse causality.
\end{figure}

This approach gives necessary, but not sufficient, identification of healthcare as a mediating mechanism.
It is not sufficient because it provides no evidence for the effect of healthcare on health and well-being, so does not identify the causal mechanism.
Studying this mechanism with suggestive evidence require an additional, hidden assumption that healthcare positively affects health outcomes.
While this assumption is not unreasonable in general, it may not apply within the data under study; subjective well-being is measured only 12 months after the Medicaid lottery, which may not be enough time for health gains to accrue.
Second, this approach does not quantify the mechanism effects.
Healthcare could have a very large effect on subjective health and well-being, or possibly a very small effect --- it is a priori unclear.
In addition, the mediator mechanism effect refers not to the average effect of healthcare usage, but to the effect for Oregon residents who were induced to use more healthcare after winning the wait-list lottery (mediator compliers).
This local effect could differ substantially from a population average, and potentially mislead conclusions about the magnitude or generality of the mechanism.
Together, these concerns mean that a suggestive analysis of healthcare as a mediating mechanism are not dispositive; without additional assumptions, they neither identify nor quantify the mediator mechanism channel.


CM offers a compelling alternative framework, explicitly defining the average direct and indirect effects and clear assumptions under which they are identified.
Moreover, it delivers quantitative answers to the key question: how much of a treatment effect operates through a specific mediator mechanism?
CM is widely used in fields such as epidemiology, psychology, and medicine where researchers regularly decompose treatment effects into component pathways.
However, CM methods have not yet been examined from an economic perspective to assess their applicability in observational causal research, such as natural experiments.

\section{Causal Mediation (CM)}
\label{sec:mediation}
CM decomposes causal effects into two channels, through a mediating mechanism (indirect effect) and through all other paths (direct effect).
To develop notation, write $Z_i = 0, 1$ for a binary treatment, $D_i = 0, 1$ a binary mediator mechanism, and $Y_i$ a continuous outcome for individuals $i = 1, \hdots, n$.\footnote{
    This paper exclusively focuses on the binary case.
    See \cite{huber2020direct} or \cite{frolich2017direct} for a discussion of CM with continuous treatment and/or mediator, and the assumptions required.
}
$D_i$ and $Y_i$ are a sum of their potential outcomes,
\begin{align*}
    D_i &= (1 - Z_i) D_i(0)
        +   Z_i      D_i(1),  \\
    Y_i &= (1 - Z_i) Y_i(0, D_i(0))
        +   Z_i      Y_i(1, D_i(1)).
\end{align*}
Assume treatment $Z_i$ is quasi-randomly assigned,\footnote{
    This assumption can hold conditional on a covariate vector, $\vec X_i$.
    To simplify notation in this section, leave the conditional part unsaid, as it changes no part of the identification framework.
}
\[ Z_i \indep  D_i(z), Y_i(z', d), \text{ for } z, z', d = 0, 1. \]

There are only two average effects which are identified without additional assumptions.
\begin{enumerate}
    \item The average first-stage refers to the effect of the treatment on mediator, $Z_i$ on $D_i$:
    \[ \Egiven{D_i}{Z_i = 1} - \Egiven{D_i}{Z_i = 0}
        = \Prob{D_i(0) \neq D_i(1)}. \]
    It is common in the economics literature to assume that $Z_i$ influences $D_i$ in at most one direction, $\Prob{D_i(0) \leq D_i(1)} = 1$ --- monotonicity \citep{imbens1994identification}.
    \item The Average Treatment Effect (ATE) refers to the effect of the treatment on outcome, $Z_i$ on $Y_i$, and is also known as the average total effect or intent-to-treat effect in social science settings, or reduced-form effect in the instrumental variables literature:
    \[ \Egiven{Y_i}{Z_i = 1} - \Egiven{Y_i}{Z_i = 0}
        = \E{Y_i(1, D_i(1)) - Y_i(0, D_i(0))}. \]
\end{enumerate}

$Z_i$ affects outcome $Y_i$ directly, and indirectly via the $D_i(Z_i)$ channel, with no reverse causality.
\autoref{fig:scm-model} visualises the design, where the direction arrows denote the causal direction.
CM aims to decompose the ATE of $Z_i$ on $Y_i$ into these two separate pathways:
\begin{align*}
    \text{Average Direct Effect (ADE): } \;\;\;&
        \E{Y_i(1, D_i(Z_i)) - Y_i(0, D_i(Z_i))}, \\
    \text{Average Indirect Effect (AIE): } \;\;\;&
            \E{Y_i(Z_i, D_i(1)) - Y_i(Z_i, D_i(0))}.
\end{align*}

\begin{figure}[h!]
    \centering
    \singlespacing
    \caption{Structural Causal Model for CM.}
    \label{fig:scm-model}
    \begin{tikzpicture}
        \node[state,thick,ForestGreen] (mediator) at (0,0) {$D_i$};
        \node[state,thick,blue] (treatment) [left=3cm of mediator] {$Z_i$};
        \node[state,thick,red] (outcome) [right=3cm of mediator] {$Y_i$};
        \node[color=ForestGreen] [above=0.1cm of mediator] {Mediator};
        \node[color=blue] [left=0.1cm of treatment] {Treatment};
        \node[text width=0.1cm, color=red] [right=-0.01cm of outcome] {Outcome};
        \path[->, thick] (treatment) edge (mediator);
        \path[->, thick] (mediator) edge (outcome);
        \path[->, thick] (treatment) edge[bend right=37.5] (outcome);
        \node[color=orange] [below left=-0.3cm and 0.5cm of mediator] {First-stage};
        \node[color=orange] [below right=-0.3cm and 0.1cm of mediator] {Switcher AIE};
        \node[color=orange] [below=1.125cm of mediator] {ADE};
        \node[state,thick,dashed,RoyalBlue] (confounderU) [above=0.75cm of outcome] {$\vec U_i$};
        \path[->,thick,dashed,color=RoyalBlue] (confounderU) edge (mediator);
        \path[->,thick,dashed,color=RoyalBlue] (confounderU) edge (outcome);
    \end{tikzpicture}
    \justify
    \footnotesize
    \textbf{Note}:
    This figure shows the structural causal model behind CM.
    The Switcher AIE refers to the AIE local to $D_i(Z_i)$ switchers, the group with $D_i(0) \neq D_i(1)$, so that AIE $=$ Average First-stage $\times$ Switcher AIE.
    $\vec U_i$ shows an unobserved confounding variable for the causal effect $D_i \to Y_i$, representing this paper's focus on the case that $D_i$ is not quasi-randomly assigned.
    \autoref{sec:regression} defines $\vec U_i$ in an applied setting.
\end{figure}

Estimating the AIE answers the following question: how much of the causal effect $Z_i$ on $Y_i$ goes through the $D_i$ channel?
When studying the health gains of winning the Medicaid wait-list lottery \citep{finkelstein2008oregon}, the AIE represents how much of the effect comes from using the hospital more often.
Estimating the ADE answers the following equation: how much is left over after accounting for the $D_i$ channel?\footnote{
    In a non-parametric setting it is not necessary that ADE $+$ AIE $=$ ATE.
    See \cite{imai2010identification} for this point in full.
}
For the example, how much of the wait-list lottery effect is a direct effect, other than increased healthcare usage --- e.g., income effects of lower medical debt, or less worry over health shocks thanks to government support.
An Instrumental Variables (IV) approach assumes this direct effect is zero for everyone (the exclusion restriction).
CM is a similar, yet distinct, framework attempting to explicitly model the direct effect, and not assuming it is zero.

The ADE and AIE are not separately identified without further assumptions.

\subsection{Identification of CM Effects}
The conventional approach to estimating direct and indirect effects assumes both $Z_i$ and $D_i$ are quasi-randomly assigned, conditional on a vector of control variables $\vec X_i$.
\begin{definition}
    \label{dfn:seq-ign}
    Sequential quasi-random assignment \citep{imai2010identification}
    \begin{align}
        \label{eqn:seq-ign-Z}
        Z_i \indep  D_i(z), Y_i(z', d) \;\; &| \;\; \vec X_i,
            &\textnormal{ for } z, z', d = 0, 1 \\
        \label{eqn:seq-ign-D}
        D_i \indep Y_i(z, d) \;\; &| \;\; \vec X_i, Z_i = z, 
            &\textnormal{ for } z, d = 0, 1.
    \end{align}
\end{definition}
Sequential quasi-random assignment assumes that the initial treatment $Z_i$ is quasi-randomly assigned conditional on $\vec X_i$ (as has already been assumed above).
It then also assumes that, after $Z_i$ is assigned, that $D_i$ is quasi-randomly assigned conditional on $\vec X, Z_i$ (hereafter, mediator quasi-random assignment).
If \ref{dfn:seq-ign}\eqref{eqn:seq-ign-Z} and \ref{dfn:seq-ign}\eqref{eqn:seq-ign-D} hold, then the ADE and AIE are identified by two-stage mean differences conditioning on $\vec X_i$.\footnote{
    In addition, a common support condition for both $Z_i, D_i$ (across $\vec X_i$) is necessary.
    \cite{imai2010identification} show a general identification statement; I show identification in terms of two-stage regression.
    See \appendixref{appendix:identification}.
}
\vspace{0.1cm}

\makebox[\textwidth]{\parbox{1.25\textwidth}{
\[ \E[D_i, \vec X_i]{
    \underbrace{\Egiven{Y_i}{Z_i = 1, D_i, \vec X_i} - \Egiven{Y_i}{Z_i = 0, D_i, \vec X_i}}_{\text{Second-stage regression, $Y_i$ on $Z_i$ holding $D_i, \vec X_i$ constant}}}
    = \underbrace{\E{Y_i(1, D_i(Z_i)) - Y_i(0, D_i(Z_i))}}_{\text{Average Direct Effect (ADE)}} \]
\[ \E[Z_i, \vec X_i]{ \underbrace{\Big(
    \Egiven{D_i}{Z_i = 1, \vec X_i} - \Egiven{D_i}{Z_i = 0, \vec X_i} \Big)}_{\text{First-stage regression, $D_i$ on $Z_i$}}
    \times \underbrace{\Big(
    \Egiven{Y_i}{Z_i, D_i = 1, \vec X_i} - \Egiven{Y_i}{Z_i, D_i = 0, \vec X_i} \Big)}_{\text{Second-stage regression, $Y_i$ on $D_i$ holding $Z_i, \vec X_i$ constant}} } \]
\[ = \underbrace{\E{Y_i(Z_i, D_i(1)) - Y_i(Z_i, D_i(0))}}_{\text{Average Indirect Effect (AIE)}} \]
}}

I refer to the estimands on the left-hand side as CM estimands, which are typically estimated by a composition of two-stage Ordinary Least Squares (OLS) estimates \citep{imai2010identification}.
While this is the most common approach in the applied literature, I do not assume the linear model for my identification analysis.
Linearity assumptions are not necessary for identification, and it suffices to note that heterogeneous treatment effects and non-linear confounding can bias OLS estimates of CM estimands in the same manner that is well documented elsewhere (see e.g., \citealt{angrist1998estimating,sloczynski2022interpreting}).
This section focuses on problems that plague conventional CM, regardless of estimation method.

\subsection{Non-identification of CM Effects}
Applied research often uses a natural experiment to study settings where treatment $Z_i$ is quasi-randomly assigned, justifying assumption \ref{dfn:seq-ign}\eqref{eqn:seq-ign-Z}.
Rarely do they also have access to an additional, overlapping natural experiment to isolate random variation in $D_i$ --- to justify mediator quasi-random assignment \ref{dfn:seq-ign}\eqref{eqn:seq-ign-D}.
One might consider conventional CM methods in such a setting to learn about the mechanisms behind the causal effect $Z_i$ on $Y_i$.
This approach leads to estimates at risk of bias, contaminating inference on direct and indirect effects.

\begin{theorem}
    \label{thm:selection-bias}
    Absent an identification strategy for the mediator, conventional CM estimates are at risk of selection bias.
    If \ref{dfn:seq-ign}\eqref{eqn:seq-ign-Z} holds, and \ref{dfn:seq-ign}\eqref{eqn:seq-ign-D} does not, then conventional CM estimands are contaminated by selection bias and group differences.
    Proof: see \appendixref{appendix:mediation-bias}.
\end{theorem}
Below I present the relevant selection bias and group difference terms, omitting the conditional on $\vec X_i$ notation for brevity.

\noindent
For the direct effect: conventional CM estimand $=$ ADE $+$ selection bias $+$ group differences.\footnote{
    The bias terms here mirror those in \cite{heckman1998characterizing,angrist2009mostly} for a single $D_i$ on $Y_i$ treatment effect, when $D_i$ is not quasi-randomly assigned:
    \vspace{-0.25cm}
    \[ \Egiven{ Y_i}{D_i =1} - \Egiven{ Y_i}{D_i =0}
        = \text{ATE}
        + \underbrace{\Big( \Egiven{ Y_i(.,0)}{D_i =1} - \Egiven{ Y_i(.,0)}{D_i =0} \Big)}_{
            \text{Selection Bias}}
        + \underbrace{ \Prob{D_i=0} (\text{ATT} - \text{ATU}) }_{
            \text{Group-differences Bias}}. \]
}
\begin{align*}
    & \mathbb E_{D_i} \Big[
        \Egiven{Y_i}{Z_i = 1, D_i} - \Egiven{Y_i}{Z_i = 0, D_i} \Big] \\
    & = \E{Y_i(1, D_i(Z_i)) - Y_i(0, D_i(Z_i))} \\
    & \;\;\;\; + \mathbb E_{D_i = d} \Big[
        \Egiven{Y_i(0, D_i(Z_i))}{D_i(1) = d} 
        - \Egiven{Y_i(0, D_i(Z_i))}{D_i(0) = d} \Big] \\
    & \;\;\;\; + \E[D_i = d]{
        \Big(1 - \Prob{D_i(1) = d} \Big)
        \left( \begin{aligned}
            &\Egiven{Y_i(1, D_i(Z_i)) - Y_i(0, D_i(Z_i))}{D_i(1) = 1-d} \\ 
            &  - \Egiven{Y_i(1, D_i(Z_i)) - Y_i(0, D_i(Z_i))}{D_i(1) = d}
            \end{aligned} \right) }
\end{align*}

\noindent
For the indirect effect: conventional CM estimand $=$ AIE $+$ selection bias $+$ group differences.
\begin{align*}
    &\E[Z_i]{
        \Big( \Egiven{D_i}{Z_i = 1} - \Egiven{D_i}{Z_i = 0} \Big) \times
        \Big( \Egiven{Y_i}{Z_i, D_i = 1} - \Egiven{Y_i}{Z_i, D_i = 0} \Big) } \\
    & = \E{Y_i(Z_i, D_i(1)) - Y_i(Z_i, D_i(0))} \\
    & \;\;\;\; + \Prob{D_i(0) \neq D_i(1)} \Big(
        \Egiven{Y_i(Z_i, 0)}{D_i = 1} - \Egiven{Y_i(Z_i, 0)}{D_i = 0} \Big) \\
    & \;\;\;\; + \Prob{D_i(0) \neq D_i(1)} \times \\
    & \;\;\;\; \;\; \left[ \begin{aligned}
        &\Big( 1 - \Prob{D_i=1} \Big)
        \left( \begin{aligned}
            &\Egiven{Y_i(Z_i, 1) - Y_i(Z_i, 0)}{D_i = 1} \\ 
            &  - \Egiven{Y_i(Z_i, 1) - Y_i(Z_i, 0)}{D_i = 0}
        \end{aligned} \right) \\
        &- \left( \frac{1 - \Prob{D_i(0) \neq D_i(1)} }{
            \Prob{D_i(0) \neq D_i(1)}} \right)
        \left( \begin{aligned}
            &\Egiven{Y_i(Z_i, 1) - Y_i(Z_i, 0)}{D_i(1) = 0 \text{ or } D_i(0)=1} \\ 
            &  - \E{Y_i(Z_i, 1) - Y_i(Z_i, 0)}
        \end{aligned} \right)
    \end{aligned} \right]
\end{align*}

The selection bias terms come from systematic differences between the groups taking or refusing the mediator ($D_i = 1$ versus $D_i = 0$), differences not fully unexplained by $\vec X_i$.
These selection bias terms would equal zero if the mediator had been quasi-randomly assigned \ref{dfn:seq-ign}\eqref{eqn:seq-ign-D}, but do not necessarily average to zero if not.
In the Oregon Health Insurance Experiment, the wait-list gave random variation in the treatment (the Medicaid wait-list lottery) but there was not a similar natural experiment for healthcare usage; correspondingly, the selection-on-observables approach to CM has selection bias.

The group differences represent the fact that a matching approach gives an average effect on the treated group, which is systematically different from the average effect if selection-on-observables does not hold.
These terms are a non-parametric framing of the bias from controlling for intermediate outcomes, previously studied only in a linear setting (i.e., bad controls in \citealt{cinelli2024crash}, or M-bias in \citealt{ding2015adjust}).

The AIE group differences term is longer, because the indirect effect is comprised of the effect of $D_i$ local to $D_i(Z_i)$ switchers.
\begin{align*}
    \text{AIE}
    &= \E{Y_i(Z_i, D_i(1)) - Y_i(Z_i, D_i(0))} \\
    &= \Prob{D_i(0) \neq D_i(1)} \; 
        \underbrace{\Egiven{Y_i(Z_i, 1) - Y_i(Z_i, 0)}{D_i(0) \neq D_i(1)}}_{
            \text{Average $D_i$ on $Y_i$ effect among $D_i(Z_i)$ switchers}
        }
\end{align*}
It is important to acknowledge the mediator switchers here, because the AIE is the treatment effect going through the $D_i(Z_i)$ channel, thus only refers to individuals pushed into mediator $D_i$ by initial treatment $Z_i$.
If we had been using a population average effect for $D_i$ on $Y_i$, then this is losing focus on the definition of the AIE; it is not about the causal effect $D_i$ on $Y_i$, it is about the causal effect $D_i(Z_i)$ on $Y_i$.

The group difference bias term arises because the convention approach to CM assumes the switcher average effect is equal to the population average effect, which does not hold true if the mediator is not quasi-randomly assigned.

\section{CM in Applied Settings}
\label{sec:applied}
Unobserved confounding is particularly problematic when studying the mechanisms behind treatment effects.
For example, in studying health gains from the Oregon wait-list lottery, we might expect that health gains came about because those who won access to Medicaid started visiting their healthcare provider more often, when in past they avoided it over financial concerns.
Applying conventional CM methods to investigate this expectation would be dismissing unobserved confounders for how often individuals visit healthcare providers, leading to biased results.

The wider population does not have one uniform bill of health; many people are born predisposed to ailments, due to genetic variation or other unrelated factors.
These conditions can exist for years before being diagnosed.
People with severe underlying conditions may visit healthcare providers more often than the rest of the population, to investigate or begin treating the ill-effects.
It stands to reason that people with more serve underlying conditions may gain more from more often attending healthcare providers once given health insurance.
These underlying causes cannot be controlled for by researchers, as we cannot hope to observe and control for health conditions that are yet to even be diagnosed.
This means underlying health conditions are an unobserved confounder, and will bias estimates of the ADE and AIE in this setting.

In this section, I further develop the issue of selection on unobserved factors in a general CM setting.
First, I show the non-parametric bias terms from \autoref{sec:mediation} can be written as omitted variables bias in a random coefficients regression framework.
Second, I show how selection bias operates in a basic model for selection-into-mediator based on costs and benefits.

\subsection{Regression Framework}
\label{sec:regression}
Inference for CM effects can be written in a regression framework with random coefficients, showing how correlation between unobserved error terms and the mediator disrupts identification.

Start by writing potential outcomes $Y_i(., .)$ as a sum of observed and unobserved factors, following the notation of \cite{heckman2005structural}.
For each $z,d = 0,1$, put $\mu_{d}(z; \vec X_i) = \Egiven{Y_i(z, d)}{\vec X_i}$ and the corresponding error terms, $U_{d, i} = Y_i(z, d) - \mu_{d}(z; \vec X_i)$, so we have the following expressions:
\[ Y_i(Z_i, 0)  = \mu_{0}(Z_i; \vec X_i) + U_{0,i}, \;\;
    Y_i(Z_i, 1) = \mu_{1}(Z_i; \vec X_i) + U_{1,i}. \]

With this notation, observed data $Z_i, D_i, Y_i, \vec X_i$ have the following random coefficient outcome formulae --- which characterise direct effects, indirect effects, and selection bias.
\begin{align}
    \label{eqn:parametric-firststage}
    D_i &= \theta + \bar \pi Z_i + \zeta(\vec X_i) + \eta_i,  \\
    \label{eqn:parametric-secondstage}
    Y_i &= \alpha + \beta D_i + \gamma Z_i + \delta Z_i D_i
    + \varphi(\vec X_i)
    + \underbrace{ \left(1 - D_i \right) U_{0,i} + D_i U_{1,i}}_{
        \text{Correlated error term.}}
\end{align}

This is not consequence  of linearity assumptions; the outcome formulae allow for unconstrained heterogeneous treatment effects, because the coefficients are random.
If either $Z_i, D_i$ were continuously distributed, then this representative would not necessarily hold true.
First-stage \eqref{eqn:parametric-firststage} is identified, with $\theta + \zeta(\vec X_i)$ the intercept, and $\bar \pi$ the first-stage average compliance rate (conditional on $\vec X_i$).
Second-stage \eqref{eqn:parametric-secondstage} has the following definitions, and is not identified thanks to omitted variables bias.
See \appendixref{appendix:regression-model} for the derivation.
\begin{enumerate}[label=\textbf{(\alph*)}]
    \item $\alpha = \E{\mu_0(0; \vec X_i)}$ and $\varphi(\vec X_i) = \mu_0(0; \vec X_i) - \alpha$ are the intercept terms.
    \item $\beta = \mu_1(0; \vec X_i) - \mu_0(0; \vec X_i)$ is the AIE conditional on $Z_i = 0, \vec X_i$.
    \item $\gamma = \mu_0(1; \vec X_i) - \mu_0(0; \vec X_i)$ is the ADE conditional on $D_i = 0, \vec X_i$.
    \item $\delta = \mu_1(1; \vec X_i) - \mu_0(1; \vec X_i) - \big( \mu_1(0; \vec X_i) - \mu_0(0; \vec X_i) \big)$ is the average interaction effect conditional on $\vec X_i$.
    \item $\left( 1 - D_i \right) U_{0,i} + D_i U_{1,i}$ is the disruptive error term.
\end{enumerate}

The ADE and AIE are averages of the random coefficients:
\begin{align*}
    \text{ADE}
        &= \E{\gamma + \delta D_i}, \\
    \text{AIE}
        &= \E{ \bar \pi \big( \beta +  \delta Z_i + \tilde U_i \big)},
        \;\;\;\; \text{ with } \tilde U_i
            = \underbrace{\Egiven{ U_{1,i} - U_{0,i}}{
                \vec X_i, D_i(0) = 0, D_i(1) = 1}}_{
                    \text{Unobserved complier gains.}}.
\end{align*}
The ADE is a simple sum of the coefficients, while the AIE includes a group differences term because it only refers to $D_i(Z_i)$ compliers.

By construction, $\vec U_i \coloneqq \left(U_{0, i}, U_{1, i} \right)$ is an unobserved confounder.
The regression estimates of $\beta, \gamma, \delta$ in second-stage \eqref{eqn:parametric-secondstage} give unbiased estimates only if $D_i$ is also conditionally quasi-randomly assigned: $D_i \indep  \vec U_{i} $.
If not, then estimates of CM effects suffer from omitted variables bias from failing to adjust for the unobserved confounder, $\vec U_i$.

\subsection{Selection on Costs and Benefits}
CM is at risk of bias because $D_i \indep  \vec U_i$ 
is unlikely to hold in applied settings.
A separate identification strategy could disrupt the selection-into-$D_i$ based on unobserved factors, and lend credibility to the mediator quasi-random assignment assumption.
Without it, bias will persist, given how we conventionally think of selection-into-treatment.

Consider a model where individual $i$ selects into a mediator based on costs and benefits (in terms of outcome $Y_i$), after $Z_i, \vec X_i$ have been assigned.
In a natural experiment setting, an external factor has disrupted individuals selecting $Z_i$ by choice (thus $Z_i$ is quasi-randomly assigned), but it has not disrupted the choice to take mediator (thus $D_i$ is not quasi-randomly assigned).
In the Oregon Health Insurance Experiment, the treatment variation comes from the wait-list lottery, while healthcare usage was not subject to a similar lottery.
Write $C_i$ for individual $i$'s costs of taking mediator $D_i$, and $\indicator{.}$ for the indicator function.
The Roy model has $i$ taking the mediator if the benefits exceed the costs,
\begin{equation}
    \label{eqn:roy-model}
    D_i \left( z \right) = \indicator{ \;
    \underbrace{C_i}_{\text{Costs}} \;\; \leq \;\;
        \underbrace{
            Y_i\left( z, 1 \right) - Y_i\left( z, 0 \right)}_{\text{Benefits}}
    \;}, \;\;\; \text{for } z=0,1.
\end{equation}

The Roy model provides an intuitive framework for analysing selection mechanisms because it captures the fundamental economic principle of decision-making based on costs and benefits in terms of the outcome under study \citep{roy1951some,heckman1990empirical}.
In the Oregon Health Insurance Experiment, this models choice to visit the doctor in terms of health and well-being benefits relative to costs.\footnote{
    If the choice is considered over a sum of outcomes, then a simple extension to a utility maximisation model maintains this same framework with expected costs and benefits.
    See \cite{heckman1990empirical,eisenhauer2015generalized}.
}
This makes the Roy model useful as a base case for CM, where selection-into-mediator may be driven by private information (unobserved by the researcher).

By using the Roy model as a benchmark, I explore the practical limits of the mediator quasi-random assignment assumption.
If selection follows a Roy model, and the mediator is quasi-randomly assigned, then unobserved benefits can play no part in selection.
The only driver of selection are individuals' differences in costs (and not benefits).
If there are any selection-into-$D_i$ benefits unobserved to the researcher, then mediator quasi-random assignment cannot hold.
\newtheorem{proposition}{Proposition}
\begin{proposition}
    \label{prop:roy-seq-ig}
    Suppose mediator selection follows a Roy model \eqref{eqn:roy-model}, and selection is not fully explained by costs and observed gains.
    Then mediator quasi-random assignment does not hold.
\end{proposition}

This is an equivalence statement: selection based on costs and benefits is only consistent with mediator quasi-random assignment if the researcher observed every single source of mediator benefits.
See \appendixref{appendix:roy-seq-ig} for the proof.
This means than the vector of control variables $\vec X_i$ must be incredibly rich.
Together, $\vec X_i$ and unobserved cost differences $U_{C,i}$ must explain selection-into-$D_i$ one hundred percent.
In the Roy model framework, however, individuals make decisions about mediator take-up based on gains --- whether the researcher observes them or not.
The unobserved gains are unlikely to be fully captured by an observed control set $\vec X_i$, except in very special cases.

In practice, the best setting to believe in the mediator quasi-random assignment assumption is to study a setting where the researcher has two causal research designs, one for treatment $Z_i$ and another for mediator $D_i$, at the same time.
Absent this, mediator quasi-random assignment become hard to believe, and the corresponding conventional CM estimates are at risk of selection bias.

%
%
%

\section{Solving Identification via the Mediator MTE}
\label{sec:selectionmodel}
If your goal is to estimate CM effects, and you could control for unobserved selection terms $U_{0,i}, U_{1,i}$, then you would.
This ideal (but infeasible) scenario would yield unbiased estimates for the ADE and AIE.
Identification via the mediator Marginal Treatment Effect (MTE) takes this insight seriously, providing conditions to model the implied confounding by $U_{0,i}, U_{1,i}$, and then controlling for it.

The main problem is that second-stage regression equation \eqref{eqn:parametric-secondstage} is not identified, because $U_{0,i},U_{1,i}$ are unobserved, and lead to omitted variables bias.
\begin{align}
    \Egiven{Y_i}{Z_i, D_i, \vec X_i} \;\; =& \;\;
        \alpha
        + \beta D_i
        + \gamma Z_i
        + \delta Z_i D_i
        + \varphi(\vec X_i) \nonumber \\
        \label{eqn:secondstage-reg}
        & \;\; +\underbrace{\left( 1 - D_i
            \right) \Egiven{ U_{0,i} }{D_i = 0, \vec X_i}
                + D_i \Egiven{ U_{1,i} }{D_i = 1, \vec X_i}}_{
                    \text{Unobserved confounding.}}
\end{align}

My approach to identifying CM effects models the contaminating terms in \eqref{eqn:secondstage-reg} via the mediator MTE, avoiding the bias terms derived in \autoref{sec:mediation}.
Thinking on MTEs began with corrections for sample selection problems \citep{heckman1974shadow}, and were extended to a general selection problem of the same form as \autoref{eqn:secondstage-reg} in parametric settings \citep{heckman1979sample,bjorklund1987estimation} and a general case \citep{heckman2005structural}.
The approach works in the following manner: (1) assume that the variable of interest follows a selection model, where unexplained first-stage selection informs unobserved second-stage confounding; (2) extract information about unobserved confounding from the first-stage; and (3) incorporate this information as control terms in the second-stage equation to adjust for selection-into-mediator.
Identification in MTE methods typically relies on identifying the corresponding Control Functions (CFs) with an external instrument or distributional assumptions; this paper focuses exclusively on the case that an instrument is available.
By explicitly accounting for the information contained in the first-stage selection model, the MTE approach enables consistent estimation of causal effects in the second-stage even when selection is driven by unobserved factors.

In the example of analysing health gains from the Oregon Health Insurance Experiment, the MTE-based approach addresses the unobserved confounding by modelling unobserved effects of underlying health conditions.
It does so by assuming that unobserved selection-into-healthcare use is informative for underlying health conditions, assuming people with more severe underlying conditions visit the doctor more often than those without.
Then it uses this information in the second-stage estimation of how much the effect goes through increased healthcare usage, estimating the ADE and AIE after controlling for this confounding.

\subsection{Re-identification of CM Effects}
The following assumptions are sufficient to model the correlated error terms, identifying $\beta, \gamma, \delta$ in the second-stage regression \eqref{eqn:parametric-secondstage}, and thus both the ADE and AIE.

\theoremstyle{definition}
\newtheorem{assumptionMTE}{Assumption}
\renewcommand\theassumptionMTE{MTE--\arabic{assumptionMTE}}
\begin{assumptionMTE}
    \label{mte:monotonicity}
    Mediator monotonicity, conditional on $\vec X_i$.
    \[ \Probgiven{ D_i(0) \leq D_i(1) }{\vec X_i} = 1. \]
\end{assumptionMTE}
\noindent
Assumption \ref{mte:monotonicity} is the monotonicity condition first used in an IV context \citep{imbens1994identification}.
Here, it is assuming that people respond to treatment, $Z_i$, by consistently taking or refusing the mediator $D_i$ (always or never-mediators), or taking the mediator $D_i$ if and only if assigned to the treatment $Z_i=1$ (mediator compliers).
There are no mediator defiers, so that the mediator switcher group ($D_i(0) \neq D_i(1)$) is comprised entirely of mediator compliers ($D_i(0) =0$, $D_i(1)=1$).

The main implication of Assumption \ref{mte:monotonicity} is that selection-into-mediator can be written as a selection model with ordered threshold crossing values that describe selection-into-$D_i$ \citep{vytlacil2002independence}.
\[ D_i(z) = \indicator{V_i \leq \psi \big( z; \vec X_i \big)},
    \;\text{ for } z=0,1 \]
where $V_i$ is a latent variable with continuous distribution and conditional cumulative density function $F_V(. \,|\vec X_i)$, and $\psi(. \,;\vec X_i)$ collects observed sources of mediator selection.
$V_i$ could be assumed to follow a known distribution; the canonical Heckman selection model assumes $V_i$ is normally distributed (a ``Heckit'' model).
The identification strategy here applies to the general case that the distribution of $V_i$ is unknown, without parametric restrictions.

I focus on the equivalent transformed model of \cite{heckman2005structural},
\[ D_i(z) = \indicator{U_i \leq \pi(z; \vec X_i)},
    \;\;\; \text{for } z=0,1 \]
where $U_i \coloneqq F_V\left( V_i \mid \vec X_i \right)$ follows a uniform distribution, and $\pi(z; \vec X_i) = F_V\big(\psi(z; \vec X_i)\big) = \Probgiven{D_i = 1}{Z_i = z, \vec X_i}$ is the mediator propensity score.
$U_i$ are the unobserved mediator take-up costs.
Note the maintained assumption that treatment $Z_i$ is quasi-randomly assigned conditional on $\vec X_i$ implies $Z_i \indep U_i$ conditional on $\vec X_i$.

This selection model setup is equivalent to the monotonicity condition, and is importing a well-known equivalence result from the IV literature to the CM setting.
The main conceptual difference is not assuming $Z_i$ is a valid instrument for identifying the $D_i$ on $Y_i$ effect among compliers; it is using the selection model representation to correct for selection bias.
See \appendixref{appendix:mte-monotonicity} for a validation of the general \cite{vytlacil2002independence} equivalence result in a CM setting, with conditioning covariates $\vec X_i$.

\begin{assumptionMTE}
    \label{mte:identification}
    Selection on mediator benefits.
    \[ \Egiven{U_{0,i}}{U_i, \vec X_i} \neq 0, \;\;\;\;
        \Egiven{U_{1,i}}{U_i, \vec X_i} \neq 0. \]
\end{assumptionMTE}
\noindent
Assumption~\ref{mte:identification} is a relevance condition;
if there is unobserved confounding in $Y_i$, then it can be measured in $D_i$.
It states that that unobserved selection in mediator take-up ($U_i$) informs second-stage confounding, when taking or refusing the mediator ($U_{0,i}$ or $U_{1,i}$, respectively).

This is a relevance condition for the MTE model, analogously to instrument relevance in IV estimation.
Just as a weak instrument provides no identifying variation in the first-stage, a violation of Assumption~\ref{mte:identification} implies that the control functions $\lambda_0, \lambda_1$
enter the second-stage regression but fail to absorb any selection bias.
If individuals select into $D_i$ according to a Roy model, then this assumption holds because $V_i = U_{C,i} - \big( U_{1,i} - U_{0,i} \big)$.
Individuals may make decisions based on other outcomes, but as long as mediator benefits guide at least part of the decision (i.e., the dependence is bounded away from zero), then this assumption will hold.

For notation purposes, suppose the vector of control variables $\vec X_i$ has at least two entries;
denote $\vec X_i^{\text{IV}}$ as one entry in the vector, and $\vec X_i^-$ as the remaining.
\begin{assumptionMTE}
    \label{mte:instrument}
    Mediator take-up cost instrument.
    \[ \vec X_i^{\text{IV}} \textnormal{ satisfies } \;
    \partialdiff{\vec X_i^{\text{IV}}}\Big\{
        \mu_{d}(z, \vec X_i) \Big\} = 0
        < \partialdiff{\vec X_i^{\text{IV}}}
        \Big\{
            \Egiven{D_i(z)}{\vec X_i}\Big\},
        \textnormal{ for } z, d = 0, 1. \]
\end{assumptionMTE}
\noindent
Assumption \ref{mte:instrument} is requiring at least one control variable guides selection-into-$D_i$ --- an IV.
It assumes an instrument exists, which satisfies an exclusion restriction (i.e., not impacting mediator gains $\mu_1-\mu_0$), and has a non-zero influence on the mediator (i.e., strong IV first-stage).
The exclusion restriction is untestable, and must be guided by domain-specific knowledge; IV first-stage strength is testable, and must be justified with data by methods common in the IV literature.

The most compelling example of a mediator IV is using data on the cost of mediator take-up as a first-stage IV, if it varies between individuals for unrelated reasons and is strong in explaining mediator take-up.

The ideal instrument $\vec X_i^{\text{IV}}$ for identification is continuous, and varies
$\pi(z;\vec X_i)$ between 0 and 1 for every possible value of $z,\vec X_i^-$
(identification at infinity).
This is sufficient, but stronger than necessary.
The ADE and AIE do not require identifying the mediator MTE over every
$p \in (0,1)$.
They require the MTE-associated CFs only over the mediator-complier interval,
\[
    \pi(0;\vec X_i) < U_i \leq \pi(1;\vec X_i).
\]
Thus, the identifying requirement is not full-support identification of the mediator MTE, but identification of the relevant CFs over the support between
the complier region between $\pi(0;\vec X_i)$ and $\pi(1;\vec X_i)$.

With finite support, I instead impose a restricted-CF condition.
Let $K$ denote the number of distinct values taken by the mediator propensity score $\pi(Z_i;\vec X_i)$ on the relevant support.
The CFs $\lambda_0(.), \lambda_1(.)$ are assumed to lie in a $K-1$ dimensional class over this propensity-score support.
After observed heterogeneity is absorbed by $\mu_d(z;\vec X_i)$, the remaining dependence between observed mediator take-up cost $U_i$ and outcome gains $U_{0,i},U_{1,i}$ is common across $\vec X_i$ and lied in a $K-1$ dimensional basis over the support of $\pi(Z_i;\vec X_i)$.
Equivalently, the instrument identifies the CFs only up to the degrees of freedom supplied by relevant propensity-score support.
For example, with only two relevant propensity-score values, this is a linear restriction on the CFs, corresponding to the restricted-MTE logic of \cite{brinch2017beyond}.
This restricted support assumption is weaker than relying on identification at infinity, with testable implications --- described in \appendixref{appendix:mte-restricted}.

\begin{proposition}
    \label{proposition:secondstage}
    If assumptions \ref{mte:monotonicity}, \ref{mte:identification}, \ref{mte:instrument}, and the restricted-CF support condition hold, then second-stage regression equation \eqref{eqn:parametric-secondstage} is identified via the MTE-associated CFs.
    \begin{align*}
        \Egiven{Y_i}{Z_i, D_i, \vec X_i} \;\; =& \;\;
            \alpha
            + \beta D_i
            + \gamma Z_i
            + \delta Z_i D_i
            + \varphi\big(\vec X_i^-\big) \\
            & \;\; + \rho_0 \left( 1 - D_i \right) \lambda_0 \big( \pi(Z_i ; \vec X_i) \big)
                + \rho_1 D_i \lambda_1 \big( \pi(Z_i ; \vec X_i) \big),
            \label{eqn:mte-secondstage}
    \end{align*}
    where $\lambda_0, \lambda_1$ are the corresponding CFs, $\rho_0, \rho_1$ are linear parameters, and mediator propensity score $\pi(z;\vec X_i)$ is separately identified in the first-stage \eqref{eqn:parametric-firststage}.
    Proof: see \appendixref{appendix:mte-secondstage}.
\end{proposition}
Again, this set-up required no linearity assumptions, and treatment effects vary, because $Z_i, D_i$ are categorical and  $\beta, \gamma, \delta , \varphi(\vec X_i)$ vary with $\vec X_i$.
The CFs are functions which measure unobserved mediator gains, for those with unobserved mediator costs above or below a propensity score value.
Following the IV notation of \cite{kline2019heckits}, put $\mu_V = \E{F_V^{-1}\left( U_i \mid \vec{X}_i \right)}$, to give the following representation for the CFs:
\begin{align*}
    \lambda_0\big(p\big) &=
        \Egiven{ F_V^{-1}\left( U_i \mid \vec{X}_i \right) - \mu_V}{p < U_i}, \\
    \lambda_1\big(p\big) &=
        \Egiven{ F_V^{-1}\left( U_i \mid \vec{X}_i \right) - \mu_V}{U_i \leq p} = 
            -\lambda_0\big( p \big) \left(\frac{1-p}{p}\right), \text{ for } p \in (0,1).
\end{align*}

All relevant parameters --- $\alpha, \beta, \gamma, \delta, \varphi(.)$ --- are identified once we control for selection bias through the CFs $\lambda_0, \lambda_1$, with $\pi(z;\vec X_i)$ identified separately in the first-stage thanks to the instrument(s) $\vec X_i^{\text{IV}}$.
In the case that the CFs have an assumed functional form, then identification is complete.
For example, in the canonical Heckman selection model, the error terms follow a normal distribution, so that $\lambda_0, \lambda_1$ are the inverse Mills ratio.
If we do not know the distribution of $\big(U_{0,i}, U_{1,i}\big)$, then $\lambda_0, \lambda_1$ can be estimated separately with semi-parametric methods to avoid relying on parametric assumptions.\footnote{
    This comes at the cost of $\alpha$ and $\rho_0, \rho_1$ no longer being separately identified from $\lambda_0, \lambda_1$.
    However, this does not jeopardise identification and estimation of the ADE and AIE --- see \autoref{sec:semiparametric-mte}.
}

This identification strategy is an MTE approach \citep{bjorklund1987estimation,heckman2005structural} applied to a CM setting.
One can see this by noting the connection to the marginal effect of the mediator,
\begin{align*}
    & \Egiven{Y_i(z, 1) - Y_i(z, 0)}{Z_i = z, \vec X_i, U_i = p} \\
    &= \beta + \delta z +
    \underbrace{\Egiven{U_{1,i} - U_{0,i}}{\vec X_i, U_i = p}}_{
        = \rho_1\lambda_1(p) + \rho_1 \lambda_1'(p)p
            - \left[\rho_0\lambda_0(p) - \rho_0\lambda_0'(p)(1-p)\right]},
    \;\;\;\; \text{ for } z = 0,1,\;p \in (0,1).
\end{align*}
The marginal effect of the mediator is identified under the \ref{thm:mte-identification} assumptions, thanks to instrumental variation in $\vec X_i^{\text{IV}}$.
The final step uses the corresponding CFs to extrapolate from $\vec X_i^{\text{IV}}$ compliers to mediator compliers, and thus identify the ADE and AIE.

\newtheorem{theoremMTE}{Theorem}
\renewcommand\thetheoremMTE{MTE Approach}
\begin{theoremMTE}
    \label{thm:mte-identification}
    If assumptions \ref{mte:monotonicity}, \ref{mte:identification}, \ref{mte:instrument}, and the restricted-CF support condition hold, then the ADE and AIE are identified as a function of the parameters in Proposition \ref{proposition:secondstage}.
    \begin{align*}
    \text{ADE}
        &= \E{\gamma + \delta D_i}, \\
    \text{AIE}
        &= \mathbb E \Bigg[ \, \bar \pi \,
            \Big( \beta +  \delta Z_i +
                \underbrace{ (\rho_1 - \rho_0) \,
                \Gamma \big(\pi(0; \vec X_i), \, \pi(1; \vec X_i) \big)}_{
                    \text{Mediator compliers adjustment}} \Big) \Bigg]
    \end{align*}
    where $\Gamma\left(p,p'\right) 
    = \Egiven{ F_V^{-1}\left( U_i \mid \vec{X}_i \right) - \mu_V}{p < U_i \leq p'}
    = \frac{p'\lambda_1\left(p'\right) - p\lambda_1\left(p\right)}{p' - p}$ is the average unobserved net gains for those with unobserved costs between $p < p'$,\footnote{
        The complier adjustment term was first written in this manner by \cite{kline2019heckits} for an IV setting.
    } and $\bar\pi = \pi(1; \vec X_i) - \pi(0; \vec X_i)$ is the mediator complier score.
    Proof: see \appendixref{appendix:mte-ade-aie}.
\end{theoremMTE}

This theorem provides a solution to the identification problem for CM effects when facing selection;
rather than assuming away selection problems, it explicitly models them.
The ADE is straightforward to calculate as an average of the direct effect parameters, while the AIE also includes an adjustment for unobserved complier gains to the mediator.
Again, this is because the AIE only refers to individuals who were induced by treatment $Z_i$ into taking mediator $D_i$ (mediator compliers).
The MTE approach measures both selection bias and complier differences, and thus purges these persistent bias terms to identify CM effects.

In a simulation with Roy selection-into-mediator based on unobserved error terms, the MTE approach pushes conventional CM estimates back to the true value. 
\autoref{fig:cm-normal-dist} shows how the MTE-based approach corrects unadjusted CM effect estimates.

\section{MTE-based Estimation of CM Effects}
\label{sec:controlfun}
A conventional approach to estimating CM effects involves a two-stage approach to estimating the ADE and the AIE: the first-stage ($Z_i$ on $D_i$), and the second-stage ($Z_i, D_i$ on $Y_i$).
An MTE approach is a simple and intuitive addition to this approach: including the CF terms $\lambda_0, \lambda_1$ in the second-stage regression to address selection-into-mediator.

This section presents two practical estimation strategies.
First, I demonstrate how to estimate CM effects with an assumed distribution of error terms, focusing on the Heckman selection model as the leading case.
Second, I consider a more flexible semi-parametric approach that avoids distributional assumptions --- at the cost of semi-parametrically estimating the corresponding CFs.
While both methods effectively address the selection bias issues detailed in previous sections, they differ in their implementation complexity, efficiency, and underlying assumptions.

\subsection{Parametric MTE}
A parametric approach to MTEs solves the identification problem by assuming a distribution for the unobserved error terms in the first-stage selection model, and modelling selection based on this distribution.
The Heckman selection model is the most pertinent example, assuming the normal distribution for unobserved errors \citep{heckman1979sample}.
Assuming distributions other than the bivariate normal works in exactly the same manner, replacing the relevant density functions for those of an alternative distribution.
As such, this section focuses exclusively on the Heckman selection model.
This estimation approach is the same as the original parametric selection model definition of MTEs, in \cite{bjorklund1987estimation}.

The Heckman selection model assumes unobserved errors $V_i$ follow a normal distribution, so estimates the first-stage using a probit model.
\[ \Probgiven{D_i = 1}{Z_i, \vec X_i}
    = \Phi \big( \theta + \bar\pi Z_i + \vec\zeta' \vec X_i \big), \]
where $\Phi(.)$ is the cumulative density function for the standard normal distribution, and $\theta, \bar\pi, \vec\zeta$ are parameters estimated with maximum likelihood.
In the parametric case, an excluded instrument ($\vec X_i^{\text{IV}}$) is not technically necessary in the first-stage equation --- though not including one exposes the method to bias from misspecification if the errors are not normally distributed.
Thus, it is best practice to use this method with access to an instrument.

From this probit first-stage, construct the inverse Mills ratio terms to serve as the MTE-associated CFs.
These terms capture the correlation between unobserved factors influencing both mediator selection and outcomes, when the errors are normally distributed.
\[ \lambda_0(p) =
        \frac{\phi( - \Phi^{-1}(p) )}{\Phi( -\Phi^{-1}(p) )}, \;\;\;\;
    \lambda_1(p) =
        \frac{\phi( \Phi^{-1}(p) )}{\Phi( \Phi^{-1}(p) )},
        \;\;\;\; \text{ for } p \in (0,1) \]
where $\phi(.)$ is the probability density function for the standard normal distribution.

Lastly, the second-stage is estimated with OLS, including the MTE-associated CFs with plug-in estimates of the mediator propensity score, and $\vec \varphi'$ a linear approximation of nuisance function $\varphi(.)$.
\begin{align*}
    \Egiven{Y_i}{Z_i, D_i, \vec X_i} \;\; =& \;\;
        \alpha
        + \beta D_i
        + \gamma Z_i
        + \delta Z_i D_i
        + \vec \varphi' \vec X_i^- \\
        & \;\; + \rho_0 (1 - D_i) \lambda_0 \big( \hat \pi(Z_i; \vec X_i) \big)
    + \rho_1 D_i \lambda_1\big( \hat \pi(Z_i; \vec X_i) \big)
    + \varepsilon_i,
\end{align*}
where $\hat\pi \big(z;\vec X_i \big)$ are the predictions from the probit first-stage.

The resulting ADE and AIE estimates are composed from sample estimates of the terms in Theorem \ref{thm:mte-identification},
\[ \hat{\text{ADE}}
    = \hat{\gamma} + \hat{\delta}\,\bar D, \;\;\;\;
    \hat{\text{AIE}}
    = \hat{\bar\pi}\; \Big(
        \hat{\beta} + \hat{\delta}\,\bar Z
        + \big(\hat \rho_1 - \hat \rho_0 \big)
        \frac1n \sum_{i=1}^n \Gamma \big( \hat\pi(0;\vec X_i), \hat\pi(1;\vec X_i)\big) \Big) \]
where $\bar D = \frac1n \sum_{i=1}^n D_i$, $\bar Z = \frac1n \sum_{i=1}^n Z_i$,
$\hat{\bar\pi}$ is the estimate of the mean compliance rate, and $\frac1n \sum_{i=1}^n \Gamma(.,.)$ is the average of the complier adjustment term as a function of $\lambda_1$ with $\hat\pi \big(0; \vec X_i \big), \hat\pi \big(1; \vec X_i \big)$ values plugged in.

The standard errors for estimates can be computed using the delta method.
Specifically, accounting for sampling variability in both first-stage mediator propensity score estimation and second-stage causal effects estimation.
This approach yields $\sqrt{n}$-consistent estimates when the underlying error terms follow a bivariate normal distribution --- i.e., when $\pi(Z_i; \vec X_i)$ is correctly modelled by the probit first-stage.
Errors can also be estimated by the bootstrap, by including estimation of both the first and second-stage within each bootstrap iteration.

\begin{figure}[!h]
    \caption{The MTE Approach Addresses Persistent Bias in Conventional CM Estimates.}
    \begin{subfigure}[c]{0.475\textwidth}
        \centering
        \caption{$\hat{\text{ADE}} - \text{ADE}$.}
        \includegraphics[width=\textwidth]{
            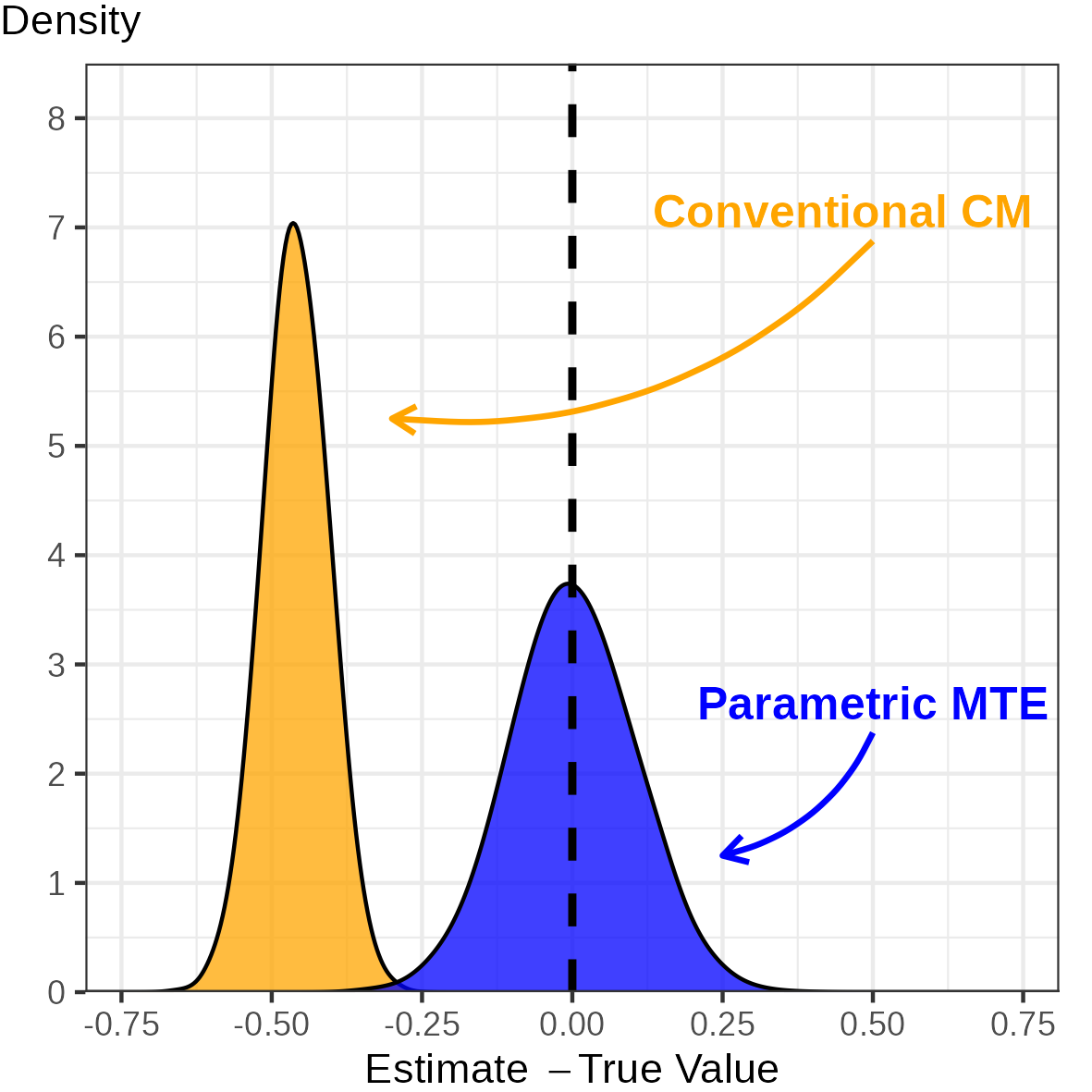}
    \end{subfigure}
    \begin{subfigure}[c]{0.475\textwidth}
        \centering
        \caption{$\hat{\text{AIE}} - \text{AIE}$.}
        \includegraphics[width=\textwidth]{
            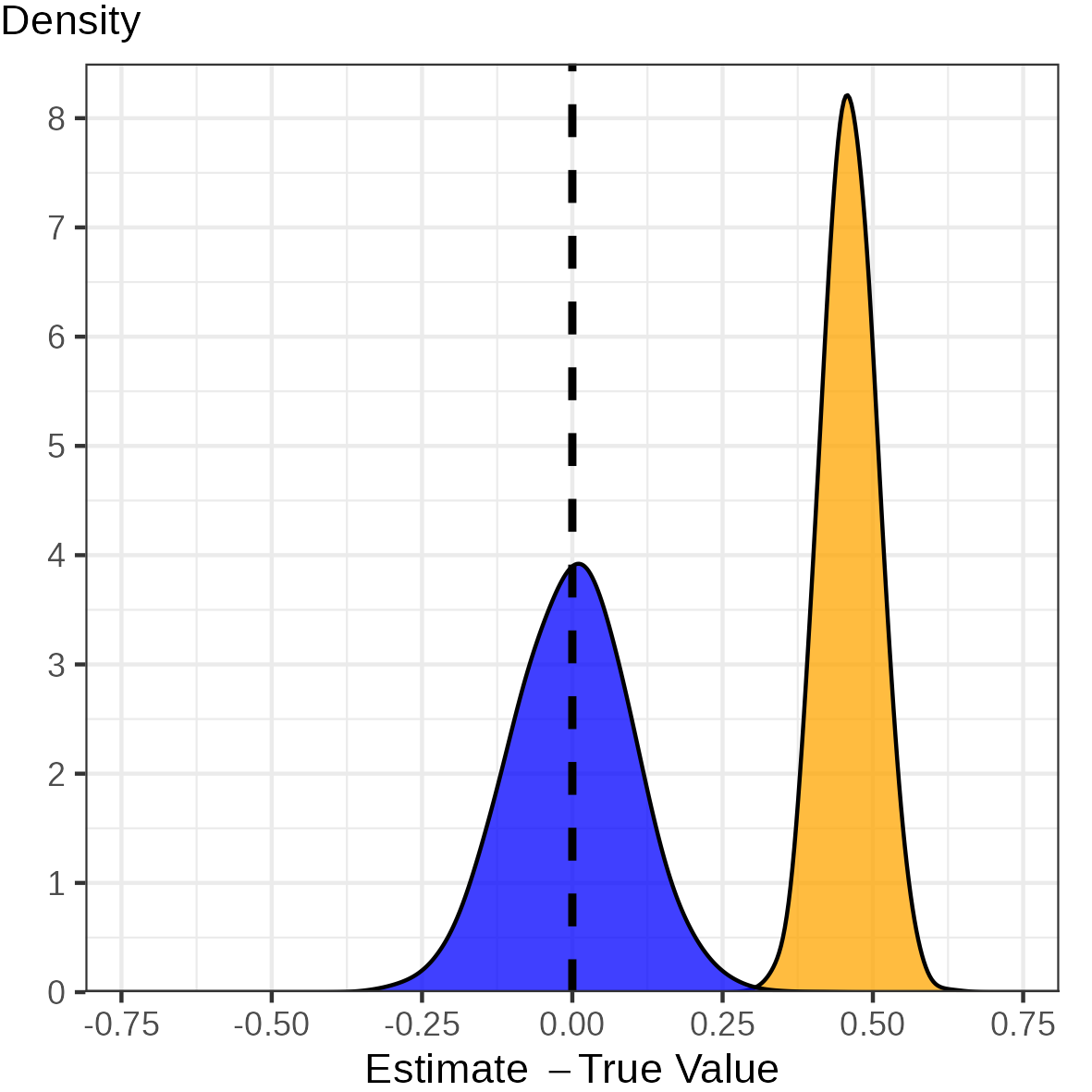}
    \end{subfigure}
    \label{fig:cm-normal-dist}
    \justify
    \footnotesize
    \textbf{Note:}
    These figures show the empirical density of point estimates minus the true average effect, for 10,000 different datasets generated from a Roy model with normally distributed error terms (with both correlated error terms and heteroscedasticity, further described in \autoref{sec:simulations}).
    The black dashed line is the true value;
    orange is the distribution of conventional CM estimates from two-stage OLS \citep{imai2010identification},
    and blue estimates with a two-stage Heckman selection adjustment.
\end{figure}

In practice, a parametric MTE approach is simple to implement using standard statistical packages.
The key advantage is computational simplicity and efficiency, particularly in moderate-sized samples.
However, this comes at the cost of strong distributional assumptions.
For example, if the error terms deviate substantially from joint normality, the estimates may be biased.\footnote{
    While this concern is immaterial in an IV setting estimating the LATE \citep{kline2019heckits}, it is pertinent in this setting as the CF extrapolates from IV compliers to mediator compliers.
}

\subsection{Semi-parametric MTE}
\label{sec:semiparametric-mte}
For settings where researchers are not comfortable specifying a specific distribution for the error terms, a semi-parametric MTE approach will nonetheless consistently estimate CM effects.
This method maintains the same identification strategy but avoids assuming a specific error distribution.
This estimation approach is used in the modern semi-parametric approach to estimating the distribution of MTEs, for example in \cite{brinch2017beyond,heckman2007econometric}.

The semi-parametric approach begins with flexible estimation of the first-stage, estimating the mediator propensity score,
\[ \Probgiven{D_i = 1}{Z_i, \vec X_i}
    = \pi\left(Z_i; \vec X_i \right), \]
where $\vec X_i$ must include the instrument(s) $\vec X_i^{\text{IV}}$.
This can be estimated using flexible methods, as long as the first-stage is estimated $\sqrt{n}$-consistently.\footnote{
    If an estimate of the first-stage that is not $\sqrt{n}$-consistent is used (e.g., a modern machine learning estimator), then the resulting second-stage estimate will not be $\sqrt{n}$-consistent.
}
An attractive option is the \cite{klein1993efficient} semi-parametric binary response model, which avoids relying on an assumed distribution of first-stage errors though requires a linear specification.
If it is important to avoid a linear specification, then a probability forest avoids linearity assumptions \citep{athey2019generalized} --- though is best used for cases with many columns in the $\vec X_i$ variables.

The second-stage is estimated with semi-parametric methods.
Consider the subsamples of mediator refusers and takers separately,
\begin{align*}
    \Egiven{Y_i}{Z_i, D_i = 0, \vec X_i} &=
        \alpha + \gamma Z_i + \varphi_0\big( \vec X_i^- \big)
        + \rho_0 \lambda_0 \big( \pi(Z_i ; \vec X_i) \big), \\
    \Egiven{Y_i}{Z_i, D_i = 1, \vec X_i} &=
        (\alpha + \beta) + (\gamma + \delta) Z_i + \varphi_1\big( \vec X_i^- \big)
        + \rho_1 \lambda_1 \big( \pi(Z_i ; \vec X_i) \big).
\end{align*}
The separated subsamples can be estimated, each individually, with semi-parametric methods.
The linear parameters (including linear approximations $\vec \varphi'_{d}$ of nuisance functions $\varphi_{d}(.)$)\footnote{
    Appropriate interactions between $Z_i, D_i$ and $\vec X_i$ can also flexibly control for $\vec X_i$, again avoiding linearity assumptions.
} can be estimated with OLS, while $\rho_0 \lambda_0$ and $\rho_1 \lambda_1$ take a flexible semi-parametric specification with first-stage estimates $\hat \pi(Z_i; \vec X_i)$ plugged in.
An attractive option is a series estimator, such as a spline specification, as this estimates the function without assuming a functional form but maintains $\sqrt n$-consistency.

The ADE is estimated by this approach as follows.
Take $\hat \gamma$, the $D_i = 0$ subsample estimate of $\E \gamma$, and $(\hat{\gamma + \delta})$, the $D_i = 1$ subsample estimate of $\E{\gamma + \delta}$, to give
\[ \hat{\text{ADE}}^{\text{Semi}}
    = (1 - \bar D) \, \hat \gamma + \bar D \, (\hat{\gamma + \delta}). \]

The AIE is less simple, for two reasons that differ from the parametric MTE setting.
First, the intercepts for each subsample, $\alpha$ and $(\alpha + \beta)$, are not separately identified from the MTE-associated CFs if the $\lambda_0, \lambda_1$ functions are flexibly estimated.
Second, a semi-parametric specification for the CFs mean $\rho_0$ and $\lambda_0$ are no longer separately identified from each other (and same for $\rho_1,\lambda_1$).
As such, it is not possible to directly use $\hat \lambda_0, \hat \lambda_1$ in estimating the complier adjustment term (as is done in the parametric case).

These problems can be avoided by estimating the AIE using its relation to the ATE.
Write $\hat{\text{ATE}}$ for the point-estimate of the ATE, and 
$\hat\delta = (\hat{\gamma + \delta}) - \hat\gamma$ for the point estimate of $\E \delta$, to give the following estimator,
\[ \hat{\text{AIE}}^{\text{Semi}}
    = \hat{\text{ATE}}
    - (1 - \bar Z) \, \left( 
        \frac 1n \sum_{i = 1}^n \hat\gamma + \hat \delta \, \hat\pi(1; \vec X_i) \right)
    - \bar Z \, \left(
        \frac 1n \sum_{i = 1}^n \hat\gamma + \hat \delta \, \hat\pi(0; \vec X_i)  \right), \]
where $\frac 1n \sum_{i = 1}^n \hat\gamma + \hat \delta \, \hat\pi(0; \vec X_i)$ estimates the ADE conditional on $Z_i = 0$, $\E{\gamma + \delta D_i(0)}$, and $\frac 1n \sum_{i = 1}^n \hat\gamma + \hat \delta \, \hat\pi(1; \vec X_i)$ estimates the ADE conditional on $Z_i = 1$, $\E{\gamma + \delta D_i(1)}$.
\appendixref{appendix:semi-parametric} describes the reasoning for this estimator of the AIE, relative to estimates of the ATE and ADE, in further detail.

This semi-parametric approach achieves valid estimation of the CM effects, without specifying the distribution behind unobserved error terms, and achieves desirable properties as long as the first-stage correctly estimates the mediator propensity score, and the structural assumptions hold true.
The standard errors for estimates can again be computed using the delta method, or estimated by the bootstrap --- again, across both first and second-stages within each bootstrap iteration.
Note that relying on propensity score estimation requires assumptions that can be found wanting in real-world settings; a common support condition for the mediator is required, and a semi-/non-parametric first-stage may become cumbersome if there are many control variables or many rows of data.

\subsection{Simulation Evidence}
\label{sec:simulations}
The following simulation gives an example to show how these methods work in practice.
Suppose data observed to the researcher $Z_i, D_i, Y_i, \vec X_i$ are drawn from the following data generating processes, for $i = 1, \hdots, N$, with 
$n = 5,000$ for this simulation.
\[ Z_i \sim \text{Binom}\left(0.5 \right),
    \;\; \vec X_i^- \sim N(4, 1),
    \;\; \vec X_i^{\text{IV}} \sim \text{Uniform}\left( -1, 1 \right),
    \;\; \left( U_{0,i}, U_{1,i}, U_{C,i} \right) \sim
    N\left( \vec 0, \mat \Sigma \right) \]
$\mat \Sigma$ is the matrix of parameters which controls the level of confounding from unobserved costs and benefits.\footnote{
    The correlation and relative standard deviations for $U_{0,i}, U_{1,i}$ affect how large selection bias in conventional CM estimates; correlation for these with unobserved costs $U_{C,i}$ does not particularly matter, though increased variance in unobserved costs makes estimates less precise for both OLS and MTE methods.
}

Each $i$ chooses to take mediator $D_i$ by a Roy model, with following mean definitions for each $z, d = 0, 1$
\begin{align*}
    D_i(z) &= \indicator{C_i \leq Y_i(z, 1) - Y_i(z, 0)},  \\
    \mu_{d}\left(z ; \vec X_i \right) &= \left( z + d + z d \right) + \vec X_i^-,
    \;\; \mu_{C}\left(z ; \vec X_i \right) = 3z + \vec X_i^- - \vec X_i^{\text{IV}}.
\end{align*}
Following \autoref{sec:regression}, these data have the following first and second-stage equations:
\begin{align*}
    D_i &= \indicator{U_{C,i} - \big( U_{1,i} - U_{0,i} \big)
    \leq -3Z_i + \vec X_i^- - \vec X_i^{\text{IV}}},  \\
    Y_i &= Z_i + D_i + Z_i D_i + \vec X_i^-
        + \left( 1 - D_i \right) U_{0,i} + D_i U_{1,i}.
\end{align*}
Treatment $Z_i$ has a causal effect on outcome $Y_i$, and it operates partially through mediator $D_i$.
Outcome mean $\mu_{D_i}\left( Z_i; \vec X_i \right)$ contains an interaction term, $Z_i D_i$, so while $Z_i,D_i$ have constant partial effects, the ATE depends on how many $i$ choose to take the mediator and there is treatment effect heterogeneity.

After $Z_i$ is assigned, $i$ chooses to take mediator $D_i$ by considering the costs and benefits --- which vary based on $Z_i$, demographic controls $\vec X_i$, and the (non-degenerate) unobserved error terms $U_{i,0}, U_{1,i}$.
As a result, sequential quasi-random assignment does not hold; the mediator is not conditionally ignorable.
Thus, a conventional approach to CM does not give an estimate for how much of the ATE goes through mediator $D$, but is contaminated by selection bias thanks to the unobserved error terms.

I simulate this data generating process 10,000 times, using $\mat\Sigma =
\left(\begin{smallmatrix} 1 & 0.75 & 0 \\ 0.75 & 2.25 & 0 \\ 0 & 0 & 0.25 \end{smallmatrix}\right)$,\footnote{
    This choice of parameters has $\Var{U_{0,i}} = 1, \Var{U_{1,i}} = 2.25, \text{Corr}\big(U_{0,i}, U_{1,i}\big) = 0.5$ so that unobserved errors meaningfully confound conventional CM methods, with notable heteroscedasticity.
    Unobserved costs are uncorrelated with $U_{0,i}, U_{1,i}$ (although non-zero correlation would not meaningfully change the results), and $\Var{U_{C,i}} = 0.25$ maintains uncertainty in unobserved costs.
}
and estimate CM effects with conventional CM methods (two-stage OLS) and the introduced MTE methods.
In this simulation $\Prob{D_i = 1} = 0.379$, and $65.77\%$ of the sample are mediator compliers (for whom $D_i(0)=0$ and $D_i(1)=1$).
This gives an ATE value of 2.60, ADE 1.38, and AIE 1.22, respectively.\footnote{
    Note that ATE $=$ ADE $+$ AIE in this setting.
    $\Prob{Z_i=1} = 0.5$ ensures this equality, but it is not guaranteed in general.
    See \appendixref{appendix:semi-parametric}.
}

\begin{figure}[h!]
    \caption{Simulated Distribution of CM Effect Estimates, Semi-parametric versus OLS, Relative to True Value.}
    \begin{subfigure}[c]{0.475\textwidth}
        \centering
        \caption{$\hat{\text{ADE}} - \text{ADE}$.}
        \includegraphics[width=\textwidth]{
            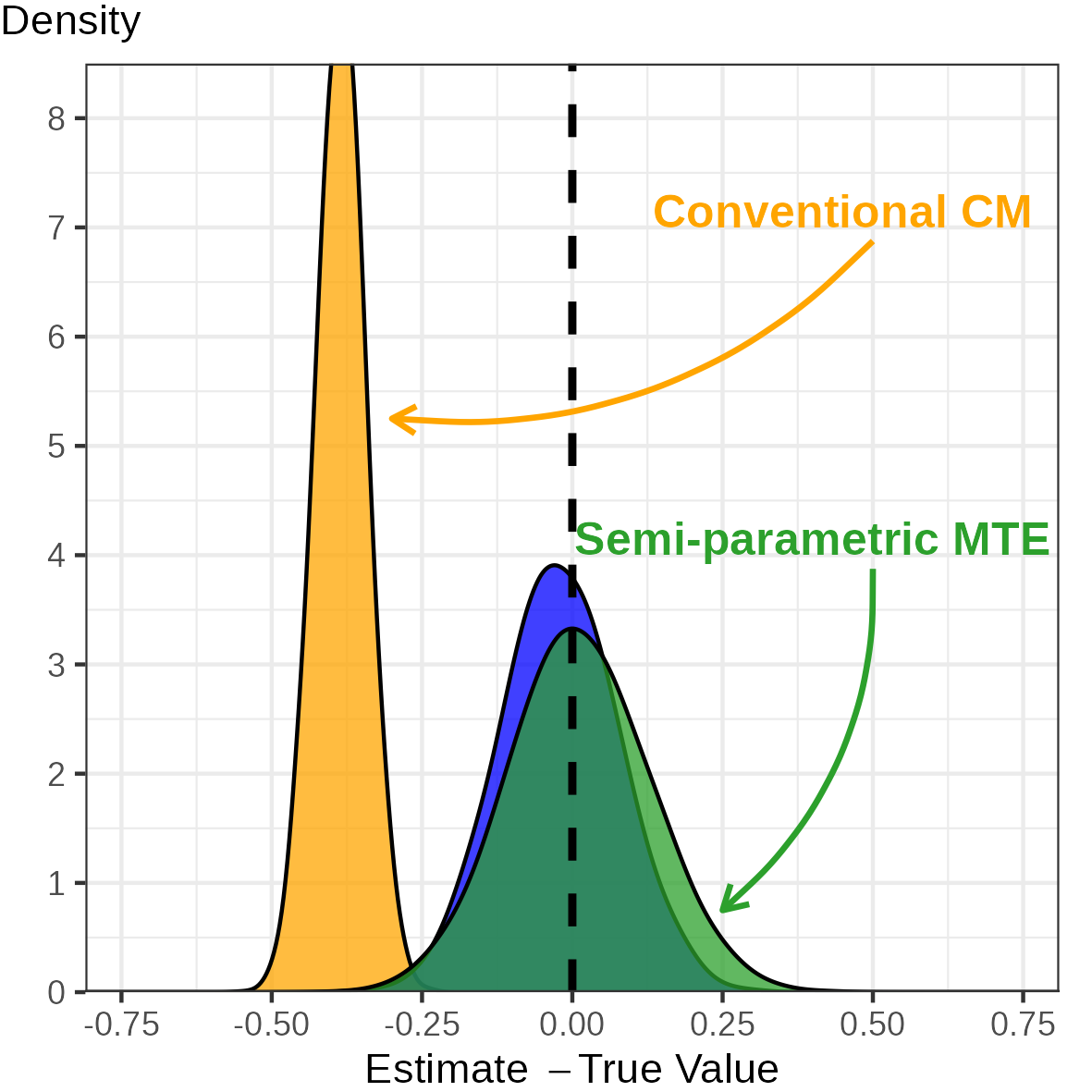}
    \end{subfigure}
    \begin{subfigure}[c]{0.475\textwidth}
        \centering
        \caption{$\hat{\text{AIE}} - \text{AIE}$.}
        \includegraphics[width=\textwidth]{
            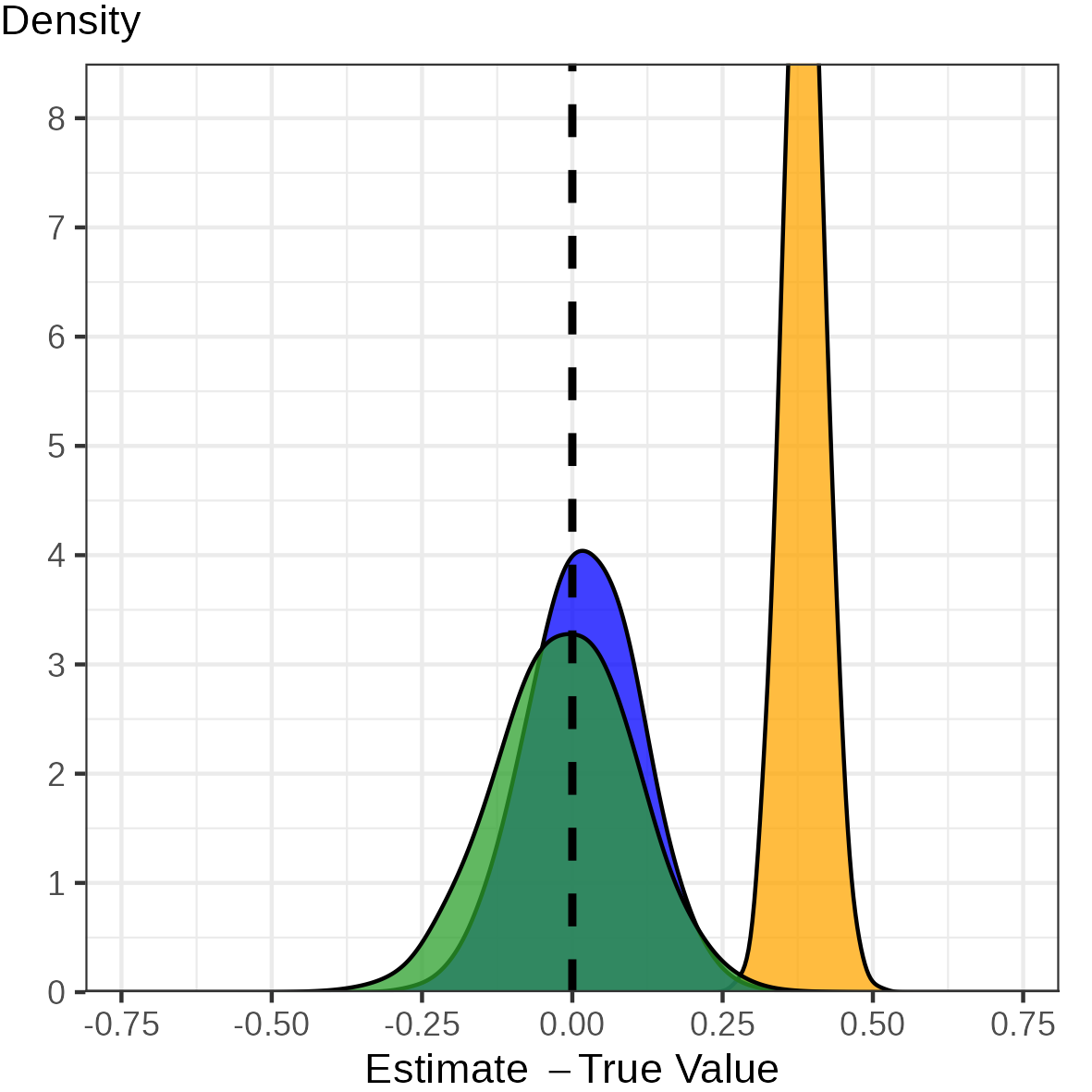}
    \end{subfigure}
    \label{fig:cm-uniform-dist}
    \justify
    \footnotesize    
    \textbf{Note:}
    These figures show the empirical density of point estimates minus the true average effect, for 10,000 different datasets generated from a Roy model with correlated uniformly distributed error terms.
    The black dashed line is the true value;
    orange is the distribution of conventional CM estimates from two-stage OLS \citep{imai2010identification},
    and green estimates with a two-stage semi-parametric MTE.
\end{figure}

\autoref{fig:cm-normal-dist} shows how these estimates perform, with a parametric MTE approach, relative to the true value.
The OLS estimates' distribution do not overlap the true values for any standard level of significance; the distance between the OLS estimates and the true values are the underlying bias terms derived in \autoref{thm:selection-bias}.
The parametric MTE approach perfectly reproduces the true values, as the probit first-stage correctly models the normally distributed error terms.
The semi-parametric approach (not shown in \autoref{fig:cm-normal-dist}) performs similarly, with a wider distribution; this is to be expected comparing a correctly specified parametric model with a semi-parametric one.

The parametric MTE may not be appropriate in setting with non-normal error terms.
I simulated the same data again, but transform $U_{0,i}, U_{1,i}$ to be correlated uniform errors (with the same standard deviations as previously).
\autoref{fig:cm-uniform-dist} shows the resulting distribution of point-estimates, relative to the truth, for the parametric and semi-parametric approaches.
The parametric MTE is slightly off target, showing persistent bias from incorrectly specifying the error term distribution.
The semi-parametric approach is centred exactly around the truth, with a slightly higher variance (as is expected).

\begin{figure}[h!]
    \caption{MTE-Based Estimates Work with Different Error Term Parameters.}
    \begin{subfigure}[c]{0.475\textwidth}
        \centering
        \caption{ADE.}
        \includegraphics[width=\textwidth]{
            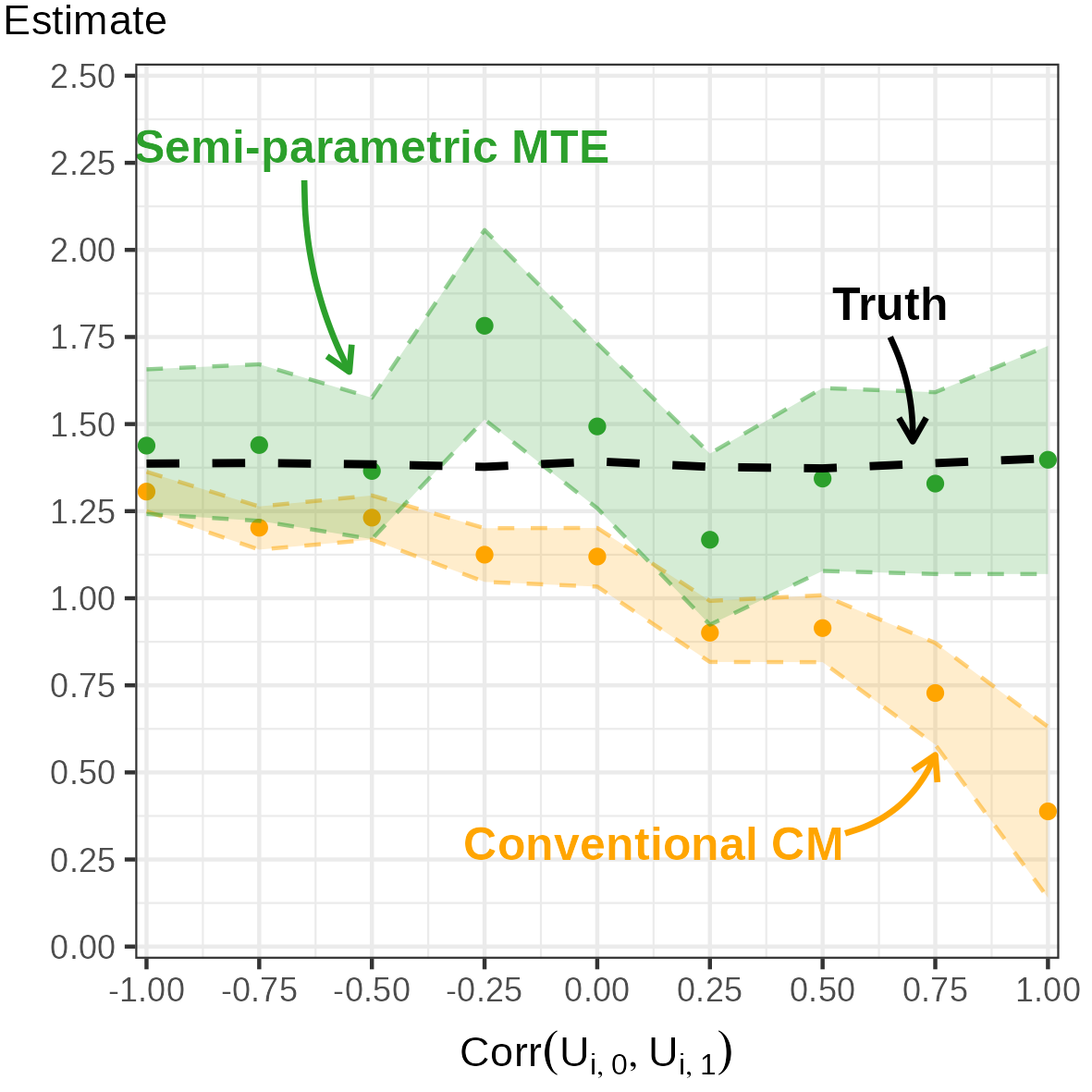}
    \end{subfigure}
    \begin{subfigure}[c]{0.475\textwidth}
        \centering
        \caption{AIE.}
        \includegraphics[width=\textwidth]{
            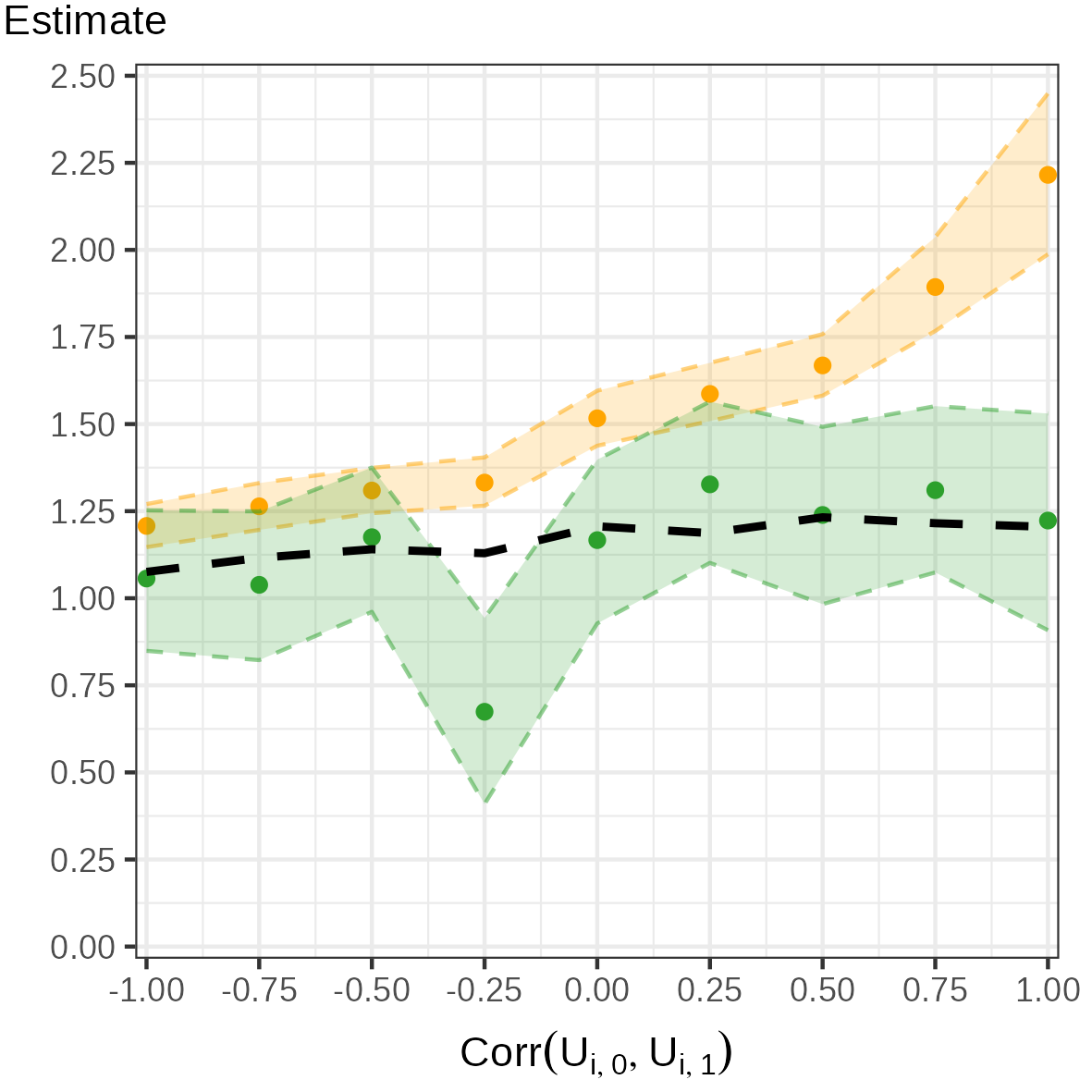}
    \end{subfigure}
    \label{fig:rho-bias}
    \justify
    \footnotesize    
    \textbf{Note:}
    These figures show the OLS and MTE-based point estimates of the ADE and AIE, for $n = 5,000$ sample size, varying $\text{Corr}\big(U_{0,i}, U_{1,i}\big)$ values with $\Var{U_{0,i}} = 1, \Var{U_{1,i}} = 1.5$ fixed.
    The black dashed line is the true value, coloured points are points estimates for the respective data generated, and shaded regions are the 95\% confidence intervals from 1,000 bootstraps each.
    Orange represents OLS estimates, green blue the semi-parametric MTE approach.
\end{figure}

The error terms determine the bias in OLS estimates of the ADE and AIE, so the bias varies for different values of the error-term parameters $\text{Corr}\big(U_{0,i}, U_{1,i}\big) \in [-1, 1]$ and $\Var{U_{0,i}}, \Var{U_{1,i}} \geq 0$.
The true AIE values vary, because $D_i(Z_i)$ compliers have higher average values of $U_{1,i} - U_{0,i}$ as $\text{Corr}\big(U_{0,i}, U_{1,i}\big)$ increases.
\autoref{fig:rho-bias} shows MTE-based estimates against estimates calculated by standard OLS, showing 95\% confidence intervals calculated from 1,000 bootstraps.
The point estimates of the MTE-based approach do not exactly equal the true values, as they are estimates from one simulation (not averages across many generated datasets, as in \autoref{fig:cm-uniform-dist}).
The MTE approach improves on OLS estimates by correcting for bias, with confidence regions overlapping the true values.\footnote{
    In the appendix, \autoref{fig:sigma-1-bias} shows the same simulation while varying $\Var{U_{1,i}}$, with fixed $\Var{U_{0,i}} = 1, \text{Corr}\big(U_{0,i}, U_{1,i}\big) = 0.5$.
    The conclusion is the same as for varying the correlation coefficient, $\rho$, in \autoref{fig:rho-bias}.
}
This correction did not come for free: the standard errors are significantly greater in an MTE approach than conventional CM estimates (based on two-stage OLS).
In this manner, this simulation shows the pros and cons of using the MTE approach to estimating CM effects in practice.

\section{CM in the Oregon Health Insurance Experiment}
\label{sec:oregon}
In the Oregon Health Insurance Experiment, winning the wait-list lottery significantly improved subjective health and well-being among participants. 
This study investigates the mechanisms behind these benefits, quantifying the extent to which improvements are mediated through increased healthcare usage.

To address concerns for unobserved selection-into-healthcare, I use the respondents' regular healthcare provider before the wait-list lottery as an IV for healthcare usage.
Approximately 73.2\% reported visiting a healthcare provider within the past year, but rates vary notably depending on their provider type: those who reported attended hospital emergency rooms (A\&E) and urgent care clinics before the wait-list lottery reported significantly lower healthcare visitation rates after the lottery (8.4 and 11 percentage points lower, respectively) than the 40\% who attended private clinics.\footnote{
    The combined $F$ statistic for the categorical variable for healthcare usual location (before the wait-list lottery) on healthcare usage (following the lottery) is 38.4.
}
The IV validity arises from differential costs faced by individuals based on their usual care provider.
Private clinics generally charge through health insurance and are more expensive without coverage, while A\&E and urgent care often provide costly services but rarely follow up on unpaid bills, effectively creating variation in healthcare attendance costs.
Additionally, individuals' choice of provider likely depend on neighbourhood-based access.

\begin{figure}[h!]
    \caption{Two-Stage Estimates in the Oregon Health Insurance Experiment.}
    \begin{subfigure}[c]{0.475\textwidth}
        \centering
        \caption{First-stage Propensity Score Estimates.}
        \includegraphics[width=\textwidth]{
            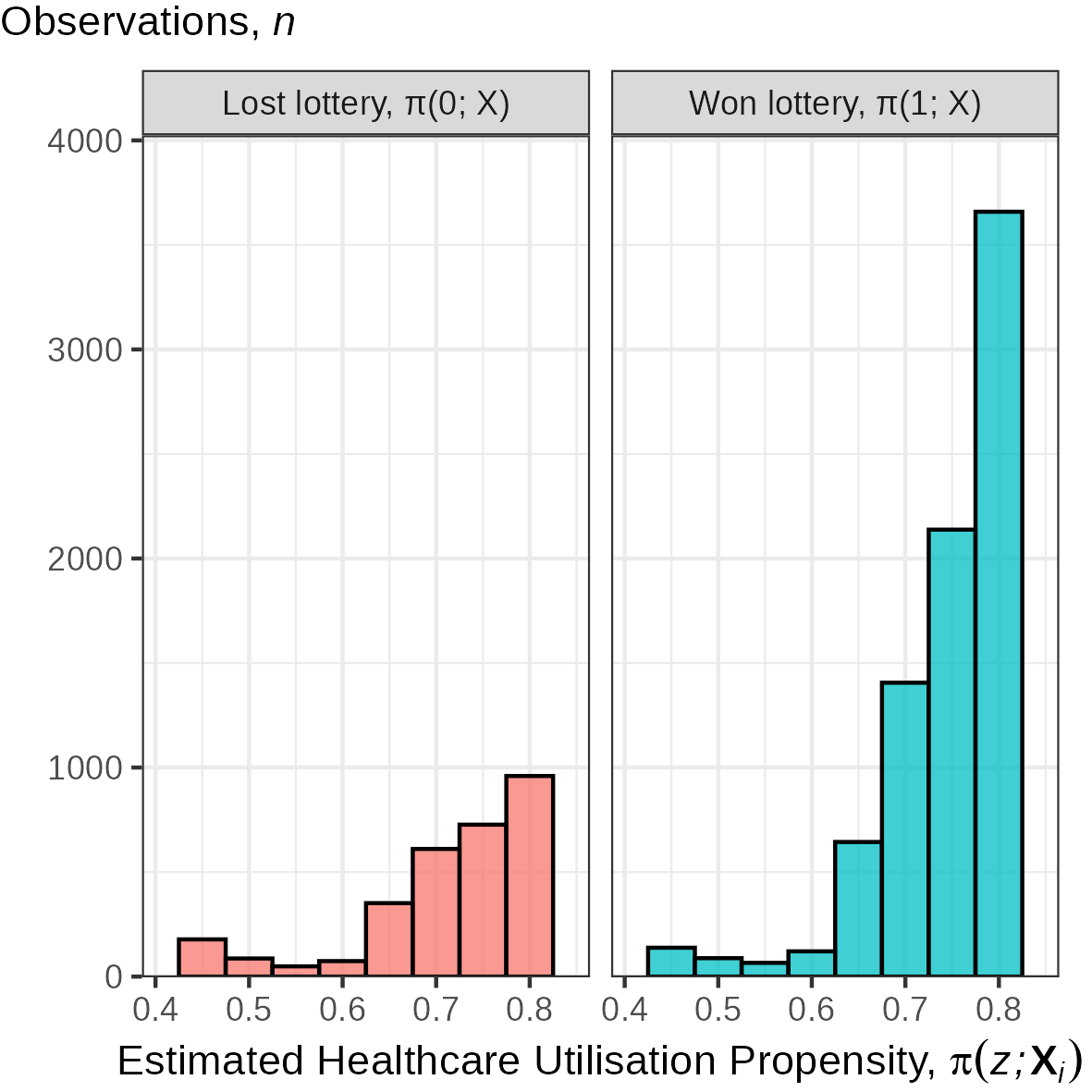}
    \end{subfigure}
    \begin{subfigure}[c]{0.475\textwidth}
        \centering
        \caption{MTE Estimates.}
        \includegraphics[width=\textwidth]{
            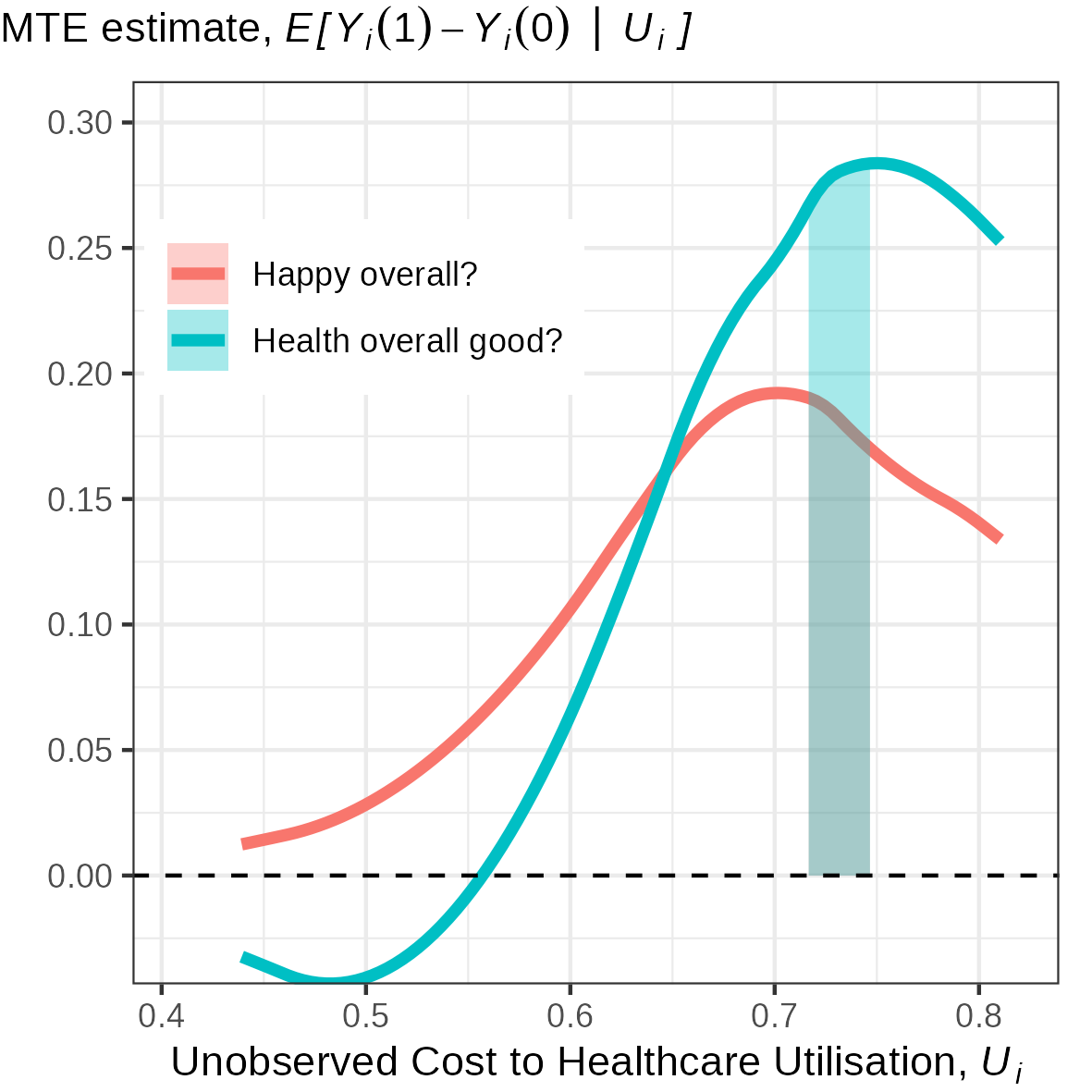}
    \end{subfigure}
    \label{fig:oregon-est}
    \justify
    \footnotesize    
    \textbf{Note:}
    Panel A shows the estimates of the first-stage propensity to use healthcare in the year following the health insurance experiment.
    It uses a logit specification, controlling for household size and whether losing the lottery $\pi(0; \vec X_i)$, or winning the lottery $\pi(1; \vec X_i)$.
    Average utilisaition after having lost the lottery is roughly $\Egiven{D_i}{Z_i = 0} = \E{\pi(0; \vec X_i)} = 0.715$, and after winning $\Egiven{D_i}{Z_i = 1} = \E{\pi(1; \vec X_i)} = 0.750$.
    Panel B shows the estimated MTE, using the spline specification in \autoref{sec:semiparametric-mte}.
    The shaded region refers to the AIE, where the average causal effect is calculated by integrating across the region of mediator compliers, with $\pi(0; \vec X_i) \leq U_i < \pi(1; \vec X_i)$ --- i.e., between 0.715 and 0.750 at the average covariate values.
\end{figure}

Initial results with unadjusted CM estimates suggest almost no mediating role for healthcare usage; the unadjusted estimates of the AIE are close to zero for both outcomes, contradicting intuitive suggestive evidence.
These estimates control for diagnosis of serious health conditions (before the wait-list lottery), such as kidney disease or diabetes.
However, this approach would not be considering possible unobserved confounding coming from underlying health conditions.
My MTE-based approach attempts to model this unobserved confounding, and applying them to these data reveals a much larger, positive AIE, restoring the mediating mechanism of healthcare usage in line with suggestive intuition.
This is because a correlational estimate of health and well-being gains to healthcare visits are practically zero, while the IV estimates restore positive gains and the MTE-based methods pick this up with a larger AIE estimate.
These numbers are reported in \autoref{tab:cm-oregon}, where panel A shows the CM effects with the binary outcome of subjective health, and panel B the binary outcome of subjective well-being.

\begin{table}[h!]
    \singlespacing
    \centering
    \small
    \caption{CM Effect Estimates for Wait-list Lottery Effects on Health and Well-being.}
    \begin{tabular}{l c c c c c}
        \\[-1.8ex]\hline \hline \\[-1.8ex] 
        & First-stage & ATE & ADE & AIE & AIE / ATE \\
        \cmidrule(lr){2-6}
        & (1) & (2) & (3) & (4) & (5) \\
        \midrule
        \multicolumn{1}{l}{\textbf{Panel A:} Health overall good?} \\
 Unadjusted &  4.700 &  4.500 &  4.700 & -0.190 & -0.043 \\ 
   & (0.880) & (0.910) & (0.910) & (0.063) & (0.018) \\ 
  Parametric MTE & 4.70 & 5.00 & 2.50 & 1.80 & 0.36 \\ 
   & (0.88) & (0.95) & (1.20) & (0.55) & (0.14) \\ 
  Semi-parametric MTE & 4.70 & 5.00 & 2.00 & 3.10 & 0.61 \\ 
   & (0.84) & (0.97) & (1.30) & (0.87) & (0.22) \\ 
  
        \\[-1.8ex]\hline \\[-1.8ex]
        \multicolumn{1}{l}{\textbf{Panel B:} Happy overall?} \\
 Unadjusted & 4.7000 & 7.1000 & 7.0000 & 0.0670 & 0.0094 \\ 
   & (0.8600) & (0.9600) & (0.9600) & (0.0520) & (0.0077) \\ 
  Parametric MTE & 4.70 & 7.50 & 5.00 & 1.90 & 0.25 \\ 
   & (0.870) & (0.990) & (1.100) & (0.510) & (0.075) \\ 
  Semi-parametric MTE & 4.70 & 7.50 & 5.00 & 2.50 & 0.34 \\ 
   & (0.84) & (0.93) & (1.20) & (0.74) & (0.11) \\ 
  
        \\[-1.8ex]\hline \\[-1.8ex]
    \end{tabular}
    \vspace{-0.125cm}
    \label{tab:cm-oregon}
    \justify
    \footnotesize
    \textbf{Note:}
    This table shows the point estimates (and SEs in brackets) of applying the proposed CM methods to replication data from the Oregon Health Insurance Experiment \citep{icspr2014oregon}.
    The first-stage column refers to the average effect of winning the wait-list lottery on healthcare usage (mediator first-stage), ATE average effect on surveyed health and well-being, ADE and AIE to respective CM effects through and absent healthcare usage.
    SEs were calculated with 1,000 bootstrap replications.
    The numbers are pp increases in the binary outcome, so an estimate of 4.7 in row 1 column 1 means an increase in 4.7 pp of using healthcare in the last 12 months after winning the wait-list lottery.
\end{table}

This reversal in conclusions highlights the importance of correcting for negative selection into healthcare usage.
A conventional approach to CM fails to account for the fact that individuals with poorer underlying health tend to visit healthcare providers more frequently, generating negative selection bias that obscures the true positive AIE, and clouds inference on the mediator mechanism.
By explicitly adjusting for this bias using the MTE approach, I isolate a credible positive indirect effect of healthcare usage on subjective health and well-being.

These findings offer credible evidence that improved healthcare access yields meaningful subjective health and well-being benefits, despite previous research emphasising negligible effects on objective health measures such as blood pressure \citep{baicker2013oregon}.
Subjective measures likely reflect broader psychological and financial relief associated with reduced healthcare-related anxiety and diminished risk of catastrophic medical debt, thus producing more noticeable short-term subjective improvements.

Nevertheless, this analysis is subject to notable limitations.
The IV is not ideal, and potentially more important mediators (such as explicit health insurance status) would require additional IVs beyond the wait-list lottery itself, presenting a challenging identification issue.
Furthermore, the 95\% confidence intervals for both ADE and AIE estimates (based on bootstrapped SEs) remain large, though statistically significant and excluding zero.
This uncertainty underscores common challenges in applied CM analyses, where statistical precision can be limited by data constraints.

\section{Summary and Concluding Remarks}
\label{sec:conclusion}
This paper has studied a selection-on-observables approach to CM in a natural experiment setting.
I have shown the pitfalls of using the most popular methods for estimating direct and indirect effects without a clear case for the mediator being quasi-randomly assigned.
Using the Roy model as a benchmark, a mediator is unlikely to be quasi-randomly assigned in natural experiment settings, and the bias terms likely crowd out inference regarding CM effects.

This paper has also contributed to the growing CM literature in economics, connecting to MTE methods and developing a compelling way of estimating direct and indirect effects in a natural experiment setting.
It has also recognised limitations in the common practice of suggestive evidence for mechanisms, and given credible CM estimates in a famous natural experiment setting well-known in the economics field.
Further research could build on the approaches presented by suggesting efficiency improvements, adjustments for common statistical irregularities (say, cluster dependence), or integrating the MTE approach to the growing double robustness literature on CM \citep{farbmacher2022causal,bia2024double}.

These findings do not provide a blanket endorsement for applied researchers to use CM methods.
There are strong structural assumptions for adjusting identifying CM effects despite unobserved selection-into-mediator, and inference requires an IV for mediator take-up.
If these assumptions do not hold true, then selection-adjusted estimates of CM effects will also be biased, and will not improve on an unadjusted conventional approach.

Yet, there are likely settings in which the structural assumptions are credible.
Mediator monotonicity aligns well with economic theory in many cases, and it is plausible for researchers to study big data settings with external variation in mediator take-up costs.
In these cases, this paper opens the door to identifying mechanisms behind treatment effects in natural experiment settings.


\singlespacing
\bibliographystyle{agsm}
\bibliography{sections/09-bibliography-doi.bib}

@article{angrist1998estimating,
 author = {Angrist, Joshua D},
 journal = {Econometrica},
 note = {\url{https://doi.org/10.2307/2998558}},
 number = {2},
 pages = {249--288},
 publisher = {Blackwell Publishing Inc.},
 title = {Estimating the labor market impact of voluntary military service using social security data on military applicants},
 volume = {66},
 year = {1998}
}

@book{angrist2009mostly,
 author = {Angrist, Joshua D and Pischke, J{\"o}rn-Steffen},
 publisher = {Princeton university press},
 title = {Mostly harmless econometrics: An empiricist's companion},
 year = {2009}
}

@article{athey2019generalized,
 author = {Athey, Susan and Tibshirani, Julie and Wager, Stefan},
 journal = {The Annals of Statistics},
 note = {\url{https://doi.org/10.1214/18-aos1709}},
 number = {2},
 pages = {1148--1178},
 title = {Generalized random forests},
 volume = {47},
 year = {2019}
}

@article{baicker2013oregon,
 author = {Baicker, Katherine and Taubman, Sarah L and Allen, Heidi L and Bernstein, Mira and Gruber, Jonathan H and Newhouse, Joseph P and Schneider, Eric C and Wright, Bill J and Zaslavsky, Alan M and Finkelstein, Amy N},
 journal = {New England Journal of Medicine},
 note = {\url{https://doi.org/10.1056/nejmsa1212321}},
 number = {18},
 pages = {1713--1722},
 publisher = {Mass Medical Soc},
 title = {The Oregon experiment—effects of Medicaid on clinical outcomes},
 volume = {368},
 year = {2013}
}

@article{bia2024double,
 author = {Bia, Michela and Huber, Martin and Laff{\'e}rs, Luk{\'a}{\v{s}}},
 journal = {Journal of Business \& Economic Statistics},
 note = {\url{https://doi.org/10.1080/07350015.2023.2271071}},
 number = {3},
 pages = {958--969},
 publisher = {Taylor \& Francis},
 title = {Double machine learning for sample selection models},
 volume = {42},
 year = {2024}
}

@article{bjorklund1987estimation,
 author = {Bj{\"o}rklund, Anders and Moffitt, Robert},
 journal = {The Review of Economics and Statistics},
 note = {\url{https://doi.org/10.2307/1937899}},
 pages = {42--49},
 title = {The estimation of wage gains and welfare gains in self-selection models},
 year = {1987}
}

@article{blackwell2024assumption,
 author = {Blackwell, Matthew and Ma, Ruofan and Opacic, Aleksei},
 journal = {arXiv preprint arXiv:2407.07072},
 note = {\url{https://doi.org/10.48550/arXiv.2407.07072}},
 title = {Assumption Smuggling in Intermediate Outcome Tests of Causal Mechanisms},
 year = {2024}
}

@article{brinch2017beyond,
 author = {Brinch, Christian N and Mogstad, Magne and Wiswall, Matthew},
 journal = {Journal of Political Economy},
 note = {\url{https://doi.org/10.1086/692712}},
 number = {4},
 pages = {985--1039},
 publisher = {University of Chicago Press Chicago, IL},
 title = {Beyond LATE with a discrete instrument},
 volume = {125},
 year = {2017}
}

@article{cinelli2024crash,
 author = {Cinelli, Carlos and Forney, Andrew and Pearl, Judea},
 journal = {Sociological Methods \& Research},
 note = {\url{https://doi.org/10.2139/ssrn.3689437}},
 number = {3},
 pages = {1071--1104},
 publisher = {Sage Publications Sage CA: Los Angeles, CA},
 title = {A crash course in good and bad controls},
 volume = {53},
 year = {2024}
}

@article{deuchert2019direct,
 author = {Deuchert, Eva and Huber, Martin and Schelker, Mark},
 journal = {Journal of Business \& Economic Statistics},
 note = {\url{https://doi.org/10.1080/07350015.2017.1419139}},
 number = {4},
 pages = {710--720},
 publisher = {Taylor \& Francis},
 title = {Direct and indirect effects based on difference-in-differences with an application to political preferences following the Vietnam draft lottery},
 volume = {37},
 year = {2019}
}

@article{ding2015adjust,
 author = {Ding, Peng and Miratrix, Luke W},
 journal = {Journal of Causal Inference},
 note = {\url{https://doi.org/10.1515/jci-2013-0021}},
 number = {1},
 pages = {41--57},
 publisher = {De Gruyter},
 title = {To adjust or not to adjust? Sensitivity analysis of M-bias and butterfly-bias},
 volume = {3},
 year = {2015}
}

@article{eisenhauer2015generalized,
 author = {Eisenhauer, Philipp and Heckman, James J and Vytlacil, Edward},
 journal = {Journal of Political Economy},
 note = {\url{https://doi.org/10.1086/679498}},
 number = {2},
 pages = {413--443},
 publisher = {University of Chicago Press Chicago, IL},
 title = {The generalized Roy model and the cost-benefit analysis of social programs},
 volume = {123},
 year = {2015}
}

@article{farbmacher2022causal,
 author = {Farbmacher, Helmut and Huber, Martin and Laff{\'e}rs, Luk{\'a}{\v{s}} and Langen, Henrika and Spindler, Martin},
 journal = {The Econometrics Journal},
 note = {\url{https://doi.org/10.1093/ectj/utac003}},
 number = {2},
 pages = {277--300},
 publisher = {Oxford University Press},
 title = {Causal mediation analysis with double machine learning},
 volume = {25},
 year = {2022}
}

@article{finkelstein2008oregon,
 author = {Finkelstein, Amy and Taubman, Sarah and Wright, Bill and Bernstein, Mira and Gruber, Jonathan and Newhouse, Joseph P. and Allen, Heidi and Baicker, Katherine and Oregon Health Study Group},
 eprint = {https://academic.oup.com/qje/article-pdf/127/3/1057/30456845/qjs020.pdf},
 issn = {0033-5533},
 journal = {The Quarterly Journal of Economics},
 month = {07},
 note = {\url{https://doi.org/10.1093/qje/qjs020}},
 number = {3},
 pages = {1057-1106},
 title = {The Oregon Health Insurance Experiment: Evidence from the First Year*},
 volume = {127},
 year = {2012}
}

@article{florens2008identification,
 author = {Florens, Jean-Pierre and Heckman, James J and Meghir, Costas and Vytlacil, Edward},
 journal = {Econometrica},
 note = {\url{https://doi.org/10.3982/ecta5317}},
 number = {5},
 pages = {1191--1206},
 publisher = {Wiley Online Library},
 title = {Identification of treatment effects using control functions in models with continuous, endogenous treatment and heterogeneous effects},
 volume = {76},
 year = {2008}
}

@article{flores2009identification,
 author = {Flores, Carlos A and Flores-Lagunes, Alfonso},
 journal = {IZA Discussion paper},
 note = {\url{https://doi.org/10.2139/ssrn.1423353}},
 publisher = {IZA Discussion paper},
 title = {Identification and estimation of causal mechanisms and net effects of a treatment under unconfoundedness},
 year = {2009}
}

@article{frolich2017direct,
 author = {Fr{\"o}lich, Markus and Huber, Martin},
 journal = {Journal of the Royal Statistical Society Series B: Statistical Methodology},
 note = {\url{https://doi.org/10.1111/rssb.12232}},
 number = {5},
 pages = {1645--1666},
 publisher = {Oxford University Press},
 title = {Direct and indirect treatment effects--causal chains and mediation analysis with instrumental variables},
 volume = {79},
 year = {2017}
}

@article{green2010enough,
 author = {Green, Donald P and Ha, Shang E and Bullock, John G},
 journal = {The Annals of the American Academy of Political and Social Science},
 note = {\url{https://doi.org/10.1177/0002716209351526}},
 number = {1},
 pages = {200--208},
 publisher = {Sage Publications Sage CA: Los Angeles, CA},
 title = {Enough already about “black box” experiments: Studying mediation is more difficult than most scholars suppose},
 volume = {628},
 year = {2010}
}

@article{heckman1974shadow,
 author = {Heckman, James},
 journal = {Econometrica: journal of the econometric society},
 note = {\url{https://doi.org/10.2307/1913937}},
 pages = {679--694},
 publisher = {JSTOR},
 title = {Shadow prices, market wages, and labor supply},
 year = {1974}
}

@article{heckman1979sample,
 author = {Heckman, James J},
 journal = {Econometrica: Journal of the econometric society},
 note = {\url{https://doi.org/10.2307/1912352}},
 pages = {153--161},
 publisher = {JSTOR},
 title = {Sample selection bias as a specification error},
 year = {1979}
}

@article{heckman1990empirical,
 author = {Heckman, James J and Honore, Bo E},
 journal = {Econometrica: Journal of the Econometric Society},
 note = {\url{https://doi.org/10.2307/2938303}},
 pages = {1121--1149},
 publisher = {JSTOR},
 title = {The empirical content of the Roy model},
 year = {1990}
}

@article{heckman1998characterizing,
 author = {Heckman, James and Ichimura, Hidehiko and Smith, Jeffrey and Todd, Petra},
 journal = {Econometrica},
 note = {\url{https://doi.org/10.2307/2999630}},
 number = {5},
 pages = {1017--1098},
 title = {Characterizing Selection Bias Using Experimental Data},
 volume = {66},
 year = {1998}
}

@article{heckman2004using,
 author = {Heckman, James and Navarro-Lozano, Salvador},
 journal = {Review of Economics and statistics},
 note = {\url{https://doi.org/10.3386/w9497}},
 number = {1},
 pages = {30--57},
 publisher = {MIT Press 238 Main St., Suite 500, Cambridge, MA 02142-1046, USA journals~…},
 title = {Using matching, instrumental variables, and control functions to estimate economic choice models},
 volume = {86},
 year = {2004}
}

@article{heckman2005structural,
 author = {Heckman, James J and Vytlacil, Edward},
 journal = {Econometrica},
 note = {\url{https://doi.org/10.1111/j.1468-0262.2005.00594.x}},
 number = {3},
 pages = {669--738},
 publisher = {Wiley Online Library},
 title = {Structural equations, treatment effects, and econometric policy evaluation 1},
 volume = {73},
 year = {2005}
}

@incollection{heckman2007econometric,
 author = {Heckman, James and Vytlacil, Edward},
 booktitle = {Handbook of Econometrics},
 chapter = {71},
 doi = {10.1016/S1573-4412(07)06071-0},
 edition = {1},
 editor = {Heckman, J.J. and Leamer, E.E.},
 publisher = {Elsevier},
 title = {Econometric Evaluation of Social Programs, Part II: Using the Marginal Treatment Effect to Organize Alternative Econometric Estimators to Evaluate Social Programs, and to Forecast their Effects in New Environments},
 volume = {6B},
 year = {2007}
}

@article{heckman2013understanding,
 author = {Heckman, James and Pinto, Rodrigo and Savelyev, Peter},
 journal = {American economic review},
 note = {\url{https://doi.org/10.3386/w18581}},
 number = {6},
 pages = {2052--2086},
 publisher = {American Economic Association},
 title = {Understanding the mechanisms through which an influential early childhood program boosted adult outcomes},
 volume = {103},
 year = {2013}
}

@article{heckman2015econometric,
 author = {Heckman, James J and Pinto, Rodrigo},
 journal = {Econometric reviews},
 note = {\url{https://doi.org/10.3386/w19314}},
 number = {1-2},
 pages = {6--31},
 publisher = {Taylor \& Francis},
 title = {Econometric mediation analyses: Identifying the sources of treatment effects from experimentally estimated production technologies with unmeasured and mismeasured inputs},
 volume = {34},
 year = {2015}
}

@article{huber2019review,
 author = {Huber, Martin},
 journal = {Handbook of labor, human resources and population economics},
 note = {\url{https://doi.org/10.1007/978-3-319-57365-6_162-1}},
 pages = {1--38},
 publisher = {Springer},
 title = {Mediation analysis},
 year = {2020}
}

@article{huber2020direct,
 author = {Huber, Martin and Hsu, Yu-Chin and Lee, Ying-Ying and Lettry, Layal},
 journal = {Journal of Applied Econometrics},
 note = {\url{https://doi.org/10.1002/jae.2765}},
 number = {7},
 pages = {814--840},
 publisher = {Wiley Online Library},
 title = {Direct and indirect effects of continuous treatments based on generalized propensity score weighting},
 volume = {35},
 year = {2020}
}

@misc{icspr2014oregon,
 author = {Finkelstein, Amy and Baicker, Katherine},
 note = {\url{https://doi.org/10.3886/ICPSR34314.v3}},
 publisher = {Inter-university Consortium for Political and Social Research},
 title = {Oregon Health Insurance Experiment, 2007-2010},
 year = {2014}
}

@article{imai2010identification,
 author = {Imai, Kosuke and Keele, Luke and Yamamoto, Teppei},
 journal = {Statistical Science},
 note = {\url{https://doi.org/10.1214/10-sts321}},
 pages = {51--71},
 title = {Identification, Inference and Sensitivity Analysis for Causal Mediation Effects},
 year = {2010}
}

@article{imai2013experimental,
 author = {Imai, Kosuke and Tingley, Dustin and Yamamoto, Teppei},
 journal = {Journal of the Royal Statistical Society Series A: Statistics in Society},
 note = {\url{https://doi.org/10.1111/j.1467-985x.2012.01032.x}},
 number = {1},
 pages = {5--51},
 publisher = {Oxford University Press},
 title = {Experimental designs for identifying causal mechanisms},
 volume = {176},
 year = {2013}
}

@article{imbens1994identification,
 author = {Imbens, Gw and Angrist, Jd},
 journal = {Econometrica},
 note = {\url{https://doi.org/10.2307/2951620}},
 number = {2},
 pages = {467--475},
 publisher = {John Wiley and Sons},
 title = {Identification and Estimation of Local Average Treatment Effects},
 volume = {62},
 year = {1994}
}

@article{klein1993efficient,
 author = {Klein, Roger W and Spady, Richard H},
 journal = {Econometrica},
 note = {\url{https://doi.org/10.2307/2951556}},
 pages = {387--421},
 publisher = {JSTOR},
 title = {An efficient semiparametric estimator for binary response models},
 year = {1993}
}

@article{kline2019heckits,
 author = {Kline, Patrick and Walters, Christopher R},
 journal = {Econometrica},
 note = {\url{https://doi.org/10.3982/ecta15444}},
 number = {2},
 pages = {677--696},
 publisher = {Wiley Online Library},
 title = {On Heckits, LATE, and numerical equivalence},
 volume = {87},
 year = {2019}
}

@article{kwon2024testing,
 author = {Kwon, Soonwoo and Roth, Jonathan},
 journal = {ArXiv preprint},
 note = {\url{https://doi.org/10.48550/arXiv.2404.11739}},
 title = {Testing Mechanisms},
 year = {2024}
}

@article{ludwig2011mechanism,
 author = {Ludwig, Jens and Kling, Jeffrey R and Mullainathan, Sendhil},
 journal = {Journal of economic Perspectives},
 note = {\url{https://doi.org/10.1257/jep.25.3.17}},
 number = {3},
 pages = {17--38},
 publisher = {American Economic Association},
 title = {Mechanism experiments and policy evaluations},
 volume = {25},
 year = {2011}
}

@manual{R2023,
 address = {Vienna, Austria},
 author = {{R Core Team}},
 note = {\url{https://www.R-project.org/}},
 organization = {R Foundation for Statistical Computing},
 title = {R: A Language and Environment for Statistical Computing},
 year = {2025}
}

@article{roy1951some,
 author = {Roy, Andrew Donald},
 journal = {Oxford economic papers},
 note = {\url{https://doi.org/10.1093/oxfordjournals.oep.a041827}},
 number = {2},
 pages = {135--146},
 publisher = {Oxford University Press},
 title = {Some thoughts on the distribution of earnings},
 volume = {3},
 year = {1951}
}

@article{sloczynski2022interpreting,
 author = {S{\l}oczy{\'n}ski, Tymon},
 journal = {Review of Economics and Statistics},
 note = {\url{https://doi.org/10.1162/rest_a_00953}},
 number = {3},
 pages = {501--509},
 publisher = {MIT Press One Rogers Street, Cambridge, MA 02142-1209, USA journals-info~…},
 title = {Interpreting OLS estimands when treatment effects are heterogeneous: Smaller groups get larger weights},
 volume = {104},
 year = {2022}
}

@article{tidyverse,
 author = {Hadley Wickham and Mara Averick and Jennifer Bryan and Winston Chang and Lucy D'Agostino McGowan and Romain François and Garrett Grolemund and Alex Hayes and Lionel Henry and Jim Hester and Max Kuhn and Thomas Lin Pedersen and Evan Miller and Stephan Milton Bache and Kirill Müller and Jeroen Ooms and David Robinson and Dana Paige Seidel and Vitalie Spinu and Kohske Takahashi and Davis Vaughan and Claus Wilke and Kara Woo and Hiroaki Yutani},
 journal = {Journal of Open Source Software},
 note = {\url{https://doi.org/10.21105/joss.01686}},
 number = {43},
 pages = {1686},
 title = {Welcome to the {tidyverse}},
 volume = {4},
 year = {2019}
}

@article{tingley2014mediation,
 author = {Tingley, Dustin and Yamamoto, Teppei and Hirose, Kentaro and Keele, Luke and Imai, Kosuke},
 journal = {Journal of statistical software},
 note = {\url{https://doi.org/10.18637/jss.v059.i05}},
 pages = {1--38},
 title = {Mediation: R package for causal mediation analysis},
 volume = {59},
 year = {2014}
}

@article{vytlacil2002independence,
 author = {Vytlacil, Edward},
 journal = {Econometrica},
 note = {\url{https://doi.org/10.1111/1468-0262.00277}},
 number = {1},
 pages = {331--341},
 publisher = {JSTOR},
 title = {Independence, monotonicity, and latent index models: An equivalence result},
 volume = {70},
 year = {2002}
}

@article{wood2016mgcv,
 author = {S.N. Wood and {N.} and {Pya} and B. S{"a}fken},
 journal = {Journal of the American Statistical Association},
 note = {\url{https://doi.org/10.1080/01621459.2016.1180986}},
 pages = {1548-1575},
 title = {Smoothing parameter and model selection for general smooth models (with discussion)},
 volume = {111},
 year = {2016}
}

\newpage
\onehalfspacing
\appendix
\setcounter{table}{0}
\renewcommand{\thetable}{A\arabic{table}}
\setcounter{figure}{0}
\renewcommand{\thefigure}{A\arabic{figure}}

\section{Supplementary Appendix}
\label{appendix}
This section is for supplementary information, and validation of presented propositions and theorems.
It is not meant for publication.

Any comments or suggestions may be sent to me at \href{mailto:seh325@cornell.edu}{\nolinkurl{seh325@cornell.edu}},
or raised as an issue on the Github project,
\url{https://github.com/shoganhennessy/mediation-natural-experiment}.


%
%

\subsection{Identification in Causal Mediation}
\label{appendix:identification}
\citet[Theorem~1]{imai2010identification} states that the ADE and AIE are identified under sequential quasi-random assignment, at each level of $Z_i = 0,1$.
For $z = 0,1$: \\
\makebox[\textwidth]{\parbox{1.25\textwidth}{
\begin{align*}
    \E{ Y_i(1, D_i(z)) - Y_i(0, D_i(z))}
    &= \int \int 
    \Big( \Egiven{ Y_i }{Z_i = 1, D_i, \vec X_i}
        - \Egiven{ Y_i }{Z_i = 0, D_i, \vec X_i} \Big)
            dF_{D_i \, | \, Z_i = z, \vec X_i} dF_{\vec X_i}, \\
    \E{ Y_i(z, D_i(1)) - Y_i(z, D_i(0))}
    &= \int \int \Egiven{ Y_i }{Z_i = z, D_i, \vec X_i}
    \Big( dF_{D_i \, | \, Z_i = 1, \vec X_i}
        - dF_{D_i \, | \, Z_i = 0, \vec X_i} \Big) dF_{\vec X_i}.
\end{align*}
}}
I focus on the averages, which are identified by consequence of the above.
\begin{align*}
    \E{ Y_i(1, D_i(Z_i)) - Y_i(0, D_i(Z_i))}
    &= \E[Z_i]{\Egiven{ Y_i(1, D_i(z)) - Y_i(0, D_i(z))}{Z_i = z}} \\
    \E{ Y_i(Z_i, D_i(1)) - Y_i(Z_i, D_i(0))}
    & = \E[Z_i]{\Egiven{ Y_i(z, D_i(1)) - Y_i(z, D_i(0))}{Z_i = z}}
\end{align*}
My estimand for the ADE is a simple rearrangement of the above.
The estimand for the AIE relies on a different sequence, relying on (1) sequential quasi-random assignment, (2) conditional monotonicity.
These give (1) identification equivalence of AIE local to mediator compliers conditional on $\vec X_i$ and AIE conditional on $\vec X_i$, LAIE $=$ AIE, (2) identification of the complier score.
\begin{align*}
    & \Egiven{ Y_i(Z_i, D_i(1)) - Y_i(Z_i, D_i(0))}{\vec X_i} \\
    & = \Probgiven{D_i(0) \neq D_i(1)}{\vec X_i}
        \Egiven{ Y_i(Z_i, 1) - Y_i(Z_i, 0)}{D_i(0) \neq D_i(1), \vec X_i} \\
    & = \Probgiven{D_i(0) \neq D_i(1)}{\vec X_i}
        \Egiven{ Y_i(Z_i, 1) - Y_i(Z_i, 0)}{\vec X_i} \\
    & = \Probgiven{D_i(0) \neq D_i(1)}{\vec X_i}
        \; \Big( \Egiven{Y_i}{Z_i, D_i = 1, \vec X_i}
            - \Egiven{Y_i}{Z_i, D_i = 0, \vec X_i} \Big) \\
    & = \Big( \Egiven{D_i}{Z_i = 1, \vec X_i} - \Egiven{D_i}{Z_i = 0, \vec X_i}
        \Big) \;
        \Big( \Egiven{Y_i}{Z_i, D_i = 1, \vec X_i}
            - \Egiven{Y_i}{Z_i, D_i = 0, \vec X_i} \Big)
\end{align*}
Monotonicity is not technically required for the above.
Breaking monotonicity would not change the identification in any of the above; it would be the same except replacing the complier score with a complier/defier score, $\Probgiven{D_i(0) \neq D_i(1)}{\vec X_i} = \Egiven{D_i}{Z_i = 1, \vec X_i} - \Egiven{D_i}{Z_i = 0, \vec X_i}$.


%
%

\subsection{Bias in Causal Mediation (CM) Estimands}
\label{appendix:mediation-bias}
Suppose that $Z_i$ is ignorable conditional on $\vec X_i$, but $D_i$ is not.

\subsubsection{Bias in the Average Direct Effect (ADE)}
To show that the conventional approach to mediation gives an estimate for the ADE with selection and group difference-bias, start with the components of the conventional estimands.
This proof starts with the relevant expectations, conditional on a specific value of $\vec X_i$ and $d \in \{0, 1 \}$.
\begin{align*}
    \Egiven{Y_i}{Z_i = 1, D_i = d, \vec X_i}
    =& \Egiven{Y_i(1, D_i(Z_i))}{D_i(1) = d, \vec X_i}, \\
    \Egiven{Y_i}{Z_i = 0, D_i = d, \vec X_i}
    =& \Egiven{Y_i(0, D_i(Z_i))}{D_i(0) = d, \vec X_i}
\end{align*}
And so,
\begin{align*}
    &  \Egiven{Y_i}{Z_i = 1, D_i = d, \vec X_i}
    - \Egiven{Y_i}{Z_i = 0, D_i = d, \vec X_i} \\
    &= \Egiven{Y_i(1, D_i(Z_i))}{D_i(1) = d, \vec X_i}
    - \Egiven{Y_i(0, D_i(Z_i))}{D_i(0) = d, \vec X_i} \\
    &= \Egiven{Y_i(1, D_i(Z_i)) - Y_i(0, D_i(Z_i))}{D_i(1) = d, \vec X_i} \\
    &\;\;\;\;+ \Egiven{Y_i(0, D_i(Z_i))}{D_i(1) = d, \vec X_i}
        - \Egiven{Y_i(0, D_i(Z_i))}{D_i(0) = d, \vec X_i}.
\end{align*}
The final term is a sum of the ADE, conditional on $D_i(1) = d$, and a selection bias term --- difference in baseline outcomes between the (partially overlapping) groups for whom $D_i(1) = d$ and $D_i(0) = d$.

To reach the final term, note the following.
\begin{align*}
    &\Egiven{Y_i(1, D_i(Z_i)) - Y_i(0, D_i(Z_i))}{\vec X_i} \\    
    &= \Egiven{Y_i(1, D_i(Z_i)) - Y_i(0, D_i(Z_i))}{D_i(1) = d, \vec X_i} \\
    &\;\;\;\;+ \Big(1 - \Probgiven{D_i(1) = d}{\vec X_i}\Big)
    \left( \begin{aligned}
        &\Egiven{Y_i(1, D_i(Z_i)) - Y_i(0, D_i(Z_i))}{D_i(1) = d, \vec X_i} \\ 
        & - \Egiven{Y_i(1, D_i(Z_i)) - Y_i(0, D_i(Z_i))}{D_i(1) = 1 - d, \vec X_i}
    \end{aligned} \right) 
\end{align*}
The second term is the difference between the ADE and LADE local to relevant complier groups.

Collect everything together, as follows.
\begin{align*}
    &  \Egiven{Y_i}{Z_i = 1, D_i = d, \vec X_i}
    - \Egiven{Y_i}{Z_i = 0, D_i = d, \vec X_i} \\
    =& \underbrace{
        \Egiven{Y_i(1, D_i(Z_i)) - Y_i(0, D_i(Z_i))}{\vec X_i}}_{
            \text{ADE, conditional on }\vec X_i} \\
    &+ \underbrace{
        \Egiven{Y_i(0, D_i(Z_i))}{D_i(1) = d, \vec X_i}
            - \Egiven{Y_i(0, D_i(Z_i))}{D_i(0) = d, \vec X_i}}_{
                \text{Selection bias}} \\
    &+ \underbrace{\Big(1 - \Probgiven{D_i(1) = d}{\vec X_i}\Big)
    \left( \begin{aligned}
        &\Egiven{Y_i(1, D_i(Z_i)) - Y_i(0, D_i(Z_i))}{D_i(1) = 1 - d, \vec X_i} \\ 
        & - \Egiven{Y_i(1, D_i(Z_i)) - Y_i(0, D_i(Z_i))}{D_i(1) = d, \vec X_i}
    \end{aligned} \right)}_{
        \text{group difference-bias}}
\end{align*}
The proof is achieved by applying the expectation across $D_i = d$, and $\vec X_i$.

\subsubsection{Bias in the Average Indirect Effect (AIE)}
To show that the conventional approach to mediation gives an estimate for the AIE with selection and group difference-bias, start with the definition of the ADE --- the direct effect among compliers times the size of the complier group.

This proof starts with the relevant expectations, conditional on a specific value of $\vec X_i$.
\begin{align*}
    &\Egiven{ Y_i(Z_i, D_i(1)) - Y_i(Z_i, D_i(0))}{\vec X_i} \\
    =& \Probgiven{D_i(0) \neq D_i(1)}{\vec X_i}
        \Egiven{ Y_i(Z_i, 1) - Y_i(Z_i, 0)}{D_i(0) \neq D_i(1), \vec X_i}
\end{align*}
When $D_i$ is not ignorable, the bias comes from estimating the second term,
\\ $\Egiven{ Y_i(Z_i, 1) - Y_i(Z_i, 0)}{D_i(0) \neq D_i(1), \vec X_i}$, the indirect effect among mediator compliers.

Let $z \in \{ 0, 1 \}$.
Again, note the mean outcomes in terms of average potential outcomes,
\begin{align*}
    \Egiven{Y_i}{Z_i = z, D_i = 1, \vec X_i}
    =& \Egiven{Y_i(z, 1)}{D_i = 1, \vec X_i}, \\
    \Egiven{Y_i}{Z_i = z, D_i = 0, \vec X_i}
    =& \Egiven{Y_i(z, 0)}{D_i = 0, \vec X_i}.
\end{align*}
Compose the selection bias term, as follows.
\begin{align*}
    & \Egiven{Y_i}{Z_i = z, D_i = 1, \vec X_i}
    - \Egiven{Y_i}{Z_i = z, D_i = 0, \vec X_i} \\
    &= \Egiven{Y_i(z, 1)}{D_i = 1, \vec X_i}
        - \Egiven{Y_i(z, 0)}{D_i = 0, \vec X_i} \\
    &= \Egiven{Y_i(z, 1) - Y_i(z, 0)}{D_i = 1, \vec X_i}
    + \Egiven{Y_i(z, 0)}{D_i = 1, \vec X_i} - \Egiven{Y_i(z, 0)}{D_i = 0, \vec X_i}
\end{align*}
The final term is a sum of the AIE, among the treated group $D_i = 1$, and a selection bias term --- difference in baseline potential outcomes between the groups for whom $D_i = 1$ and $D_i = 0$.

The AIE is the indirect effect among compliers times the size of the complier group, so we need to compensate for the difference between the treated group $D_i = 1$ and complier group $D_i(0) \neq D_i(1)$.

Start with the difference between treated groups average and overall average.
\begin{align*}
    & \Egiven{Y_i(z, 1) - Y_i(z, 0)}{D_i = 1, \vec X_i} \\
    =& \Egiven{Y_i(z, 1) - Y_i(z, 0)}{\vec X_i} \\
    &+ \Big(1 - \Probgiven{D_i = 1}{\vec X_i} \Big)
    \left( \begin{aligned}
        &\Egiven{Y_i(z, 1) - Y_i(z, 0)}{D_i = 1, \vec X_i} \\ 
        &  - \Egiven{Y_i(z, 1) - Y_i(z, 0)}{D_i = 0, \vec X_i}
    \end{aligned} \right)
\end{align*}
Then the difference between the compliers' average and the overall average.
\begin{align*}
    & \Egiven{ Y_i(z, 1) - Y_i(z, 0)}{D_i(0) \neq D_i(1), \vec X_i} \\
    =& \Egiven{Y_i(z, 1) - Y_i(z, 0)}{\vec X_i} \\
    & + \frac{1 - \Probgiven{D_i(0) \neq D_i(1)}{\vec X_i} }{
        \Probgiven{D_i(0) \neq D_i(1)}{\vec X_i}}
    \left( \begin{aligned}
        &\Egiven{Y_i(z, 1) - Y_i(z, 0)}{D_i(1) = 0 \text{ or } D_i(0)=1, \vec X_i} \\ 
        &  - \Egiven{Y_i(z, 1) - Y_i(z, 0)}{\vec X_i}
    \end{aligned} \right)
\end{align*}

Collect everything together, as follows.
\begin{align*}
    &  \Egiven{Y_i}{Z_i = z, D_i = 1, \vec X_i}
    - \Egiven{Y_i}{Z_i = z, D_i = 0, \vec X_i} \\
    =& \underbrace{
        \Egiven{Y_i(z, 1) - Y_i(z, 0)}{D_i(1) =1, D_i(0)=0,\vec X_i}}_{
            \text{AIE among compliers, conditional on }\vec X_i, Z_i = z} \\
    &+ \underbrace{
        \Egiven{Y_i(z, 0)}{D_i = 1, \vec X_i}
            - \Egiven{Y_i(z, 0)}{D_i = 0, \vec X_i}}_{
                \text{Selection bias}} \\
    &+ \underbrace{\left[ \begin{aligned}
        &\Big( 1 - \Probgiven{D_i = 1}{\vec X_i} \Big)
        \left( \begin{aligned}
            &\Egiven{Y_i(z, 1) - Y_i(z, 0)}{D_i = 1, \vec X_i} \\ 
            &  - \Egiven{Y_i(z, 1) - Y_i(z, 0)}{D_i = 0, \vec X_i}
        \end{aligned} \right) \\
        &- \frac{1 - \Probgiven{D_i(0) \neq D_i(1)}{\vec X_i} }{
            \Probgiven{D_i(0) \neq D_i(1)}{\vec X_i}} 
        \left( \begin{aligned}
            &\Egiven{Y_i(z, 1) - Y_i(z, 0)}{D_i(1) = 0 \text{ or } D_i(0)=1, \vec X_i} \\ 
            &  - \Egiven{Y_i(z, 1) - Y_i(z, 0)}{\vec X_i}
        \end{aligned} \right)
    \end{aligned} \right] }_{
        \text{group difference-bias}}
\end{align*}
The proof is finally achieved by multiplying by the complier score, 
$\Probgiven{D_i(0) \neq D_i(1)}{\vec X_i}$
$= \Egiven{D_i}{Z_i = 1, \vec X_i} - \Egiven{D_i}{Z_i = 0, \vec X_i}$,
then applying the expectation across $Z_i = z$, and $\vec X_i$.

\subsection{A Regression Framework for Direct and Indirect Effects}
\label{appendix:regression-model}
Put $\mu_{d}(z; \vec X) = \Egiven{Y_i(z, d)}{\vec X}$ and $U_{d, i} = Y_i(z, d) - \mu_{d}(z; \vec X)$ for each $z,d=0,1$, so we have the following expressions:
\[ Y_i(Z_i, 0)
        = \mu_{0}(Z_i; \vec X_i) + U_{0,i}, \;\;
    Y_i(Z_i, 1)
        = \mu_{1}(Z_i; \vec X_i) + U_{1,i}. \]
$U_{0,i}, U_{1,i}$ are error terms with unknown distributions, mean independent of $Z_i, \vec X_i$ by definition --- but possibly correlated with $D_i$.
$Z_i$ is conditionally independent of potential outcomes, so that $U_{0,i}, U_{1,i} \indep Z_i$.

The first-stage regression of $Z \to Y$ has unbiased estimates, since $Z_i \indep D_i(.) \big| \vec X_i$.
Put $\pi(z; \vec X) = \Egiven{D_i(z)}{\vec X}$, and $\eta_{z, i} = D_i(z) - \pi(z; \vec X)$ the first-stage error terms.
\begin{align*}
    D_i &= Z_i D_i(1) + (1 - Z_i) D_i(0) \\
        &= D_i(0) +
            Z_i \left[ D_i(1) - D_i(0) \right] \\
        &= \underbrace{\pi(0; \vec X_i) 
        }_{\text{Intercept, } \coloneqq \theta + \zeta(\vec X_i)} +
            \underbrace{Z_i \big( \pi(1; \vec X_i) - \pi(0; \vec X_i) \big)}_{
                \text{Regressor, } \coloneqq \bar\pi Z_i }
        + \underbrace{(1- Z_i) \eta_{0,i} + Z_i \eta_{1,i}}_{
            \text{Errors, } \coloneqq \eta_i} \\
    \implies \Egiven{D_i}{Z_i, \vec X_i}
        &= \theta + \bar \pi Z_i + \zeta(\vec X_i).
\end{align*}
Since the quasi-random assignment assumption gives $\Egiven{Z_i \eta_{z,i}}{\vec X_i} = \Egiven{Z_i}{\vec X_i} \Egiven{\eta_{z,i}}{\vec X_i} = 0$, for each $z =0,1$.
By the same argument $Z_i$ is also assumed independent of potential outcomes $Y_i(.,.)$, so that $U_{0,i}, U_{1,i} \indep Z_i$.
Thus, the reduced form regression $Z \to Y$ also leads to unbiased estimates for the ATE.

The same cannot be said of the regression that estimates direct and indirect effects, without further assumptions.
\begin{align*}
    Y_i &= Z_i Y_i(1, D_i(1)) + (1 - Z_i) Y_i(0, D_i(0)) \\
        &= Z_i D_i Y_i(1, 1) \\
        & \;\;\;\; + (1 - Z_i) D_i Y_i(0, 1) \\
        & \;\;\;\; + Z_i (1 - D_i) Y_i(1, 0) \\
        & \;\;\;\; + (1 - Z_i) (1 - D_i) Y_i(0, 0) \\
        &= Y_i(0, 0) \\
        & \;\;\;\; + Z_i \left[Y_i(1, 0) - Y_i(0, 0) \right] \\
        & \;\;\;\; + D_i \left[Y_i(0, 1) - Y_i(0, 0) \right] \\
        & \;\;\;\; + Z_i D_i \left[Y_i(1, 1) - Y_i(1, 0)
            - \left( Y_i(0, 1) - Y_i(0, 0) \right)\right]
\end{align*}
And so $Y_i$ can be written as a regression equation in terms of the observed factors and error terms.
\begin{align*}
    Y_i &= \mu_0(0; \vec X_i) \\
        & \;\;\;\; + D_i \left[\mu_1(0; \vec X_i) - \mu_0(0; \vec X_i) \right] \\
        & \;\;\;\; + Z_i \left[\mu_0(1; \vec X_i) - \mu_0(0; \vec X_i) \right] \\
        & \;\;\;\; + Z_i D_i \left[\mu_1(1; \vec X_i) - \mu_0(1; \vec X_i)
            - \left( \mu_1(0; \vec X_i) - \mu_0(0; \vec X_i) \right)\right] \\
        & \;\;\;\; + U_{0,i} + D_i \left( U_{1,i} - U_{0,i} \right) \\
        &=
            \alpha + \beta D_i + \gamma Z_i + \delta Z_i D_i
            + \varphi(\vec X_i)
            + \left( 1 - D_i \right) U_{0,i} + D_i U_{1,i}
\end{align*}
With the following definitions:
\begin{enumerate}[label=\textbf{(\alph*)}]
    \item $\alpha = \E{\mu_0(0; \vec X_i)}$ and $\varphi(\vec X_i) = \mu_0(0; \vec X_i) - \alpha$ are the intercept terms.
    \item $\beta = \mu_1(0; \vec X_i) - \mu_0(0; \vec X_i)$ is the indirect effect under $Z_i = 0$
    \item $\gamma = \mu_0(1; \vec X_i) - \mu_0(0; \vec X_i)$ is the direct effect under $D_i = 0$.
    \item $\delta = \mu_1(1; \vec X_i) - \mu_0(1; \vec X_i)- \left( \mu_1(0; \vec X_i) - \mu_0(0; \vec X_i) \right)$ is the interaction effect.
    \item $\left( 1 - D_i \right) U_{0,i} + D_i U_{1,i}$ is the remaining error term.
\end{enumerate}
This sequence gives us the resulting regression equation:
\begin{align*}
    \Egiven{Y_i}{Z_i, D_i, \vec X_i} \;\; =& \;\;
        \alpha
        + \beta D_i
        + \gamma Z_i
        + \delta Z_i D_i
        + \varphi(\vec X_i) \\
        & \;\; +\left( 1 - D_i \right) \Egiven{ U_{0,i} }{D_i = 0, \vec X_i}
            + D_i \Egiven{ U_{1,i} }{D_i = 1, \vec X_i}
\end{align*}
Taking the conditional expectation, and collecting for the expressions of the direct and indirect effects:
\begin{align*}
    \E{Y_i(1, D_i(Z_i)) - Y_i(0, D_i(Z_i))}
        &= \E{\gamma + \delta D_i} \\
    \E{Y_i(Z_i, D_i(1)) - Y_i(Z_i, D_i(0))}
        &= \E{\bar \pi \left( \beta +  Z_i \delta + \tilde U_i \right)}
\end{align*}
These equations have simpler expressions after assuming constant treatment effects in a linear framework;
I have avoided this as having compliers, and controlling for observed factors $\vec X_i$ only makes sense in the case of heterogeneous treatment effects.

These terms are conventionally estimated in a simultaneous regression \citep{imai2010identification}.
If sequential quasi-random assignment does not hold, then the regression estimates from estimating the mediation equations (without adjusting for the contaminated bias term) suffer from omitted variables bias.

\makebox[\textwidth]{\parbox{1.25\textwidth}{
\begin{align*}
    \E[\vec X_i]{\Egiven{Y_i}{Z_i = D_i = 0, \vec X_i}}
        &= \E\alpha + \Egiven{ U_{0,i} }{D_i = 0} \\
    \E[\vec X_i]{\Egiven{Y_i}{Z_i = 0, D_i = 1, \vec X_i}
        - \Egiven{Y_i}{Z_i = 0, D_i = 0, \vec X_i}}
        &= \E\beta + 
            \left( \Egiven{ U_{1,i} }{D_i = 1} - \Egiven{ U_{0,i} }{D_i = 0} \right) \\
    \E[\vec X_i]{\Egiven{Y_i}{Z_i = 1, D_i = 0, \vec X_i}
        - \Egiven{Y_i}{Z_i = 0, D_i = 0, \vec X_i}}
        &= \E\gamma + \Egiven{ U_{0,i} }{D_i = 0} \\
    \E[\vec X_i]{\begin{aligned}
        &\Egiven{Y_i}{Z_i = 1, D_i = 1, \vec X_i}
            - \Egiven{Y_i}{Z_i = 1, D_i = 0, \vec X_i} \\
            &- \left( \Egiven{Y_i}{Z_i = 0, D_i = 1, \vec X_i}
                - \Egiven{Y_i}{Z_i = 0, D_i = 0, \vec X_i} \right)
        \end{aligned}}
    &= \E\delta
\end{align*}
}}
And so the ADE and AIE estimates are contaminated by these bias terms.
Additionally, the AIE estimates refers to gains from the mediator among $D(z)$ compliers (not the entire average), so will be biased when not accounting for $\tilde U_i$, too.

\subsection{Roy Model and Sequential quasi-random assignment}
\label{appendix:roy-seq-ig}
\textit{Proof of Proposition \ref{prop:roy-seq-ig}.}

Suppose $Z_i$ is ignorable, and selection-into-$D_i$ follows a Roy model, with the definitions in \autoref{sec:applied}.
If selection-into-$D_i$ is degenerate on $U_{0,i}, U_{1,i}$:
\[ \Egiven{D_i}{Z_i, \vec X_i, U_{1,i}- U_{0,i} = u}
    = \Egiven{D_i}{Z_i, \vec X_i, U_{1,i}- U_{0,i} = u'},
    \text{ for all } u, u' \text{ in the range of $U_{1,i}- U_{0,i}$.} \]
In this case, the control set $\vec X_i$ and the costs $\mu_c, U_{c,i}$ are the only determinants of selection-into-$D_i$ --- and, $U_{0,i}, U_{1,i}$ play no role.
This could be achieved by either assuming that unobserved gains are degenerate (the researcher had observed everything in $\vec X_i$), or selection-into-$D_i$ had been disrupted in some fashion (e.g., by a natural experiment design for $D_i$).

To motivate a contraposition argument, suppose $D_i$ is ignorable conditional on $Z_i, \vec X_i$.
For each $z, d = 0, 1$
\begin{align*}
    & D_i \indep Y_i(z, d) \;\; | \;\; \vec X_i, Z_i = z \\
    &\implies D_i \indep \mu_{d}(z; \vec X_i) + U_{d,i} \;\; | \;\; \vec X_i, Z_i = z \\
    &\implies D_i \indep U_{d,i} \;\; | \;\; \vec X_i, Z_i = z \\
    &\implies D_i \indep U_{1,i} - U_{0,i} \;\; | \;\; \vec X_i, Z_i = z \\
    &\implies \Egiven{D_i}{U_{1,i} - U_{0,i} = u', \vec X_i, Z_i = z}
    = \Egiven{D_i}{\vec X_i, Z_i = z} \\
    & \;\;\; \;\;\; \;\;\; \text{for all } u' \text{ in the range of $U_{1,i}- U_{0,i}$.}
\end{align*}
This final implication is that selection-into-$D_i$ is degenerate on $U_{0,i}, U_{1,i}$.
Thus, a contraposition argument has that if selection-into-$D_i$ is non-degenerate on $U_{0,i}, U_{1,i}$, then $D_i$ is not ignorable.


\subsection{Monotonicity $\implies$ Selection Model, in a CM Setting.}
\label{appendix:mte-monotonicity}
\textit{Proof that (conditional) monotonicity implies a selection model representation in a CM setting.
    This proof is an applied example of the \cite{vytlacil2002independence} equivalence result, now including conditioning covariates $\vec X_i$, and is presented merely as a validation exercise.
}

Assume condition monotonicity \ref{mte:monotonicity} holds, for any treatment values $z < z$ and any covariate value $\vec X_i = \vec x$.
\[ \Probgiven{D_i(z) \geq D_i(z)}{\vec x} = 1. \]
For each value of $\vec{X}_i = \vec{x}$ and any treatment values $z < z$, we first define:
\begin{itemize}
    \item $\mathcal{A} = \{i : D_i(z) = D_i(z) = 1\}$, always-mediators
    \item $\mathcal{N} = \{i : D_i(z) = D_i(z) = 0\}$, never-mediators
    \item $\mathcal{C} = \{i : D_i(z) = 0, D_i(z) = 1\}$, mediator-compliers.
\end{itemize}
For any mediator complier $i \in \mathcal C$, partition the set as follows.
\begin{itemize}
    \item $\mathcal{Z}_1(i) = \{z : D_i(z) = 1\}$, treatment values where $i$ takes the mediator
    \item $\mathcal{Z}_0(i) = \{z : D_i(z) = 0\}$, treatment values where $i$ doesn't take the mediator.
\end{itemize}
Note that having binary $Z_i = 0,1$ reduces this to the simple case of $\mathcal{Z}_0(i) = \{0\}$, and $\mathcal{Z}_1(i) = \{1\}$.
The equivalence result holds for continuous values of $Z_i$, so continue with the more general $\mathcal{Z}_0(i), \mathcal{Z}_1(i)$ notation.

By monotonicity, we have
\[ \sup_{z \in \mathcal{Z}_0(i)} \pi(z;\vec{x})
    \leq \inf_{z \in \mathcal{Z}_1(i)} \pi(z;\vec{x}), \;\;
    \text{ for any } i \in \mathcal{C} \]
where $\pi(z;\vec{x}) = \Probgiven{D_i = 1}{Z_i = z, \vec{X}_i = \vec{x}}$ is the mediator propensity score.
A simple proof by contradiction verifies this statement (\citealt[Lemma 1]{vytlacil2002independence}).

Now we construct $V_i$ as follows:
\[ V_i = 
    \begin{cases}
        1, & \text{if } i \in \mathcal{N} \\
        0, & \text{if } i \in \mathcal{A} \\
        \inf_{z \in \mathcal{Z}_1(i)} \pi(z;\vec{x}), & \text{if } i \in \mathcal{C}.
    \end{cases} \]
Define $\psi(z;\vec{x}) = \pi(z;\vec{x})$.
Then we can represent $D_i(z)$ as a selection model,
\[ D_i(z) = \indicator{\psi(z;\vec{X}_i) \geq V_i}, \;\text{ for } z=0,1. \]
We can verify this works:
\begin{itemize}
    \item For $i \in \mathcal{A}$: $V_i = 0$ and $\psi(z;\vec{x}) \geq 0$ for all $z$, so $D_i(z) = 1$
    
    \item For $i \in \mathcal{N}$: $V_i = 1$ and $\psi(z;\vec{x}) \leq 1$ for all $z$, with $\psi(z;\vec{x}) < 1$ for $z \in \mathcal{Z}_0(i)$, so $D_i(z) = 0$ for $z \in \mathcal{Z}_0(i)$
    
    \item For $i \in \mathcal{C}$: $V_i = \inf_{z \in \mathcal{Z}_1(i)} \pi(z;\vec{x})$
    \begin{itemize}
        \item When $z \in \mathcal{Z}_1(i)$: $\psi(z;\vec{x}) \geq \inf_{z' \in \mathcal{Z}_1(i)} \pi(z';\vec{x}) = V_i$, so $D_i(z) = 1$
        \item When $z \in \mathcal{Z}_0(i)$: $\psi(z;\vec{x}) < \inf_{z' \in \mathcal{Z}_1(i)} \pi(z';\vec{x}) = V_i$, so $D_i(z) = 0$.
    \end{itemize}
\end{itemize}
Therefore, the construction $D_i(z) = \indicator{\psi(z;\vec{X}_i) \geq V_i}$ is a valid representation of the selection process under monotonicity.

This selection model can be transformed to one with a uniform distribution, to get the general selection model of \cite{heckman2005structural}.
Let $F_V \big( .\, \big|\, \vec{X}_i \big)$ be the conditional cumulative density function of $V_i$ given $\vec{X}_i$.
Define
\begin{align*}
    U_i &= F_V \left( V_i \mid \vec{X}_i \right) \\
    \pi(z; \vec{X}_i) 
        &= F_V\left( \psi(z; \vec{X}_i) \mid \vec{X}_i \right)
        = \Probgiven{D_i = 1}{Z_i = z, \vec{X}_i}
\end{align*}
We can then equivalently represent the mediator choice as the transformed selection model
\[ D_i(z) = \indicator{\pi(z; \vec{X}_i) \geq U_i}, \;\;\; \text{for } z=0,1 \]
where $U_i \,| \,\vec{X}_i \sim \text{Uniform}(0,1)$ by the probability integral transformation.

\subsection{Restricted MTE-CF Condition}
\label{appendix:mte-restricted}
Proposition~\ref{proposition:secondstage} shows that the second-stage regression is identified once the MTE-associated CF terms
$\rho_0\lambda_0\big(\pi(Z_i;\vec X_i)\big), \; \rho_1\lambda_1\big(\pi(Z_i;\vec X_i)\big)$ are identified.
Assumption~\ref{mte:instrument} supplies excluded variation in the mediator propensity
score, $\pi(z;\vec X_i)$.
However, if the mediator IV has finite support, this excluded variation does not identify
an unrestricted CF over the whole interval $p\in(0,1)$.
This subsection states the restricted-CF condition used in the main text.

The restricted-CF condition has two parts.
First, after observed heterogeneity has been absorbed by the potential-outcome means
$\mu_d(z;\vec X_i)$, the remaining dependence between mediator resistance $U_i$ and the
outcome errors $U_{d,i}$ is assumed to be common across $\vec X_i^-$.
For $d=0,1$, define
\[
    m_d(p;\vec x)
    =
    \Egiven{U_{d,i}}{U_i=p,\vec X_i=\vec x}.
\]
The separability restriction is
\[
    m_d(p;\vec x)
    =
    m_d(p),
    \qquad d=0,1.
    \label{eqn:mte-cf-separability}
\]
Equivalently, conditional on the observed components of the potential outcomes, the shape
of selection on unobserved mediator resistance is common across values of $\vec X_i^-$.
This is the analogue, in this CM setting, of the separable restricted-MTE structure used with
finite-support instruments in \cite{brinch2017beyond}.
It rules out unrestricted interactions between $\vec X_i^-$ and $U_i$ in the unobserved
component of the mediator MTE.

Fix the non-instrument controls at $\vec X_i^-=\vec x^-$, and define the set of distinct
propensity-score values generated by $Z_i$ and the mediator IV:
\[
    \mathcal P(\vec x^-)
    =
    \left\{
        \pi(z;x^{\textnormal{IV}},\vec x^-)
        :
        z\in\{0,1\},
        x^{\textnormal{IV}}
        \in
        \operatorname{supp}\left[
            \vec X_i^{\textnormal{IV}}\mid \vec X_i^-=\vec x^-
        \right]
    \right\}.
\]
\[\text{Let } \;\;\;\;
    K(\vec x^-)
    =
    \left|\mathcal P(\vec x^-)\right|. \]
$K(\vec x^-)$ counts the number of distinct values taken by the mediator propensity score $\pi(Z_i;\vec X_i)$, not the number of raw values taken by
$\vec X_i^{\textnormal{IV}}$.
For example, if there are no additional controls and both $Z_i$ and
$\vec X_i^{\textnormal{IV}}$ are binary, then $\pi(Z_i;\vec X_i)$ may take up to four distinct values.

The restricted-CF condition then assumes that, over the relevant support, the common CFs
lie in a finite-dimensional class. Specifically, for $d=0,1$,
\[
    \rho_d\lambda_d(p)
    =
    c_d
    +
    \sum_{\ell=1}^{K(\vec x^-)-1}
        a_{d\ell}\, b_\ell(p),
    \qquad
    p\in\mathcal P(\vec x^-),
    \label{eqn:restricted-cf-basis}
\]
where $b_1,\ldots,b_{K(\vec x^-)-1}$ are known basis functions.
The constant $c_d(\vec x^-)$ is absorbed by the corresponding intercept in the
second-stage outcome equation, so the excluded variation identifies the non-constant
components of the CF.
The associated rank condition is
\[
    \operatorname{rank}
    \begin{bmatrix}
        b_1(p_1) & \cdots & b_{K(\vec x^-)-1}(p_1) \\
        \vdots   & \ddots & \vdots \\
        b_1(p_K) & \cdots & b_{K(\vec x^-)-1}(p_K)
    \end{bmatrix}
    =
    K(\vec x^-)-1,
    \qquad
    \{p_1,\ldots,p_K\}=\mathcal P(\vec x^-).
    \label{eqn:restricted-cf-rank}
\]
This condition is the finite-support analogue of identifying the CFs with a continuous
instrument.
When there are only two relevant propensity-score values, the restriction leaves one
non-constant degree of freedom, equivalent to a linear CF over that support.
With additional propensity-score values, richer restricted-MTE specifications can be used,
following the logic of \cite{brinch2017beyond}.

The support required for the AIE is narrower than the whole unit interval.
By mediator monotonicity, individuals induced into the mediator by $Z_i$ are those with
\[
    \pi(0;\vec X_i)<U_i\leq \pi(1;\vec X_i).
\]
Thus, the AIE requires the CFs only at the propensity-score values that determine this
mediator-complier interval.
Equivalently, the relevant support is
\[
    \mathcal P_C
    =
    \left\{
        p:
        \pi(0;\vec x)\leq p\leq \pi(1;\vec x)
        \textnormal{ for some } \vec x
    \right\}.
    \label{eqn:mediator-complier-support}
\]
Identification at infinity is a sufficient special case, because it identifies the CFs for all
$p\in(0,1)$.
The weaker requirement used here is that the restricted-CF specification in
\eqref{eqn:restricted-cf-basis} identifies
\[
    \rho_0\lambda_0(p)
    \quad\textnormal{and}\quad
    \rho_1\lambda_1(p)
    \qquad
    \textnormal{for } p\in\mathcal P_C.
    \label{eqn:restricted-cf-complier-support}
\]

This is enough for Theorem~\ref{thm:mte-identification}.
The ADE uses the CF-adjusted second-stage parameters from
Proposition~\ref{proposition:secondstage}.
The AIE additionally uses the average unobserved gain among mediator compliers, which is
pinned down by the CF values at the endpoints $\pi(0;\vec X_i)$ and $\pi(1;\vec X_i)$:
\[
    \Gamma\big(\pi(0;\vec X_i),\pi(1;\vec X_i)\big)
    =
    \frac{
        \pi(1;\vec X_i)\lambda_1\big(\pi(1;\vec X_i)\big)
        -
        \pi(0;\vec X_i)\lambda_1\big(\pi(0;\vec X_i)\big)
    }{
        \pi(1;\vec X_i)-\pi(0;\vec X_i)
    }.
\]
Therefore the theorem does not require pointwise identification of the full mediator MTE
over $p\in(0,1)$.
It requires only the restricted CFs on the support relevant for mediator compliers.

\subsection{MTE Identification of the Second-stage}
\label{appendix:mte-secondstage}
\textit{Proof of Proposition \ref{proposition:secondstage}.
    This proof relies heavily on the notation and reasoning of \cite{kline2019heckits} for an IV setting.}

By Assumption \ref{mte:monotonicity} (mediator monotonicity), selection-into-mediator can be represented as a threshold-crossing selection model.
\[ D_i(z) = \indicator{ \pi(z; \vec X_i) \geq U_i }, \text{ for } z = 0, 1 \]
where $U_i = F_V\left( V_i \mid \vec{X}_i \right)$ follows a uniform distribution on $[0,1]$, and $\pi(z; \vec X_i) = \Egiven{D_i}{Z_i = z, \vec X_i}$ is the mediator propensity score.

The threshold crossing selection model represents individuals who refuse the mediator as follows:
\[ D_i = 0 \implies \pi(Z_i; \vec X_i) < U_i \]
Our objective is to determine $\Egiven{U_{0,i}}{D_i=0, Z_i, \vec X_i}$, which can then be written as
\[ \Egiven{U_{0,i}}{\pi(Z_i; \vec X_i) < U_i, Z_i, \vec X_i}. \]
Since $Z_i$ is ignorable, we have:
\[ \Egiven{U_{0,i}}{\pi(Z_i; \vec X_i) < U_i, Z_i, \vec X_i}
    = \Egiven{U_{0,i}}{\pi(Z_i; \vec X_i) < U_i} \]

Assumption \ref{mte:identification} has $\text{Cov}(U_i, U_{0,i}) \neq 0$.
This non-zero covariance implies statistical dependence between the selection error and outcome error.
This dependence allows us to represent $U_{0,i}$ using a linear projection.
We use $F_V^{-1}\left( U_i \mid \vec{X}_i \right)$ rather than $U_i$ directly in the projection to allow for flexibility in how the selection error affects outcomes.
The linear projection can be written as follows
\[ U_{0,i} = \rho_0 \big( F_V^{-1}\left( U_i \mid \vec{X}_i \right) - \mu_V \big) + \varepsilon_{0,i}, \]
where
\begin{itemize}
    \item $\mu_V = \E{F_V^{-1}\left( U_i \mid \vec{X}_i \right)}$ is the mean of $F_V^{-1}\left( U_i \mid \vec{X}_i \right)$
    \item $\rho_0 = \frac{\Cov{U_{0,i}, F_V^{-1}\left( U_i \mid \vec{X}_i \right)}}{\Var{F_V^{-1}\left( U_i \mid \vec{X}_i \right)}}$ is the projection coefficient
    \item $\varepsilon_{0,i}$ is a residual with $\Egiven{\varepsilon_{0,i}}{F_V^{-1}\left( U_i \mid \vec{X}_i \right)} = 0$.
\end{itemize}

The coefficient $\rho_0$ is the slope in the best linear predictor of $U_{0,i}$ given $F_V^{-1}\left( U_i \mid \vec{X}_i \right)$, and is chosen to ensure that the residual $\varepsilon_{0,i}$ is uncorrelated with $F_V^{-1}\left( U_i \mid \vec{X}_i \right)$.
This property is crucial for the identification strategy, as it isolates the component of $U_{i}$ that is related to selection-into-$D_i$.

The non-zero covariance condition in \ref{mte:identification} ensures $\rho_0 \neq 0$, so is relevant.
Since $U_i$ and $F_V^{-1}\left( U_i \mid \vec{X}_i \right)$ are related by a monotonic transformation (the inverse cumulative density function), the covariance $\text{Cov}(U_i, U_{0,i}) \neq 0$ implies $\text{Cov}(F_V^{-1}\left( U_i \mid \vec{X}_i \right), U_{0,i}) \neq 0$.

Given the linear projection of $U_{0,i}$ onto $F_V^{-1}\left( U_i \mid \vec{X}_i \right)$, we can compute the conditional expectation:
\[ \Egiven{U_{0,i}}{\pi(Z_i; \vec X_i) < U_i} = \Egiven{\rho_0\big(F_V^{-1}\left( U_i \mid \vec{X}_i \right) - \mu_V\big) + \varepsilon_{0,i}}{\pi(Z_i; \vec X_i) < U_i} \]

Since $\Egiven{\varepsilon_{0,i}}{F_V^{-1}\left( U_i \mid \vec{X}_i \right)} = 0$ by construction, and $U_i$ is a function of $F_V^{-1}\left( U_i \mid \vec{X}_i \right)$, we have
\[ \Egiven{\varepsilon_{0,i}}{\pi(Z_i; \vec X_i) < U_i} = 0. \]

Therefore:
\[ \Egiven{U_{0,i}}{\pi(Z_i; \vec X_i) < U_i} = \rho_0 \Egiven{F_V^{-1}\left( U_i \mid \vec{X}_i \right) - \mu_V}{\pi(Z_i; \vec X_i) < U_i}. \]

This gives us the control function representation:
\[ \Egiven{U_{0,i}}{D_i=0, Z_i, \vec X_i} = \rho_0 \lambda_0\big(\pi(Z_i; \vec X_i)\big) \]
where $\lambda_0 \left(p\right) = \Egiven{F_V^{-1}\left( U_i \mid \vec{X}_i \right) - \mu_V}{p < U_i}$.
The control function $\lambda_0\left(p\right)$ captures the expected value of the transformed selection term, conditional on being above the threshold $p \in (0,1)$.

The same sequence of steps for mediator takers, $D_i = 1$, gives the other mte:
\[ \Egiven{U_{1,i}}{D_i=1, Z_i, \vec X_i} = \rho_1 \lambda_1\big(\pi(Z_i; \vec X_i)\big), \]
where $\lambda_1 \left(p\right) = \Egiven{F_V^{-1}\left( U_i \mid \vec{X}_i \right) - \mu_V}{U_i \leq p}$ for $p \in (0,1)$, and $\rho_1 = \frac{\Cov{U_{1,i}, F_V^{-1}\left( U_i \mid \vec{X}_i \right)}}{\Var{F_V^{-1}\left( U_i \mid \vec{X}_i \right)}}$ is the corresponding projection coefficient.

The relationship between $\lambda_0(p)$ and $\lambda_1(p)$ can be derived as:
\[ \lambda_1 \left(p\right) = -\lambda_0 \left(p\right) \left( \frac{1-p}{p} \right), \text{ for } p \in (0,1). \]
This relationship ensures consistency in the CF approach across the $D_i=0$ and $D_i= 1$ groups \citep{kline2019heckits}.

Assumption \ref{mte:instrument} (mediator take-up cost instrument $\vec X_i^{\text{IV}}$) ensures identification of the propensity score function $\pi(z; \vec X_i)$ in the first stage by providing valid instrumental variation.
This variation allows us to identify the propensity score, and consequently the control functions $\lambda_0$ and $\lambda_1$.

Combining all elements, the conditional expectation of $Y_i$ given $Z_i, D_i, \vec X_i$ is
\begin{align*}
    \Egiven{Y_i}{Z_i, D_i, \vec X_i} \;\; =& \;\;
        \alpha
        + \beta D_i
        + \gamma Z_i
        + \delta Z_i D_i
        + \varphi(\vec X_i) \\
        & \;\; + \left( 1 - D_i \right) \Egiven{U_{0,i}}{D_i = 0}
            + D_i \Egiven{U_{1,i}}{D_i = 1}.
\end{align*}
Substitute the CFs,
\begin{align*}
    &\;(1-D_i) \Egiven{ U_{0,i} }{Z_i, D_i=0, \vec X_i}
        + D_i \Egiven{ U_{1,i}}{Z_i, D_i=1, \vec X_i} \\
    &=\; (1-D_i)\rho_0 \lambda_0 \big(\pi(Z_i; \vec X_i) \big)
    + D_i \rho_1 \lambda_1 \big(\pi(Z_i; \vec X_i) \big).
\end{align*}
This gives the final result,
\begin{align*}
    \Egiven{Y_i}{Z_i, D_i, \vec X_i} \;\; =& \;\;
        \alpha
        + \beta D_i
        + \gamma Z_i
        + \delta Z_i D_i
        + \varphi(\vec X_i) \\
        & \;\; +  \rho_0 \left( 1 - D_i \right) \lambda_0 \big( \pi(Z_i ; \vec X_i) \big)
            + \rho_1 D_i \lambda_1 \big( \pi(Z_i ; \vec X_i) \big).
\end{align*}
All parameters --- $\alpha, \beta, \gamma, \delta, \varphi(.), \rho_0, \rho_1$ --- are identified once we control for selection bias through the CFs $\lambda_0, \lambda_1$, with $\pi(z;\vec X_i)$ identified separately in the first-stage.
$\lambda_0, \lambda_1$ can be assumed to be certain functions (say, the inverse Mills ratio in \citealt{heckman1979sample}), or treated as semi-parametric parameters to be estimated --- at cost of the constant and $\rho_0, \rho_1$ no longer being separately identified from $\lambda_0, \lambda_1$, see \appendixref{appendix:semi-parametric}.

\subsection{MTE Identification of the ADE and AIE}
\label{appendix:mte-ade-aie}
\textit{Proof of Theorem \ref{thm:mte-identification}.}

Assume \ref{mte:monotonicity}, \ref{mte:identification}, \ref{mte:instrument} hold.
Then Proposition~\ref{proposition:secondstage} has $\alpha, \beta, \gamma, \delta, \varphi(.), \rho_0, \rho_1$ identified in a regression.
The following composes the ADE and AIE from these parameters.

For the ADE,
\begin{align*}
    \E{ \gamma + \delta D_i}
        &= \E{ \Big( \mu_0(1; \vec X_i) - \mu_0(0; \vec X_i) \Big)
            + D_i \Big( \mu_1(1; \vec X_i) - \mu_0(1; \vec X_i) - \big( \mu_1(0; \vec X_i) - \mu_0(0; \vec X_i) \big) \Big) } \\
    &= \E{ D_i \Big(\mu_1(1; \vec X_i) - \mu_1(0; \vec X_i) \Big)
        + (1 - D_i) \Big( \mu_0(1; \vec X_i) - \mu_0(0; \vec X_i) \Big) } \\
    &= \E{ D_i \Big( Y_i(1,1) - U_{1,i} - \big( Y_i(0,1) - U_{1,i} \big) \Big)
        + (1 - D_i) \Big( 
            Y_i(1,0) - U_{0,i} - \big( Y_i(0,0) - U_{0,i} \big) \Big) } \\
    &= \E{ D_i \Big( Y_i(1,1) - Y_i(0,1) \Big)
        + (1 - D_i) \Big( Y_i(1,0) - Y_i(0,0) \Big) } \\
    &= \E{ Y_i(1, D_i(Z_i)) - Y_i(0, D_i(Z_i))} \\
    &= \text{ADE}.
\end{align*}

Identification is similar for the AIE, but also involves the complier adjustment term.
\begin{align*}
    (\rho_1 - \rho_0) \, \Gamma \big(\pi(0; \vec{X}_i), \, \pi(1; \vec{X}_i) \big)
    &= (\rho_1 - \rho_0) \, \frac{\pi(1; \vec{X}_i)\lambda_1(\pi(1; \vec{X}_i)) - \pi(0; \vec{X}_i)\lambda_1(\pi(0; \vec{X}_i))}{\pi(1; \vec{X}_i) - \pi(0; \vec{X}_i)} \\
    &= (\rho_1 - \rho_0) \, \Egiven{ F_V^{-1}(U_i | \vec{X}_i) - \mu_V }{\pi(0; \vec{X}_i) < U_i \leq \pi(1; \vec{X}_i), \vec{X}_i} \\
    &= (\rho_1 - \rho_0) \, \Egiven{ F_V^{-1}(U_i | \vec{X}_i) - \mu_V }{
        D_i(0) = 0, D_i(1) = 1, \vec{X}_i} \\
    &= \Egiven{ \rho_1 \big( F_V^{-1}(U_i | \vec{X}_i) - \mu_V \big) }{
        D_i(0) = 0, D_i(1) = 1, \vec{X}_i} \\
    &\;\;\; - \Egiven{ \rho_0 \big( F_V^{-1}(U_i | \vec{X}_i) - \mu_V \big) }{
        D_i(0) = 0, D_i(1) = 1, \vec{X}_i} \\
    &= \Egiven{ U_{1,i} - U_{0,i} }{D_i(0) = 0, D_i(1) = 1, \vec{X}_i}.
\end{align*}
This complier adjustment was first presented for an IV setting by \cite{kline2019heckits}.

Collecting for the AIE,
\begin{align*}
    &\E{ \bar \pi \, \Big( \beta +  \delta Z_i +
        (\rho_1 - \rho_0) \Gamma \big(\pi(0; \vec X_i), \, \pi(1; \vec X_i) \big) \Big) } \\
    &= \E{ \bar \pi \, \Bigg( \Big( \mu_1(0; \vec X_i) - \mu_0(0; \vec X_i) \Big)
        +  Z_i \Big( \mu_1(1; \vec X_i) - \mu_0(1; \vec X_i)
            - \big(\mu_1(0; \vec X_i) - \mu_0(0; \vec X_i) \big) \Big) \Bigg) } \\
    &\;\;\;\; + \mathbb E \Big[ \bar \pi \; \Egiven{ U_{1,i} - U_{0,i} }{D_i(0) = 0, D_i(1) = 1, \vec{X}_i} \Big] \\
    &= \E{ \bar \pi \, \Bigg(
        Z_i \Big( \mu_1(1; \vec X_i) - \mu_0(1; \vec X_i) \Big)
        + (1 - Z_i) \Big( \mu_1(0; \vec X_i) - \mu_0(0; \vec X_i) \Big) \Bigg) } \\
    &\;\;\;\; + \mathbb E \Big[ \bar \pi \; \Egiven{ U_{1,i} - U_{0,i} }{D_i(0) = 0, D_i(1) = 1, \vec{X}_i} \Big] \\
    &= \E{ \bar \pi \, \Bigg(
        \mu_1(Z_i, \vec X_i) - \mu_0(Z_i, \vec X_i)
        + \Egiven{ U_{1,i} - U_{0,i} }{D_i(0) = 0, D_i(1) = 1, \vec{X}_i} \Bigg) } \\
    &= \E{ \bar \pi \; \Egiven{
        \mu_1(Z_i, \vec X_i) - \mu_0(Z_i, \vec X_i) + U_{1,i} - U_{0,i} }{
            D_i(0) = 0, D_i(1) = 1, \vec{X}_i} } \\
    &= \E{ \Egiven{D_i(1) - D_i(0)}{\vec X_i} \; \Egiven{
        Y_i(Z_i, 1) - Y_i(Z_i, 0) }{ D_i(0) = 0, D_i(1) = 1, \vec{X}_i} } \\
    &= \E{ \Egiven{Y_i(Z_i,D_i(1)) - Y_i(Z_i,D_i(0))}{\vec X_i} } \\
    &= \E{ Y_i(Z_i,D_i(1)) - Y_i(Z_i,D_i(0)) } \\
    &= \text{AIE}.
\end{align*}

\subsection{Semi-parametric Estimation of the AIE}
\label{appendix:semi-parametric}
It is difficult to directly use the CFs to compose estimates of the complier adjustment term, because various intercepts lose identification, but also because trusting semi-parametric estimates at individual points across the $\hat\lambda_0(p), \hat\lambda_1(p)$ functions would increase variation more than is necessary.

This can be avoided by noting the relation between the ATE and the conditional ADE and conditional AIE.
The following showing how to identify the AIE via relation to the ATE and conditional ADE, and omits the conditional on $\vec X_i$ for brevity. 

A simple algebraic rearrangement has the following (as first noted in \citealt[Section~3.1]{imai2010identification}),
\begin{align*}
    \text{ATE} &= \E{Y_i(1, D_i(1)) - Y_i(1, D_i(1))} \\
    &= \E{ Y_i(1, D_i(1)) - Y_i(0, D_i(1))}
    +  \E{ Y_i(0, D_i(1)) - Y_i(0, D_i(0))} \\
    &= \underbrace{\Egiven{ Y_i(1, D_i(Z_i)) - Y_i(0, D_i(Z_i))}{ Z_i = 1}}_{
        \text{ADE conditional on } Z_i = 1}
    +  \underbrace{\Egiven{ Y_i(Z_i, D_i(1)) - Y_i(Z_i, D_i(0))}{ Z_i = 0}}_{
        \text{AIE conditional on } Z_i = 0}.
\end{align*}
A similar re-arrangement also has the following,
\[ \text{ATE} 
    = \underbrace{\Egiven{ Y_i(Z_i, D_i(1)) - Y_i(Z_i, D_i(0))}{ Z_i = 1}}_{
        \text{AIE conditional on } Z_i = 1}
    + \underbrace{\Egiven{ Y_i(1, D_i(Z_i)) - Y_i(0, D_i(Z_i))}{ Z_i = 0}}_{
        \text{ADE conditional on } Z_i = 0}. \]

Reverting to the regression notation, to show how the ADE conditional on $Z_i$ is identified:
\begin{align*}
    \text{ADE} &= \E{ Y_i(1, D_i(Z_i)) - Y_i(0, D_i(Z_i))} \\
        &= \E{ \gamma + \delta D_i(Z_i) }. \\
    \implies \text{ADE conditional on } Z_i = 0
        &= \Egiven{ \gamma + \delta D_i(Z_i) }{Z_i = 0} \\
        &= \E{ \gamma + \delta D_i(0) }. \\
    \text{ADE conditional on } Z_i = 1
        &= \Egiven{ \gamma + \delta D_i(Z_i) }{Z_i = 1} \\
        &= \E{ \gamma + \delta D_i(1) }.
\end{align*}

Finally achieve identification of the AIE via the ATE and conditional ADE, as follows,
\begin{align*}
    \text{AIE}
    &= \Prob{Z_i = 0}
    \underbrace{\Egiven{ Y_i(Z_i, D_i(1)) - Y_i(Z_i, D_i(0))}{ Z_i = 0}}_{
        \text{AIE conditional on } Z_i = 0} \\
    &\;\;\;\;+ \Prob{Z_i = 1}
    \underbrace{\Egiven{ Y_i(Z_i, D_i(1)) - Y_i(Z_i, D_i(0))}{ Z_i = 1}}_{
        \text{AIE conditional on } Z_i = 1} \\
    &= \Prob{Z_i = 0} \Big[
        \text{ATE} - (\text{ADE conditional on } Z_i = 1) \Big] \\
    &\;\;\;\;+ \Prob{Z_i = 1} \Big[
        \text{ATE} - (\text{ADE conditional on } Z_i = 0) \Big] \\
    &= \text{ATE} - \Prob{Z_i = 0} \E{ \gamma + \delta D_i(1) }
        - \Prob{Z_i = 1} \E{ \gamma + \delta D_i(0) }.
\end{align*}

The semi-parametric AIE estimate then uses this representation, avoiding directly interacting with the estimated CFs, by plugging in estimates $\hat{\Pr}(Z_i = 1) = \bar Z$, $\hat{\text{ATE}}$, and the estimates from each side of the $D_i =0,1$ separated samples $\hat\gamma, \hat\delta$.
\[ \hat{\text{AIE}}^{\text{CF}}
    = \hat{\text{ATE}}
    - (1 - \bar Z) \, \left( \hat\gamma +
        \frac 1n \sum_{i = 1}^N \hat \delta \, \hat\pi(1; \vec X_i) \right)
    - \bar Z \, \left(\hat\gamma +
        \frac 1n \sum_{i = 1}^N \hat \delta \, \hat\pi(0; \vec X_i)  \right), \]
where $\frac 1n \sum_{i = 1}^N \hat \delta \, \hat\pi(0; \vec X_i)$ estimates $\E{\delta D_i(0)}$, and $\frac 1n \sum_{i = 1}^N \hat \delta \, \hat\pi(0; \vec X_i)$ estimates $\E{\delta D_i(1)}$.
Everything involved is a standard point estimate, so their composition will converge to a normal distribution, too.
Standard error computation can be achieved by a bootstrap procedure.

\subsection{Implementation and Further Simulation Evidence}
\label{appendix:implement}
A number of statistical packages, for the R language \citep{R2023}, made the simulation analysis for this paper possible.
\begin{itemize}
    \item \textit{Tidyverse} \citep{tidyverse} collected tools for data analysis in the R language.
    \item \textit{Mgcv} \citep{wood2016mgcv} allows semi-parametric estimation, using splines, in the R language.
    \item \textit{Mediate} \citep{tingley2014mediation} automates the sequential quasi-random assignment estimates of CM effects \citep{imai2010identification} in the R language.
\end{itemize}

\begin{figure}[h!]
    \caption{OLS versus MTE-based Estimates of CM Effects, varying $\Var{U_{1,i}}$ relative to $\Var{U_{0,i}} = 1$.}
    \begin{subfigure}[c]{0.475\textwidth}
        \centering
        \caption{ADE.}
        \includegraphics[width=\textwidth]{
            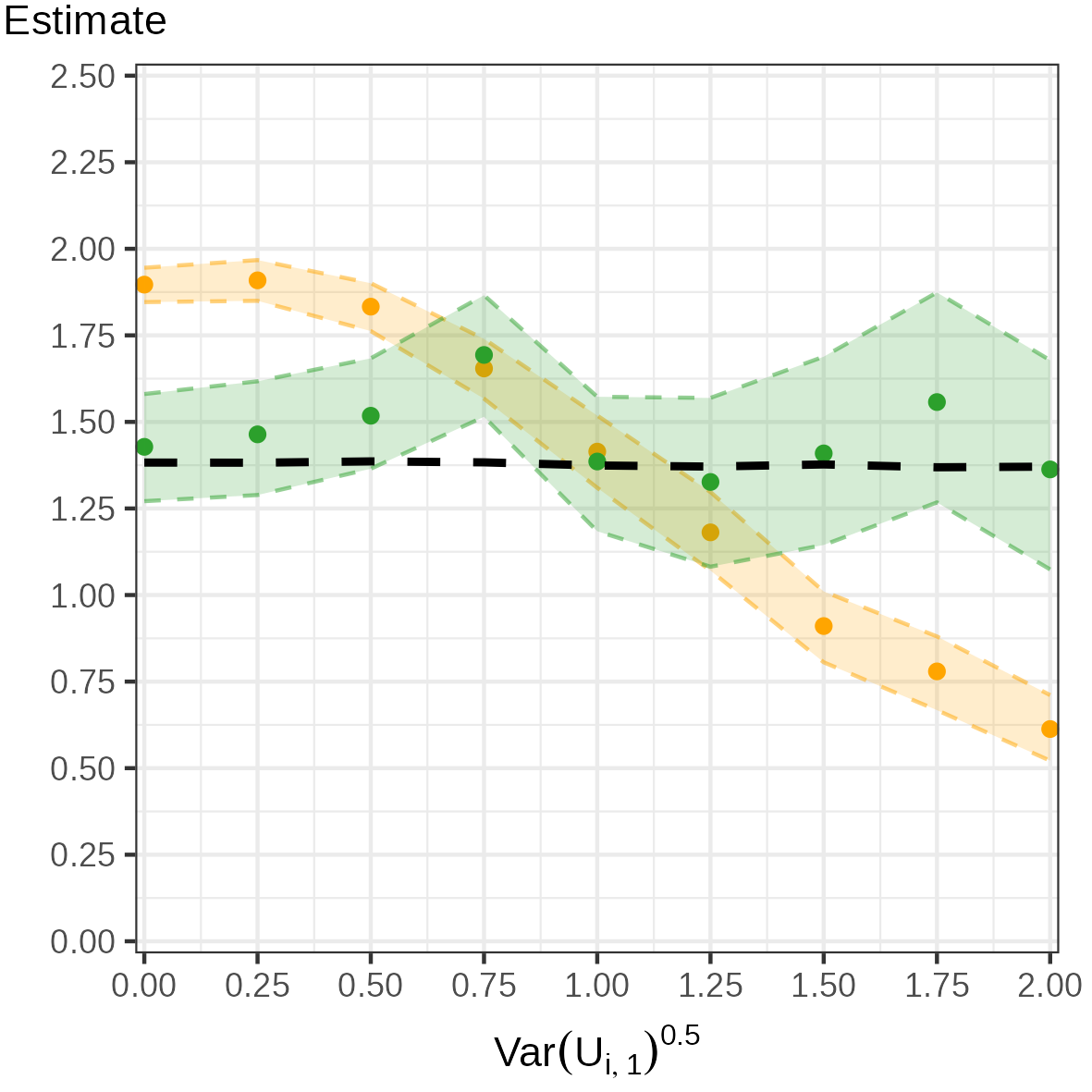}
    \end{subfigure}
    \begin{subfigure}[c]{0.475\textwidth}
        \centering
        \caption{AIE.}
        \includegraphics[width=\textwidth]{
            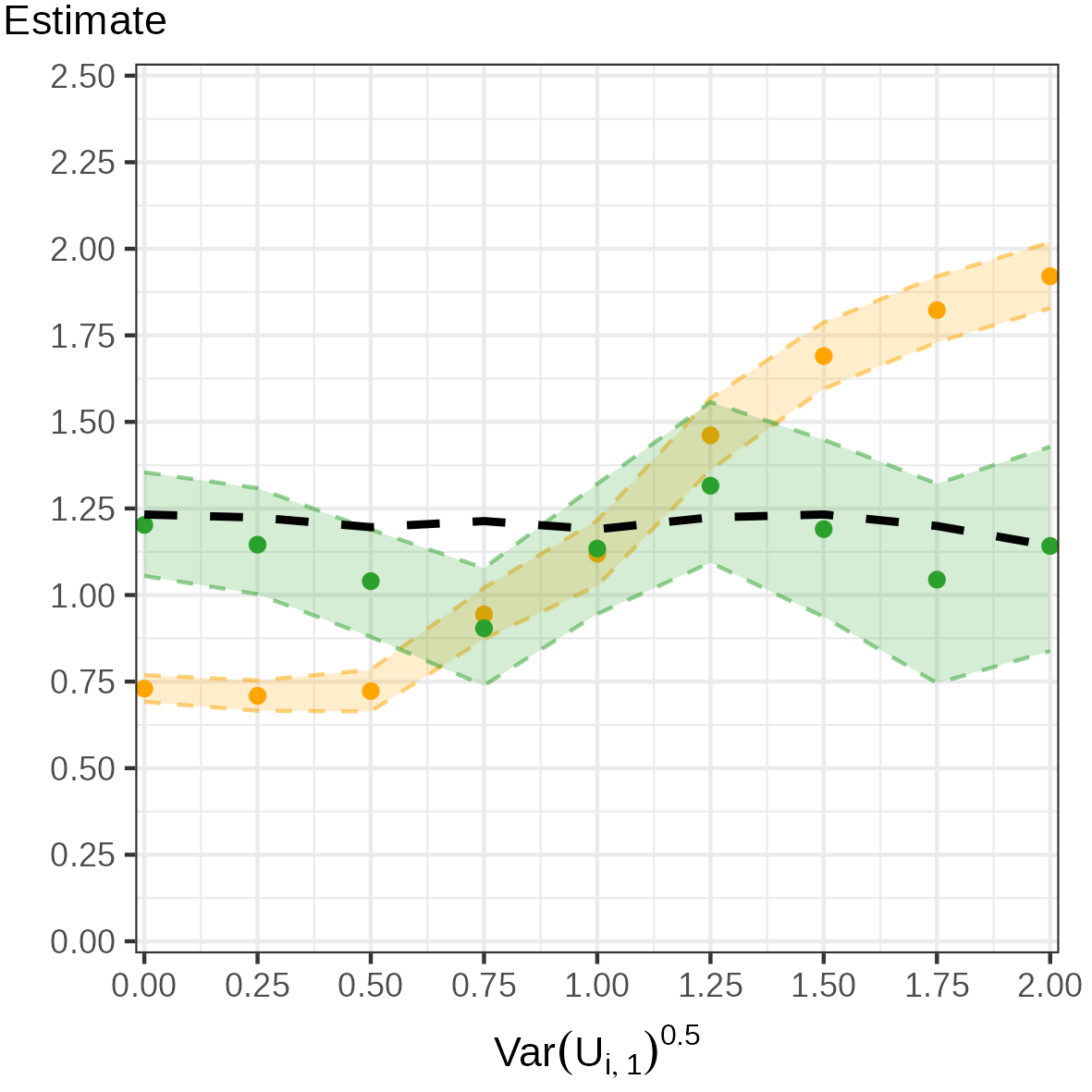}
    \end{subfigure}
    \label{fig:sigma-1-bias}
    \justify
    \footnotesize    
    \textbf{Note:}
    These figures show the OLS and MTE-based estimates of the ADE and AIE, for $n = 5,000$ sample size.
    The black dashed line is the true value, points are points estimates from data simulated with a given $\text{Corr}\big(U_{0,i}, U_{1,i}\big) =0.5$, $\Var{U_{0,i}} = 1$, and $\Var{U_{1,i}}^{\frac12}$ varied across $[0, 2]$.
    Shaded regions are the 95\% confidence intervals;
    orange are the OLS estimates, green the semi-parametric MTE-based approach.
\end{figure}

%

\end{document}